\documentclass[11pt]{article}
\usepackage[ansinew]{inputenc}
\usepackage[bbgreekl]{mathbbol}
\usepackage{amsmath,amssymb}
\usepackage{graphicx}
\usepackage{hyperref}
\usepackage{url}
\usepackage[numbers,square]{natbib}
\usepackage{multirow}
\usepackage{tikz}
\usepackage{tikz-cd}
\usetikzlibrary{arrows,matrix}
\usepackage{bbm}
\usepackage{authblk}
\usepackage{subcaption}

\usepackage{stmaryrd}
\usepackage{comment}
\usepackage{soul}

\usetikzlibrary{calc,patterns,shapes.geometric}
\usetikzlibrary{arrows,shapes,positioning}
\usetikzlibrary{decorations.markings,decorations.pathmorphing,decorations.pathreplacing}
\usetikzlibrary{decorations.text,decorations.shapes}

\usepackage[bbgreekl]{mathbbol}
\usepackage{amsmath,amssymb}
\usepackage{amsthm}  
\usepackage{bm}
\usepackage{bbm}
\usepackage{upgreek}

\def\bbb{\phantom{\big|} }

\def\mint{\int \mspace{-21.mu}\raise .8ex\hbox{\rotatebox{-80}{$|$}\,}}
\def\smint{\smallint \mspace{-13.mu}\raise -.1ex\hbox{\rotatebox{10}{--}}}

\def\bproof{\bigskip\noindent{\em Proof.}~}
\def\eproof{\hfill$\square$\nl}

\def\beq{\begin{equation}}
\def\eeq{\end{equation}}

\def\nl{\newline}

\def\<{\langle}
\def\>{\rangle}

\def\Chi{\raise .3ex \hbox{\large $\chi$}} 

\newcommand{\ii}{\ensuremath{\mathrm{i}}}

\def\dd{\, {\rm d}} 

\def\tb{\hbox{$\|\kern -.09em |$}}
\def\bigtb{\hbox{$\big\|\kern -.09em \big|$}}
\def\Bigtb{\hbox{$\Big\|\kern -.09em \Big|$}}

\def\px{\bX}  
\def\pv{\bV}  

\def\CC{\mathbbm{C}}

\def\II{\mathbbm{I}}

\def\MM{\mathbbm{M}}
\def\NN{\mathbbm{N}}

\def\QQ{\mathbbm{Q}}
\def\RR{\mathbbm{R}}

\def\ZZ{\mathbbm{Z}}

\newcommand{\mat}[1]{\mathbb{#1}}

\def\matC{\mathbb{C}}

\def\matJ{\mathbb{J}}

\def\matM{\mathbb{M}}

\def\matR{\mathbb{R}}
\def\matS{\mathbb{S}}

\def\matW{\mathbb{W}}

\def\matLambda{\mathbb{\Lambda}}

\newcommand{\arr}[1]{{\bm{\mathsf{#1}}}}

\newcommand\arrB{{\bm{\mathsf{B}}}}

\newcommand\arrE{{\bm{\mathsf{E}}}}
\newcommand\arrU{{\bm{\mathsf{U}}}}
\newcommand\arrV{{\bm{\mathsf{V}}}}

\newcommand\arrX{{\bm{\mathsf{X}}}}

\newcommand\sfk{{\sf k}}

\def\cA{\mathcal{A}}

\def\cC{\mathcal{C}}
\def\cD{\mathcal{D}}

\def\cF{\mathcal{F}}
\def\cG{\mathcal{G}}
\def\cH{\mathcal{H}}

\def\cO{\mathcal{O}}
\def\cP{\mathcal{P}}

\def\cV{\mathcal{V}}

\usepackage{mathrsfs}

\newcommand\bfx{\mathbf{x}}

\newcommand\uvec{{\be}}

\newcommand{\one}{\ensuremath{\mathbbm{1}}}

\def\Mf{\MM_{\rm f}}
\def\Mp{\MM_{\rm p}}
\def\Qp{\QQ_{\rm p}}

\def\be{{\bs e}}

\def\bk{{\bs k}}
\def\bsm{{\bs m}}

\def\br{{\bs r}}

\def\bv{{\bs v}}

\def\bx{{\bs x}}

\def\bB{{\bs B}}

\def\bE{{\bs E}}
\def\bF{{\bs F}}
\def\bG{{\bs G}}

\def\bJ{{\bs J}}

\def\bR{{\bs R}}

\def\bV{{\bs V}}

\def\bX{{\bs X}}

\def\bell{{\bs \ell}}

\def\bs{\boldsymbol}
\def\vp{{\varphi}}

\newcommand{\ee}{\ensuremath{\mathrm{e}}}

\def\Ampere{{Amp\`{e}re}}

\def\Dt{{\Delta t}}

\def\dt{{\partial_{t}}}

\DeclareMathOperator{\Span}{Span}
\DeclareMathOperator{\diag}{diag}
\DeclareMathOperator{\sinc}{sinc}

\DeclareMathOperator{\Div}{div}

\DeclareMathOperator{\curl}{curl}

\providecommand{\abs}[1]{\lvert#1\rvert}

\providecommand{\Bigabs}[1]{\Big|#1\Big|}

\providecommand{\norm}[1]{\lVert#1\rVert}

\providecommand{\range}[2]{\llbracket #1, #2 \rrbracket}

\newtheorem{theorem}{Theorem}
\newtheorem{lemma}{Lemma}
\newtheorem{remark}{Remark}
\newtheorem{definition}{Definition}
\newtheorem{assumption}{Assumption}

\title{On Geometric Fourier Particle In Cell Methods}

\author[1]{Martin Campos Pinto}
\author[1,2]{Jakob Ameres}
\author[1,2]{Katharina Kormann}
\author[1,2]{Eric Sonnendr\"ucker }
\affil[1]{Max-Planck-Institut f\"ur Plasmaphysik, Garching, Germany}
\affil[2]{Technische Universit\"at M\"unchen, Zentrum Mathematik, Garching, Germany}

\begin{document}

\maketitle

\begin{abstract}
  In this article we describe a unifying framework for variational electromagnetic particle schemes of spectral type,
  and we propose a novel spectral Particle-In-Cell (PIC) scheme that preserves a discrete Hamiltonian structure.
  Our work is based on a new abstract variational derivation of particle schemes 
  which builds on a de Rham complex where Low's Lagrangian is discretized using a 
  particle approximation of the distribution function.
  %
  In this framework, which extends the recent
  Finite Element based Geometric Electromagnetic PIC (GEMPIC) method
  to a wide variety of field solvers, the discretization of the electromagnetic potentials and fields is represented by a de Rham sequence of compatible spaces, and the particle-field coupling procedure is described by approximation operators that commute with the differential operators involved in the sequence.
  In particular, for spectral Maxwell solvers the choice of truncated $L^2$ projections using continuous Fourier transform coefficients for the commuting approximation operators yields the gridless
  Particle-in-Fourier method, 
  whereas spectral Particle-in-Cell methods are obtained by using discrete Fourier transform coefficients computed from a grid.
  By introducing a new sequence of spectral pseudo-differential approximation operators, we then obtain a novel variational
  spectral PIC method with discrete Hamiltonian structure that we call Fourier-GEMPIC. Fully discrete schemes
  are then derived using a Hamiltonian splitting procedure, leading to explicit time steps that preserve the Gauss laws
  and the discrete Poisson bracket associated with the Hamiltonian structure.
  These explicit steps are found to share many similarities
  with a standard spectral PIC method that appears as a Gauss and momentum-preserving variant of the variational method.
  As arbitrary filters are allowed in our framework, we also discuss aliasing errors and study a natural
  back-filtering procedure to mitigate the damping caused by anti-aliasing smoothing particle shapes.
\end{abstract}


\section{Introduction}

Representing the electromagnetic fields in truncated Fourier spaces has been a standard practice in plasma simulation,
from the early particle schemes \cite{Langdon.Birdsall.1970.pof, Hockney.Eastwood.1988.tf, Birdsall.Langdon.1991.iop}
to more recent parallel high-performance codes \cite{vlad2001gridless, decyk2011description, Godfrey.Vay.Haber.2014.jcp, ohana2016towards}.

In these spectral particle methods, two main approaches exist.
The simplest consists of coupling the particles to the fields via continuous Fourier transforms,
which leads to a gridless method sometimes called {\em Particle-in-Fourier} (PIF) \cite{ohana2016towards, ameres2018, Mitchell:2019}.
This method, which may include smoothing techniques through low pass filters or smoothing shape functions,
is naturally charge and energy preserving. More importantly,
it can be derived from a variational principle \cite{Evstatiev:2013, Shadwick:2014, webb2016spectral},
allowing for numerical schemes with very good stability on long time ranges.
Using the translational invariance of the discrete Fourier spaces, it has also
been shown to preserve momentum.

Another approach that is often preferred for simulations where many modes are needed,
such as turbulence in tokamak plasmas, is to use a grid for the coupling, and discrete Fourier transforms (DFT).
This leads to spectral Particle-in-Cell (PIC) methods
\cite{Langdon.1970.jcp, Langdon.1979.pof, decyk2011description, Godfrey.Vay.Haber.2014.jcp},
which we may call Fourier-PIC here.
Smooth (continuous) particle shapes are then necessary for the coupling to be well-defined,
which in Fourier space corresponds to a low-pass filtering. Using a DFT grid allows to localize
the coupling, but it also causes dispersion and aliasing errors
\cite{Langdon.1970.jcp,Langdon.1979.pof,Birdsall.Langdon.1991.iop},
which in turn have been shown to lead to numerical instabilities of various types,
including grid heating \cite{Birdsall.Maron.1980.jcp}, 
electrostatic finite grid instabilities \cite{huang2016finite} and
numerical Cherenkov instabilities \cite{Godfrey.1974.jcp, Xu.al.2013.cpc, Godfrey.Vay.Haber.2014.jcp}.
In most cases these methods are momentum preserving and in some cases they are also charge preserving,
see e.g.~\cite{Godfrey.Vay.Haber.2014.jcp}.
However, it seems that until very recently a proper derivation of variational spectral PIC methods was still missing.

This has been done in part in a new article \cite{CPKS_variational_2020} where a
flexible yet rigorous method is proposed to design variational,
gauge-free electromagnetic particle schemes with a Hamiltonian structure
relying on a non-canonical discrete Poisson bracket.
Formulated in a general framework with a minimal set of properties, this approach
essentially extends the recent Geometric Electromagnetic PIC (GEMPIC) scheme \cite{kraus2016gempic},
based on spline finite elements and point particles, to a wide range of field solvers and particle coupling techniques.
In particular, one application described in \cite{CPKS_variational_2020} consists
of a new Hamiltonian spectral PIC method, where the particle-field coupling is done
through a DFT grid. 
However, 
as this method relies on particle-field operators defined through geometric degrees of freedom, the resulting
deposition algorithms involve volume integrals of the shaped particles as they travel through space,
which somehow deviates from standard PIC scheme that are based on pointwise evaluations of the particle shape functions.


In this article we thus propose a novel variational spectral PIC method called Fourier-Gempic,
that has a discrete Hamiltonian structure and relies on particle-field coupling techniques
very similar to those encountered in standard PIC schemes.
Our method is obtained by applying the abstract derivation of \cite{CPKS_variational_2020}
to truncated Fourier spaces, combined with novel particle-field coupling operators of pseudo-differential
type. By observing that our variational derivation combined with continuous ($L^2$-orthogonal) projection operators
leads to the gridless PIF method, we show that this framework actually unifies the formulation of
variational spectral particle methods,
where the standard PIC scheme appears as a momentum-conserving variant, albeit non-variational.

By applying a Hamiltonian splitting technique in the spirit of \cite{Crouseilles:2015,He:2015} we
are able to propose fully discrete schemes for the three different methods, with explicit time steps
that provide an additional insight into their differences and similarities.
For the Hamiltonian GEMPIF and Fourier-GEMPIC methods, these fully discrete schemes preserve the total
energy within the time splitting error as guaranteed by backward error analysis \cite{HairerLubichWanner:2006}.
For the PIF and standard PIC methods, they preserve the total momentum.
All of them preserve the Gauss laws to machine accuracy.

In addition to a Hamiltonian structure which guarantees very good stability properties on long time ranges,
the Fourier-Gempic method has the ability to include various shape functions and filter coefficients in Fourier space.
In particular, we show that a natural back-filtering mechanism can be associated to the
usual low-pass filtering effect of high order spline shapes in order to strongly reduce
aliasing errors inherent in the DFT, without damping relevant modes in the computational range.
\newline
\newline
The outline is as follows. In Section~\ref{sec:FP} we recall the two main coupling approaches
for Fourier-particle methods, namely Particle-in-Fourier (PIF) and Fourier-PIC,
we discuss aliasing errors and consider a simple back-filtering technique to mitigate the smoothing
effect of anti-aliasing splines.
In Section~\ref{sec:var} we then present the general form of a variational spectral scheme
as derived in \cite{CPKS_variational_2020}, as well as that of the momentum-conserving variant,
and apply it to two particular sets of particle-field approximation operators:
using $L^2$ projections which correspond to continuous Fourier coefficients,
this leads to the gridless PIF method, which coincides with its momentum-conserving variant.
Using a novel class of pseudo-differential DFT operators, we obtain a new Fourier-PIC method, whose
Hamiltonian structure is guaranteed by the commuting de Rham diagram property of the new DFT
approximation operators and the analysis from \cite{CPKS_variational_2020}.
In Section~\ref{sec:fulldis} we then describe fully discrete schemes of arbitrary orders,
obtained by applying a Hamiltonian splitting procedure.
Both in the gridless and PIC cases, we provide explicit formulas for the different steps of the discrete schemes.
The standard Fourier-PIC coupling is then found to coincide with the momentum-conserving variant
of the Hamiltonian Fourier-GEMPIC method, while the latter differs in the computation of the pushing fields.
In Section~\ref{sec:num} we assess the basic numerical properties of the different methods.
Their accuracy is compared on standard test cases, as well as their long-time stability and conservation properties.
It is shown that all Hamiltonian methods are very stable in energy and momentum, and that back-filtered methods are very accurate
for approximating the fundamental growth and damping rates, even for low-resolution runs.

\section{Spectral particle methods, with or without a grid}
\label{sec:FP}

Electromagnetic particle models formally consist of Maxwell's equations
for the fields, 
\begin{align}
\dt \bE(\bx, t) - \curl \bB(\bx,t) &= - \bJ_N(\bx,t)  \label{Ampere}
\\
\dt \bB(\bx, t) + \curl \bE(\bx,t) &= 0 \label{Faraday}
\\
\Div \bE(\bx,t) &= \rho_N(\bx,t) \label{GaussE}
\\
\Div \bB(\bx,t) &= 0, \label{GaussM}
\end{align}
coupled with discrete particles with positions $\px_p$ and velocities $\pv_p$, $p = 1, \dots, N$,
subject to Lorentz force trajectories 
\begin{equation} \label{traj}
\frac{\dd}{\dd t} \px_p(t) = \pv_p(t),
\qquad
\frac{\dd}{\dd t} \pv_p(t) =  \frac{q_p}{m_p}\big( \bE(\px_p(t),t) + \pv_p(t) \times \bB(\px_p(t), t)   \big).
\end{equation}
Here $q_p$ and $m_p$ represent the charge and mass of the $p$-th numerical particle, and the
current and charge densities, which are the sources in Maxwell's equations,
are obtained by summing each particle contribution,
\begin{equation} \label{rhoJ_N}
  \rho_N(t,\bx) = \sum_{p = 1 \cdots N} q_p \delta(\bx-\px_p(t))
  \quad \text{ and } \quad
  \bJ_N(t,\bx) = \sum_{p=1 \cdots N} q_p \pv_p(t) \delta(\bx-\px_p(t)).
\end{equation}
In the limit $N \to \infty$ of infinitely many particles,
this model approximates the kinetic Vlasov equation \cite{Victory.Allen.1991.sinum}
where the plasma is represented by a continuous phase-space density function $f(t,\bx,\bv)$,
and the choice of Dirac densities in \eqref{rhoJ_N} 
corresponds to a pointwise evaluation of the continuous charge and current densities
$\rho(t,\bx) = \int q f(t,\bx,\bv)\dd \bv$, $\bJ(t,\bx) = \int q \bv f(t,\bx,\bv)\dd \bv$.

In spectral solvers, electromagnetic fields are represented by truncated
Fourier expansions of the form
\begin{equation} \label{EB_K}
\left\{
\begin{aligned}
  &\bE_K(t,\bx) =
  \sum_{\bk \in \range{-K}{K}^3}
  \bE_{\bk}(t) \ee^\frac{2\ii \pi \bk \cdot \bx}{L} 
  \\
  &\bB_K(t,\bx) =   \sum_{\bk \in \range{-K}{K}^3}
    \bB_{\bk}(t) \ee^\frac{2\ii \pi \bk \cdot \bx}{L} 
\end{aligned}
\right.
\qquad \text{ for } \bx \in [0,L]^3,
\end{equation}
here with $2K+1$ modes per dimension, and the sources need to be properly represented
in the same truncated Fourier spaces.
The discrete Maxwell's equations then take the form 
\begin{equation} \label{max}
\left\{
\begin{aligned}
  - &\dt \bE_\bk + \Big(\frac{2\ii\pi \bk}{L}\Big) \times \bB_\bk = \bJ^S_\bk,
  \\
  &\dt \bB_\bk + \Big(\frac{2\ii\pi \bk}{L}\Big)\times \bE_\bk = 0
\end{aligned}
\right.
\qquad \text{ for } \qquad \bk \in \range{-K}{K}^3
\end{equation}
and the trajectory equations for the particles read
\begin{equation} \label{trajS}
\left\{
\begin{aligned}
  &\tfrac{\dd}{\dd t} \px_p = \pv_p,
  \\
  &\tfrac{\dd}{\dd t} \pv_p = \frac{q_p}{m_p}({\bE^S}(\px_p) + \pv_p \times {\bB^S}(\px_p))
  \end{aligned}
\right.
\qquad \text{ for } \qquad p \in \range{1}{N}.
\end{equation}
In \eqref{max}--\eqref{trajS} we have denoted the coupling terms by
\begin{itemize}
  \item $\bJ^S_\bk$ the Fourier coefficients of the current density seen by the discrete field,
  \item $\bE^S$ and $\bB^S$, the electromagnetic field seen by the particles.
\end{itemize}
These coupling terms involve a shape function $S$ which is primarily used to define an auxiliary
particle current density
\begin{equation} \label{JSN}
  \bJ^S_N(\bx) := \bJ_N * S (\bx) = \sum_{p=1}^N q_p \pv_p S(\bx-\px_p)
\end{equation}
from which the Fourier coefficients $\bJ^S_\bk$ are derived.
This shape $S$ may either be the Dirac mass or some smooth approximation of it such as a B-spline,
in which case $\bJ^S_N$ corresponds to a convolution smoothing of $\bJ_N = \bJ_N^\delta$.

How the coupling terms are precisely defined will then characterize the numerical
method at this semi-discrete level. Spectral particle methods can essentially be divided in two classes:
gridless {\em Particle-in-Fourier} methods where the particles are directly
coupled to the fields, and {\em Particle-in-Cell} methods that use an
intermediate grid and a Discrete Fourier Transform (DFT) to localize the coupling steps.
In the remainder of this section we recall the main features of these methods
and discuss aliasing errors and anti-aliasing techniques.

\subsection{Gridless coupling: the Particle-In-Fourier (PIF) approach}
\label{sec:pif}

The simplest option consists of a gridless coupling as
in e.g.~\cite{Langdon.Birdsall.1970.pof, vlad2001gridless, decyk2011description,
Evstatiev:2013},
which leads to a method sometimes called {\em Particle-in-Fourier} \cite{ohana2016towards, ameres2018, Mitchell:2019}.
Here, the coupling current terms $\bJ^S_\bk$ are simply obtained as the Fourier coefficients of $\bJ^S_N$.
Letting
\begin{equation} \label{CFC}
\cF_\bk(G) := \Big(\frac 1L\Big)^3 \int_{[0,L]^3} G(\bx) \ee^{-\frac{2\ii\pi \bk\cdot \bx}{L}} \dd \bx 
\end{equation}
denote the $\bk$-th (continuous) Fourier coefficient of an arbitrary function $G$, this gives
\begin{equation} \label{Jk_pif}
  {\bJ_\bk^S} := \cF_\bk(\bJ^S_N)
      = \Big(\frac{1}{L}\Big)^3 \sum_{p = 1 \cdots N} q_p \pv_p \int_{[0,L^3]} S(\bx-\px_p)\ee^{-\frac{2\ii\pi \bk\cdot \bx}{L}} \dd \bx.
\end{equation}
The coupling fields are then defined by continuous convolution products
\begin{equation} \label{EBS_pif}
  {\bE^S} (\bx) = \int_{[0,L]^3} \bE_K(\tilde \bx) S(\tilde \bx-\bx)\dd \tilde \bx,
  \qquad 
  {\bB^S} (\bx) = \int_{[0,L]^3} \bB_K(\tilde \bx) S(\tilde \bx-\bx)\dd \tilde \bx
\end{equation}
evaluated at the particle positions $\bx = \px_p$.
It will sometimes be convenient to rewrite this gridless coupling in terms of the Fourier coefficients
of the function $S_{\px_p}(\bx) = S(\bx-\px_p)$, as
\begin{equation} \label{Jk_pif2}
  {\bJ_\bk^S} = \sum_{p = 1 \cdots N} q_p \pv_p \cF_\bk(S_{\px_p})
\end{equation}
and
\begin{equation} \label{EBS_pif2}
  {\bE^S} (\px_p) = L^3 \sum_{\bk \in \range{-K}{K}^3} \bE_\bk \overline{\cF_\bk(S_{\px_p})},
  \qquad 
  {\bB^S} (\px_p) = L^3 \sum_{\bk \in \range{-K}{K}^3} \bB_\bk \overline{\cF_\bk(S_{\px_p})},
\end{equation}
where $\overline{\cF}$ denotes the complex conjugate of $\cF$.
We note that here a Dirac shape can be used, $S = \delta$,
since Fourier coefficients are well defined for Dirac distributions \cite{Gasquet.Witomski.1999},
$$
\cF_\bk(\delta_{\px_p}) = \Big(\frac 1L\Big)^3 \ee^{-\frac{2\ii\pi \bk\cdot \px_p}{L}}
$$
and this is indeed a standard option in gridless methods \cite{ohana2016towards, ameres2018}.
Here we keep the possibility of using arbitrary shapes, for the sake of generality and clarity of exposition.

In \cite{Evstatiev:2013, Shadwick:2014} it was shown that this semi-discrete system can be derived from a discrete variational principle,
and that it preserves exactly the charge (namely the Gauss laws), as well as the total energy and momentum of the system.
However, the coupling is global: every particle contributes directly to every Fourier mode, and vice versa.
For problems involving a large number of Fourier modes, this leads to a computational complexity of $\cO(NK^3)$
which is prohibitive for simulations using a large number $N$ of particles.

\subsection{Coupling with a DFT grid: the spectral or Fourier-PIC approach}
\label{sec:f-pic}

A standard approach \cite{Langdon.1970.jcp, Langdon.1979.pof, Hockney.Eastwood.1988.tf, Birdsall.Langdon.1991.iop}
consists of using an intermediate grid with $M$ points per dimension, $M \ge 2K+1$, and
discrete Fourier transforms (DFT), which is more efficient for simulations where a large number of Fourier modes are needed. This approach is sometimes referred to as
spectral or pseudo-spectral PIC \cite{decyk2011description, Godfrey.Vay.Haber.2014.jcp}.
Denoting by
\begin{equation} \label{DFC}
\cF_{M,\bk}(G) := \Big(\frac 1M\Big)^3 \sum_{\bsm \in \range{1}{M}^3} G(\bsm h) \ee^{-\frac{2\ii\pi \bk\cdot \bsm}{M}}
\qquad \text{with} \qquad
h = \frac LM,
\end{equation}
the {\em discrete} Fourier coefficients associated with this grid, 
the current source is then defined as
\begin{equation} \label{Jk_pic}
  \bJ^S_\bk := \cF_{M,\bk}(\bJ_N^S) 
\end{equation}
with a smoothed current density given again by $\bJ_N^S = \bJ_N * S$, see \eqref{JSN}.
In practice this amounts to first depositing this current on the grid as in a standard Particle-in-Cell method,
$$
\bJ^{\rm pic}_\bsm := \bJ^S_N(\bsm h) = \sum_{p = 1 \cdots N} q_p \pv_p S(\bsm h -\px_p) \qquad \text{ for } ~ \bsm \in \range{1}{M}^3
$$
and then performing a 
DFT,
$$
\bJ^S_\bk = \Big(\frac 1 M \Big)^3 \sum_{\bsm \in \range{1}{M}^3} \bJ^{\rm pic}_\bsm \ee^{-\frac{2\ii\pi \bk\cdot \bsm}{M}}
 \qquad \text{ for } ~ \bk \in \range{-K}{K}^3.
$$
The pushing fields are then defined by a discrete convolution,
\begin{equation} \label{EBS_pic}
  \bE^S(\bx) := h^3 \!\!\!\sum_{\bsm \in \range{1}{M}^3}\!\! \bE_K(\bsm h) S(\bsm h -\bx),
  \qquad 
  \bB^S(\bx) := h^3 \!\!\!\sum_{\bsm \in \range{1}{M}^3}\!\! \bB_K(\bsm h) S(\bsm h -\bx)
\end{equation}
also evaluated at the particle positions $\bx = \px_p$. In practice the steps are similar, in a transposed order:
the field values \eqref{EB_K} are first computed on the grid, which corresponds to an inverse DFT
$$ 
\left\{
\begin{aligned}
&\bE^{\rm pic}_\bsm := \bE_K(\bsm h) = \sum_{\bk \in \range{-K}{K}^3} \bE_{\bk} \ee^\frac{2\ii \pi \bk \cdot \bsm}{M}
\\
&\bB^{\rm pic}_\bsm := \bB_K(\bsm h) = \sum_{\bk \in \range{-K}{K}^3} \bB_{\bk} \ee^\frac{2\ii \pi \bk \cdot \bsm}{M}
\end{aligned}
\right.
\qquad \text{ for } ~ \bsm \in \range{1}{M}^3,
$$ 
and they are gathered on the particles with the shape function $S$, as in a standard PIC method
$$ 
\left\{
\begin{aligned}
&\bE^S(\px_p) = h^3 \!\!\sum_{\bsm \in \range{1}{M}^3} \bE^{\rm pic}_{\bsm} S(\bsm h -\px_p)
\\
&\bB^S(\px_p) = h^3 \!\!\sum_{\bsm \in \range{1}{M}^3} \bB^{\rm pic}_{\bsm} S(\bsm h -\px_p).
\end{aligned}
\right.
$$ 
We note that these coupling terms can be rewritten in a form similar to
\eqref{Jk_pif2}--\eqref{EBS_pif2}, now with the discrete Fourier coefficients
of the function $S_{\px_p}(\bx) = S(\bx-\px_p)$. Indeed we have
\begin{equation} \label{Jk_pic2}
  {\bJ_\bk^S} = \sum_{p = 1 \cdots N} q_p \pv_p \cF_{M,\bk}(S_{\px_p})
\end{equation}
and
\begin{equation} \label{EBS_pic2}
  {\bE^S} (\px_p) = L^3 \sum_{\bk \in \range{-K}{K}^3} \bE_\bk \overline{\cF_{M,\bk}(S_{\px_p})},
  \qquad 
  {\bB^S} (\px_p) = L^3 \sum_{\bk \in \range{-K}{K}^3} \bB_\bk \overline{\cF_{M,\bk}(S_{\px_p})}.
\end{equation}
For these terms to be well-defined, we see that $S$ must now be at least continuous.
A common choice is to take (the periodic extension of) a tensor-product B-spline of degree $\kappa \ge 1$ scaled to the grid,
\begin{equation} \label{B-per}
  S(\bx) := \sum_{\br \in \ZZ^3} S^h_\kappa(\bx+ \br L)
  \qquad \text{ where } \qquad
  S^h_\kappa(\bx) := \Big(\frac 1h\Big)^3 \prod_{\alpha \in \range{1}{3}} \hat S_\kappa\Big(\frac{x_\alpha}{h} \Big),
\end{equation}
with cardinal univariate B-splines 
defined on the reference grid as
$\hat S_0(x) := \one_{[-\frac{1}{2}, \frac{1}{2}]}(x)$
and
$$
\hat S_\kappa(x) := \hat S_0 * \hat S_{\kappa-1}(x) = \int_{-\frac 12}^{\frac 12} \hat S_{\kappa-1}(x-y) \dd y
 \qquad \text{for} \qquad \kappa \ge 1. $$
As these shape functions have localized supports,
$
{\rm supp}(S) = \left[ - h \big(\frac{\kappa+1}{2}\big),  h \big(\frac{\kappa+1}{2}\big)\right],
$
the particles only interact with the $(\kappa+1)^3$ neighbouring grid nodes, which makes the deposition/gathering steps local.
Moreover if $M$ is a power of two, then the DFTs can be efficiently performed with an FFT algorithm,
leading to a computational complexity of $\cO(N (\kappa+1)^3) + \cO(M^3 \log(M))$ that is more affordable in
simulations involving a large number of particles and Fourier modes.
In \cite{Mitchell:2019} it is observed that combining DFT couplings with the
filtering method of \cite{Steidl:1998io} provides an efficient approximation of the gridless method,
corresponding to a {\em nonequispaced fast Fourier transform} \cite{Plonka:2018ec}.
This technique will be revisited in Section~\ref{sec:bf} as a natural back-filtering method.
More generally, we note that Fourier filtering is commonly used in modern PIC codes in order to reduce
the statistical noise inherent to particle approximations, see e.g.
\cite{McMillan:2010fn, HKKMBBS_2019_jpp}.

Fourier-PIC coupling often leads to momentum-preserving schemes.
In some cases they have been shown to be also charge-preserving,
see e.g. \cite{Godfrey.Vay.Haber.2014.jcp} where the DFT current deposition is
seen as a spectral adaptation of the classical Esirkepov method \cite{esirkepov2001exact}.
However, it does not preserve the energy and in general it cannot be derived from a variational principle.

\subsection{Shape filtering and grid aliasing in Fourier space}
\label{sec:Sf}

As is well known (see e.g.~\cite[Sec.~8-7]{Birdsall.Langdon.1991.iop}),
smooth particle shapes have a low-pass filtering effect in Fourier space.
This is most easily seen in the gridless case,
where the coupling fields \eqref{Jk_pif}--\eqref{EBS_pif} defined by continuous convolution products satisfy
\begin{equation} \label{J_Sk_pif}
  \bJ^S_\bk
    = \sigma_\bk \cF_\bk(\bJ_N) 
    \qquad \text{ for } \quad \bk \in \range{-K}{K}^3
\end{equation}
 and
\begin{equation} \label{EB_Sk_pif}
  \left\{
  \begin{aligned}
    \cF_\bk(\bE^S)  
      = \overline{\sigma_\bk} \bE_\bk
    \\
    \cF_\bk(\bB^S) 
      = \overline{\sigma_\bk} \bB_\bk
   \end{aligned} \right.
   \qquad \text{ for } \quad \bk \in \ZZ^3
 \end{equation}
where we have denoted  $\sigma_\bk := L^3 \cF_\bk(S)$ and set $\bE_\bk := \bB_\bk := 0$
for $\abs{\bk}_\infty > K$. Notice that $\overline{\sigma_\bk} = \sigma_\bk$ for symmetric shapes.
As smoother functions are associated with faster decreasing spectra,
we can clearly see the filtering effect of smooth shapes.
Specifically, with the Dirac shape $S = \delta$ we have $\sigma_\bk = 1$ for all~$\bk$, hence no filtering.
With a B-spline \eqref{B-per} of degree $\kappa \in \NN$ and scale $h = L/M$, we have
\begin{equation} \label{sigmak}
\sigma_\bk = L^3 \cF_\bk(S) = \prod_{\alpha = 1}^3 \left( \sinc\Big(\frac{\pi k_\alpha}{\mu(2K+1)}\Big)\right)^{\kappa+1}
\qquad \text{ with } \qquad \mu := \frac{M}{2K+1} \ge 1.
\end{equation}
Here $\sinc(\theta) := \frac 1\theta \sin \theta $ and $\mu$ is the {\em oversampling parameter}.
Since $\abs{\frac{\pi k_\alpha}{\mu(2K+1)}} < \frac{\pi}{2\mu} \le \frac \pi 2$ for $\bk \in \range{-K}{K}^3$,
this makes explicit how modes $\abs{\bk}_\infty \approx K$ are damped
with ``smoother'' splines, namely higher degrees and coarser grids.

In the Fourier-PIC case a similar filtering effect can be observed, but an additional phenomenon enters into play.
Indeed the use of a grid leads to the well-known {\em aliasing} effect \cite{Blackman.Tukey.1958.dover}, an $M$-periodization
of the discrete Fourier coefficients through the superposition of high frequency modes, i.e.
$$
\cF_{M,\bk}(G) = \sum_{\bell \in \ZZ^3} \cF_{\bk+\bell M}(G) \qquad \text{ for any } ~  G \in \cC^0_{\rm per}.
$$
Applying this equality to $\bJ^S_N$ and using again that $\cF_\bk(\bJ^S_N) = \sigma_\bk \cF_\bk(\bJ_N)$,
we find that in the Fourier-PIC case the modes of the coupling current \eqref{Jk_pic} read
\begin{equation} \label{J_Sk_pic}
  \bJ^S_\bk 
  = \sum_{\bell \in \ZZ^3} \sigma_{\bk+\bell M} \cF_{\bk+\bell M}(\bJ_N)
    \qquad \text{ for } \quad \bk \in \range{-K}{K}^3.
\end{equation}
Here the aliases are the modes corresponding to $\bell \neq 0$, all outside the main
range $\range{-K}{K}^3$, since $M \ge 2K+1$.
For the pushing fields \eqref{EBS_pic} the discrete convolution leads to a dual aliasing phenomenon, of the form
\begin{equation} \label{EB_Sk_pic}
  \left\{
  \begin{aligned}
    \cF_\bk(\bE^S)
      = \overline{\sigma_\bk} \sum_{\bell \in \ZZ^3} \bE_{\bk + \bell M}
    \\
    \cF_\bk(\bB^S)
      = \overline{\sigma_\bk} \sum_{\bell \in \ZZ^3} \bB_{\bk + \bell M}
   \end{aligned} \right.
   \qquad \text{ for } \quad \bk \in \ZZ^3.
\end{equation}
Again, the aliases consist of the $\bell \neq 0$ terms, which now
correspond to the main modes of $\bE_K$ and $\bB_K$ contributing to
{\em higher} frequencies of the coupling field. 
Indeed, using that $\bE_{\bk} = 0$ for $\bk \notin \range{-K}{K}^3$
we may rewrite \eqref{EB_Sk_pic} as
\begin{equation} \label{EB_Sk_pic_2}
  \cF_{\bk'}(\bE^{S}) = \begin{cases}
   \overline{\sigma_{\bk + \bell M}} \bE_{\bk}
   \qquad &\text{ for } \bk' = \bk + \bell M \in \range{-K}{K}^3 + M\ZZ^3 
   \\
   0
   \qquad &\text{ for } \bk' \notin \range{-K}{K}^3 + M\ZZ^3
 \end{cases}
\end{equation}
and similarly for $\bB^{S}$.

The repercussions of aliasing in numerical simulations 
have been studied since the early days of computational plasma modelling, either
through linearized dispersion analysis or fully nonlinear studies
\cite{Langdon.1970.jcp,Okuda.1972.jcp,Godfrey.1974.jcp,Langdon.1979.pof}. 
By introducing spurious modes which can then be coupled in the nonlinear models,
aliasing has often been recognized as the source of many issues in the simulations,
including grid heating \cite{Birdsall.Maron.1980.jcp}, 
finite grid instabilities \cite{huang2016finite} and
numerical Cherenkov instabilities \cite{Xu.al.2013.cpc, Godfrey.Vay.Haber.2014.jcp}.

\subsection{Anti-aliasing and back-filtering}
\label{sec:bf}

It is clear from \eqref{J_Sk_pic}--\eqref{EB_Sk_pic_2} that
smooth shapes with a low-pass filtering effect, such as B-splines, may be used for anti-aliasing purposes.
However, by 
filtering also some frequencies within the
computational range $\range{-K}{K}^3$, they can lead to the {\em overdamping} of relevant modes,
in particular for low-resolution discretizations.

In order to mitigate the aliasing errors and thus reduce the associated instabilities,
a successful approach has consisted in associating the anti-aliasing properties of smooth spline shapes
with additional ad-hoc filters.
In \cite{Godfrey.Vay.Haber.2014.jcp} for instance, the authors show that many instabilities
can be strongly reduced by using filters determined so as to reduce specific growing modes in the dispersion relations.
And in \cite{Mitchell:2019}, the particular DFT coupling that is proposed to reduce aliasing
is based on the nonequispaced fast Fourier transforms (NFFT) \cite{Steidl:1998io,Plonka:2018ec} which
precisely involves filter coefficients that match the low-pass filter effect of the smoothing splines.

Here we propose to interpret these filtering techniques as an effective {\em back-filtering} method.
Indeed, it is easily seen from \eqref{J_Sk_pic} that a simple solution for the overdamping issue consists
of dividing each deposited current mode with the corresponding shape filter coefficient, leading to a new current defined as
\begin{equation} \label{Jk_pic_bf}
  \bJ^{S,{\rm bf}}_\bk := \frac{1}{\sigma_\bk}\cF_{M,\bk}(\bJ_N^S). 
\end{equation}
For the pushing fields the idea is the same but we see from \eqref{EB_Sk_pic} that the back-filtering
needs to be applied on the original field, rather than on the coupling terms. This leads to setting
\begin{equation} \label{EBS_pic_bf}
    \bE^{S,{\rm bf}}(\bx) := h^3 \!\!\!\!\sum_{\bsm \in \range{1}{M}^3}\!\! \bE_K^{\rm bf}(\bsm h) S(\bsm h -\bx),
    \qquad 
    \bB^{S,{\rm bf}}(\bx) := h^3 \!\!\!\!\sum_{\bsm \in \range{1}{M}^3}\!\! \bB_K^{\rm bf}(\bsm h) S(\bsm h -\bx)
\end{equation}
with back-filtered fields defined as
\begin{equation} \label{EB_bf}
  \left\{
  \begin{aligned}
    &\bE_K^{\rm bf}(\bx) :=
    \sum_{\bk \in \range{-K}{K}^3}
     \overline{\Big(\frac{1}{\sigma_\bk}\Big)}
     \bE_{\bk} \ee^\frac{2\ii \pi \bk \cdot \bx}{L}
    \\
    &\bB_K^{\rm bf}(\bx) :=
    \sum_{\bk \in \range{-K}{K}^3}
      \overline{\Big(\frac{1}{\sigma_\bk}\Big)}
      \bB_{\bk} \ee^\frac{2\ii \pi \bk \cdot \bx}{L}.
  \end{aligned}
  \right.
\end{equation}
With this coupling, formulas~\eqref{J_Sk_pic}--\eqref{EB_Sk_pic} become
\begin{equation} \label{J_Sk_pic_bf}
  \bJ^{S,{\rm bf}}_\bk
  = \cF_{\bk}(\bJ_N)
    + \sum_{\bell \neq 0} \frac{\sigma_{\bk+\bell M}}{\sigma_{\bk}} \cF_{\bk+\bell M}(\bJ_N)
    \qquad \text{ for } \quad \bk \in \range{-K}{K}^3
\end{equation}
and
\begin{equation} \label{EB_Sk_pic_bf}
  \left\{
  \begin{aligned}
    \cF_\bk(\bE^{S,{\rm bf}})
      = \bE_{\bk} + \sum_{\bell \neq 0} \overline{\Big(\frac{\sigma_\bk}{\sigma_{\bk + \bell M}}\Big)} \bE_{\bk + \bell M}
    \\
    \cF_\bk(\bB^{S,{\rm bf}})
      = \bB_{\bk} + \sum_{\bell \neq 0} \overline{\Big(\frac{\sigma_\bk}{\sigma_{\bk + \bell M}}\Big)} \bB_{\bk + \bell M}
   \end{aligned} \right.
   \qquad \text{ for } \quad \bk \in \ZZ^3
\end{equation}
where we have separated each contribution into its main mode ($\bell = 0$) and the filtered aliases.
And again, using explicitly that $\bE_{\bk} = 0$ for $\bk \notin \range{-K}{K}^3$
we can rewrite \eqref{EB_Sk_pic_bf} as
\begin{equation} \label{EB_Sk_pic_bf2}
  \cF_{\bk'}(\bE^{S,{\rm bf}}) = \begin{cases}
  \overline{\Big(\frac{\sigma_{\bk + \bell M}}{\sigma_{\bk}}\Big)}
   \bE_{\bk}
   \qquad &\text{ if } \bk' = \bk + \bell M \in \range{-K}{K}^3 + M\ZZ^3 
   \\
   0
   \qquad &\text{ if } \bk' \notin \range{-K}{K}^3 + M\ZZ^3
 \end{cases}
\end{equation}
and similarly for $\bB^{S,{\rm bf}}$. For B-splines we can see from \eqref{sigmak} that $\sigma_\bk$ is far from 0
in the range $\bk \in \range{-K}{K}^3$. Thus, back-filtering allows to reduce
the amplitude of all the aliased modes in a similar proportion as with a standard filtering, but without
damping any mode in the computational range.

\section{Variational spectral particle discretizations}
\label{sec:var}

In this section we follow the variational structure-preserving discretization framework of
\cite{CPKS_variational_2020} and apply it to discrete Fourier spaces. This essentially allows
us to extend the Finite Element spline GEMPIC method from \cite{kraus2016gempic}
to spectral Maxwell solvers.

\subsection{Structure-preserving particle-field discretizations} 

We remind that a central feature of this framework is to preserve the
de Rham sequence of $\RR^3$,
\begin{equation}\label{CGL}
H^1(\RR^3)
\xrightarrow{ \mbox{$~ \nabla ~$} }
H(\curl;\RR^3)
\xrightarrow{ \mbox{$~ \curl ~$}}
H(\Div;\RR^3)
\xrightarrow{ \mbox{$~ \Div ~$}}
L^2(\RR^3)
\end{equation}
at the discrete level, and to admit a sequence of projection operators
$\Pi^0, \dots, \Pi^3$ mapping infinite-dimensional function spaces into the discrete ones,
such that the following diagram commutes:
\begin{equation}\label{CD}
\begin{tikzpicture}[baseline=(current  bounding  box.center)]
  \matrix (m) [matrix of math nodes,row sep=3em,column sep=4em,minimum width=2em] {
    \bbb V^0 \bbb
      & \bbb V^1 \bbb
          & \bbb V^2 \bbb%
              & \bbb V^3 \bbb
    \\
           V^0_K & V^1_K & V^2_K & V^3_K
    \\
    };
  \path[-stealth]
    (m-1-1) edge node [left]  {$\Pi^0$} (m-2-1)
    (m-1-1) edge node [above] {$\nabla$} (m-1-2)
    (m-1-2) edge node [left]  {$\Pi^1$} (m-2-2)
    (m-2-1) edge node [above] {$\nabla$} (m-2-2)
    (m-1-3) edge node [left]  {$\Pi^2$} (m-2-3)
    (m-1-4) edge node [left]  {$\Pi^3$} (m-2-4)
    (m-1-2) edge node [above] {$\curl$} (m-1-3)
    (m-2-2) edge node [above] {$\curl$} (m-2-3)
    (m-1-3) edge node [above] {$\Div$}  (m-1-4)
    (m-2-3) edge node [above] {$\Div$}  (m-2-4)
    ;
\end{tikzpicture}
\end{equation}
We point out that such commuting de Rham diagrams are a key tool in
Finite Element Exterior Calculus (FEEC), see e.g. \cite{bossavit1998computational, hiptmair2002finite,%
Arnold.Falk.Winther.2010.bams,%
buffa2010isogeometric, campos_pinto_compatible_2017_I}.
In our framework, it is these operators $\Pi^\ell$, together with the shape functions $S$,
that will encode the coupling mechanism between the particles and the discrete fields.
The bottom row thus consists of truncated Fourier spaces
$$
\begin{aligned}
  &V^0_K := V^3_K := \Span\big(\{ \Lambda_\bk : \bk \in \range{-K}{K}^3 \}\big)
  \\ 
  &V^1_K := V^2_K := \Span\big(\{\Lambda_\bk \uvec_\alpha : \bk \in \range{-K}{K}^3, ~  \alpha \in \range{1}{3} \}\big)
\end{aligned}
$$
where $\uvec_\alpha$ is the unit basis vector in the $\alpha$ dimension, and
$$
\Lambda_\bk(\bx) := \ee^\frac{2\ii \pi \bk \cdot \bx}{L}, \qquad
\bx \in [0,L]^3
$$
is the Fourier basis function of index $\bk \in \ZZ^3$.
Using the fact that the discrete spaces $V^1_K$ and $V^2_K$ coincide in this spectral
setting, we find that the weak discrete differential operators associated
with the sequence \eqref{CD} coincide with the strong (continuous) ones.
As a consequence, the variational method derived in \cite{CPKS_variational_2020}
takes the following form: the field equations read
\begin{equation} \label{gfp-solve}
  \left\{
  \begin{aligned}
    - &\dt \bE_K + \curl \bB_K = \Pi^2 \bJ^S_N
    \\
    &\dt \bB_K + \curl \bE_K = 0
  \end{aligned}
  \right.
  \qquad  \text{ with } \qquad
    \Pi^2 \bJ^S_N = \sum_{p = 1 \cdots N} q_p \Pi^2( \pv_p S_{\px_p})
  \end{equation}
  with $\bB_K$, $\bE_K$ in $V^1_K = V^2_K$, and the particle trajectories read
\begin{equation} \label{gfp-push}
\left\{
\begin{aligned}
  &\tfrac{\dd}{\dd t} \px_p = \pv_p
  \\
  &\tfrac{\dd}{\dd t} \pv_p = \frac{q_p}{m_p} \big(\bE^S(\px_p) + \pv_p \times \bB^S(\px_p)\big)
\end{aligned}
\right.
\quad  \text{ with } \quad
\left\{
\begin{aligned}
  &E^S_{\alpha} (\px_p) = \int E_{\alpha}(\bx) (\Pi^2_\alpha S_{\px_p})(\bx) \dd \bx
\\
  &B^S_{\alpha} (\px_p) = \int B_{\alpha}(\bx) (\Pi^1_\alpha S_{\px_p})(\bx) \dd \bx
\end{aligned}
\right.
\end{equation}
for $p \in \range{1}{N}$ and $\alpha \in \range{1}{3}$,
with $E_\alpha$ and $B_\alpha$ the $\alpha$-components of $\bE_K$ and $\bB_K$.

As for the operators $\Pi^\ell$, several choices can then be made that lead to a commuting diagram
and each choice results in a different coupling mechanism between the particles and the fields.
In \cite{CPKS_variational_2020} a set of projections was presented that is based on
interpolation and hispolation, which in practice amounts to performing discrete Fourier transforms (DFT)
on a grid with $M = 2K+1$ nodes but also involves surface and volume integrals of the particle shapes.
In this article we will study two types of approximation operators:
$L^2$ projection operators corresponding to continuous Fourier transforms,
as recalled in Section~\ref{sec:L2},
and new pseudo-differential operators based on discrete Fourier transforms, that we present in Section~\ref{sec:psd}.
These choices will then respectively lead to two variational schemes with Hamiltonian structure, namely
the gridless Particle-in-Fourier method, and a new spectral Fourier-GEMPIC method.

\begin{remark} \label{rem:noproj}
Although we often follow the common usage and refer to $\Pi^\ell$ as commuting projection operators,
we emphasize that 
we do not require them to be actual {\em projections}
in the sense that one would have $\Pi^\ell = I$ on $V^\ell_K$.
Indeed this property is not needed for the Hamiltonian structure of
the resulting schemes \cite{CPKS_variational_2020},
and by relaxing it we can directly extend our analysis to coupling methods
that involve filtering or back-filtering mechanisms.
\end{remark}

\subsection{Gauss and momentum-preserving variant}
\label{sec:mom-gauss-variant}

In \cite{CPKS_variational_2020}, a variant of the abstract system
\eqref{gfp-solve}--\eqref{gfp-push} was also proposed, that is a priori not Hamiltonian but
preserves both the Gauss laws and a discrete momentum.
This modified system involves the same operators from the general
commuting diagram~\eqref{CD}, and some interior products coupled
with a dimension-dependent approximation operator $\cA_{h,\alpha}$.
Specifically, the momentum-preserving variant consists of the same field
solver~\eqref{gfp-solve} as above, and of a modified particle pusher where
the Lorentz term $\bF^S(\px_p,\pv_p) = \bE^S(\px_p) + \pv_p \times \bB^S(\px_p)$
from \eqref{gfp-push} is replaced by
$$
\tilde \bF^S(\px_p,\pv_p) := \tilde \bE^S(\px_p) + \tilde \bR^S(\px_p,\pv_p)
$$
with components given by
\begin{equation} \label{momentum_push}
  \left\{
  \begin{aligned}
    &\tilde E^S_\alpha(\px_p) = \int E_{\alpha}(\bx) (\cA_{h,\alpha}\Pi^3 S_{\px_p})(\bx) \dd \bx
    \\
    &\tilde R^S_\alpha(\px_p,\pv_p)
      = \int \bB_K(\bx) \cdot \big(
          \cA_{h,\alpha} (\uvec_\alpha \times \Pi^2 (\pv_p S_{\px_p}) \big)(\bx) \dd \bx.
  \end{aligned}
  \right.
\end{equation}
In the general construction of \cite{CPKS_variational_2020} the operator
$\cA_{h,\alpha}$ was defined as a directional averaging on a grid with $M$ points, $h = L/M$,
which in Fourier space amounts to a diagonal filtering of the form
$ (\cA_{h,\alpha} G)_\bk = \sinc\big(\frac{2\pi k_{\alpha}}{M}\big) G_\bk $.
This allows for the method to be well posed in general polynomial or spline finite element settings.
With spectral solvers however, this averaging is not needed and one may simply
take the identity operator, $ (\cA_{h,\alpha} G)_\bk = G_\bk $.
We may then rewrite these pushing fields in terms of Fourier coefficients, for a clearer
comparison. The fields in the Hamiltonian pusher~\eqref{gfp-push} take the form
\begin{equation}
  \label{EBS-k}
  \left\{
  \begin{aligned}
    &E^S_\alpha(\px_p) = L^3 \sum_{\bk} E_{\alpha,\bk} \overline{(\Pi^2_\alpha S_{\px_p})_\bk}
    \\
    &R^S_\alpha(\px_p,\pv_p)
      = \sum_{\nu = \pm 1} \nu V_{p,\alpha+\nu} B_{\alpha-\nu}(\px_p)
      = L^3 \sum_{\bk}
         \sum_{\nu = \pm 1} \nu B_{\alpha-\nu,\bk} V_{p,\alpha+\nu}\overline{(\Pi^1_{\alpha-\nu} S_{\px_p})_\bk}
      , 
  \end{aligned}
  \right.
\end{equation}
whereas the momentum-preserving ones read 
\begin{equation}
  \label{EBS_mom-k}
\left\{
\begin{aligned}
  &\tilde E^S_\alpha(\px_p) = L^3 \sum_{\bk} 
    E_{\alpha,\bk} \overline{(\Pi^3 S_{\px_p})_\bk}
  \\
  &\tilde R^S_\alpha(\px_p,\pv_p)
    = L^3 \sum_{\bk} 
       \sum_{\nu = \pm 1} \nu B_{\alpha-\nu,\bk} V_{p,\alpha+\nu}\overline{(\Pi^2_{\alpha+\nu} S_{\px_p})_\bk}
    .
\end{aligned}
\right.
\end{equation}
In particular, we observe that the two schemes differ in the particle-field coupling operators
involved in the pushing fields.
After performing a convenient time discretization, we will see in Section~\ref{sec:f-pic-steps}
that this latter formulation will result in fully discrete PIC schemes with standard
particle-field coupling terms, as described in Section~\ref{sec:f-pic}.

\subsection{Semi-discrete conservation properties}

In \cite{CPKS_variational_2020} we have shown that the 
semi-discrete equations~\eqref{gfp-solve}--\eqref{gfp-push}
have a discrete Hamiltonian structure, which in particular implies that they preserve
the total energy and any discrete Casimir functional.
In the particular setting of truncated Fourier spaces, a few basic conservation properties
can be proven with a direct argument.

\begin{theorem}  \label{th:cons}
  The variational spectral particle method \eqref{gfp-solve}--\eqref{gfp-push} preserves the strong Gauss laws
  \begin{equation} \label{gfp-Gauss}
    \left\{
    \begin{aligned}
      &\Div \bE_K = \Pi^3 \rho^S_N
      \\
      &\Div \bB_K = 0
    \end{aligned}
    \right.
  \end{equation}
  as well as the discrete energy
  \begin{equation} \label{H}
    \cH(t) = \frac 12 \sum_{p = 1}^N m_p \abs{\pv_p(t)}^2 + \frac 12 \int_{[0,L]^3} \Big(\abs{\bE_K(t,\bx)}^2 + \abs{\bB_K(t,\bx)}^2\Big) \dd \bx.
  \end{equation}
  If in addition the projection operators satisfy
  \begin{equation} \label{allPi}
  \Pi^1_\alpha = \Pi^2_\alpha = \Pi^3 \qquad \text{ for } \quad 1 \le \alpha \le 3,
  \end{equation}
  then it also preserves the discrete momentum
  \begin{equation} \label{P}
  \cP(t) = \sum_{p = 1}^N m_p \pv_p(t) + \int_{[0,L]^3} \bE_K(t,\bx) \times \bB_K(t,\bx) \dd \bx.
  \end{equation}
\end{theorem}

\bproof
The preservation of the magnetic Gauss law readily follows from the strong Faraday equation
in \eqref{gfp-solve}.
As for the electric Gauss law, we compute for an arbitrary smooth function $\vp$
$$
\frac{\dd}{\dd t} \int \rho^S_N(t,\bx) \vp(\bx) \dd \bx
  = \sum_{p = 1}^N q_p \int S(\tilde \bx) \pv_p \cdot \nabla \vp(\tilde \bx+\px_p) \dd \tilde \bx
  = \int \bJ^S_N(t,\bx) \cdot \nabla \vp(\bx) \dd \bx
$$
which shows that the continuity equation
\begin{equation} \label{cont}
  \dt \rho^S_N + \Div \bJ^S_N = 0
\end{equation}
always holds in distribution's sense.
Taking next the divergence of the discrete \Ampere{} equation in \eqref{gfp-solve}
the commuting diagram property \eqref{CD} 
allows us to write
$$
\dt \Div \bE_K = -\Div \Pi^2 \bJ^S_N = -\Pi^3 \Div \bJ^S_N = \dt \Pi^3 \rho^S_N
$$
where the last equality follows from \eqref{cont} and from the time-invariance of the
operator $\Pi^3$.
Integrating over time this shows that the electric Gauss law is indeed preserved.
To show the energy conservation, we next compute using \eqref{gfp-solve}
$$
\frac{\dd}{\dd t} \Big(\frac 12 \int \abs{\bE_K}^2 + \abs{\bB_K}^2 \Big)
  =  \int \bE_K \cdot (\nabla \times \bB_K - \Pi^2 \bJ^S_N) - \bB_K \cdot \nabla \times \bE_K
  =  - \int \bE_K \cdot \Pi^2 \bJ^S_N
$$
and, using \eqref{gfp-push},
$$ 
\frac{\dd}{\dd t} \Big( \frac 12 \sum_{p = 1}^N m_p \abs{\pv_p}^2 \Big)
  = \sum_{p = 1}^N q_p \int \Pi^2(\pv_p S_{\px_p}) \cdot \bE_K
  =  \int \bE_K \cdot \Pi^2 \bJ^S_N,
$$ 
so that $\cH$ is indeed constant over time.
We finally turn to the momentum conservation and assume that \eqref{allPi} holds.
Then the coupling fields take the form
$$
\bG^S(\px_p) = \int \bG_K(\bx) (\Pi^3 S_{\px_p})(\bx) \dd \bx \qquad \text{ with } \quad \bG = \bE \text{ or } \bB,
$$
and the deposited current reads
$
\Pi^2 \bJ^S_N = \sum_p q_p \Pi^2( \pv_p S_{\px_p}) = \sum_p q_p \pv_p \Pi^3(S_{\px_p}).
$
Using \eqref{gfp-push} we compute
$$
\frac{\dd }{\dd t} \sum_{p = 1}^N m_p \pv_p
  = \sum_{p = 1}^N q_p \int \big(\bE_K \Pi^3 S_{\px_p} + \pv_p \times \bB_K \Pi^3 S_{\px_p}\big)
  = \int \bE_K \Pi^3 \rho^S_N + \int (\Pi^2 \bJ^S_N) \times \bB_K,
$$
and using \eqref{gfp-solve} together with the identity
$
\int_\Omega \bG \times (\curl \bG)
  = \int_\Omega \nabla (\tfrac 12 \bG^2) - (\bG \cdot \nabla)  \bG
  = \int_\Omega (\Div \bG) \bG
$
valid for an arbitrary function $\bG$, allows us to compute
$$
 \begin{aligned}
 \frac{\dd }{\dd t} \int_\Omega \bE_K \times \bB_K
 &= \int_\Omega (\curl \bB_K - \Pi^2 \bJ^S_N) \times \bB_K - \int_\Omega \bE_K \times (\curl \bE_K)
 \\
 &= - \int_\Omega (\Div \bB_K) \, \bB_K - \int_\Omega \Pi^2 \bJ^S_N \times \bB_K - \int_\Omega (\Div \bE_K) \, \bE_K
 \\
&= - \int_\Omega (\Pi^2 \bJ^S_N)\times \bB_K - \int_\Omega (\Pi^3 \rho^S_K) \, \bE_K
\end{aligned}
$$
where in the last equality we have used the Gauss laws \eqref{gfp-Gauss}. The result follows.
\eproof

\subsection{Commuting projections using continuous Fourier transforms}
\label{sec:L2}

Because truncated Fourier spaces have the particular property that they are stable under space differentiation,
$L^2$ projection operators can be used for the commuting diagram. This choice essentially corresponds to a
gridless Particle-in-Fourier coupling described above, and in this article we will refer to the resulting method
as a {\em Geometric Electromagnetic Particle-in-Fourier} ({GEMPIF}) method, to emphasize its natural expression
in the general GEMPIC framework.
In order to account for general filtering and back-filtering mechanisms, we consider
an arbitrary collection of Hermitian coefficients $\gamma_\bk = \overline{\gamma_{-\bk}} \in \CC$,
and set
\begin{equation} \label{Pi_CFT}
\Pi^0 := \Pi^1_\alpha := \Pi^2_\alpha := \Pi^3 
\qquad \text{ for }  \alpha \in \range{1}{3},
\end{equation}
where $\Pi^3$ is the operator that maps a function to its $\gamma$-filtered Fourier series of rank $K$,
\begin{equation} \label{Pi3_CFT}
  \Pi^3 G := \sum_{\bk \in \range{-K}{K}^3} \gamma_\bk \cF_\bk(G) \Lambda_\bk,
\end{equation}
with $\cF_\bk$ the continuous Fourier coefficient operator, see \eqref{CFC}.
We observe that with unit filters $\gamma_\bk = 1$
the operators in \eqref{Pi_CFT} coincide with the $L^2$ projection $P_K$
on $V^3 = V^0$, 
characterized by
$$
\int_{[0,L]^3} P_K G(\bx) \overline{\Lambda_\bk(\bx)} \dd \bx
= \int_{[0,L]^3} G(\bx) \overline{\Lambda_\bk(\bx)} \dd \bx \qquad \text{ for } ~ \bk \in \range{-K}{K}^3.
$$
For general filter coefficients we have the following result.

\begin{lemma}
The filtered $L^2$ projection operators $\Pi^\ell$ defined by \eqref{Pi_CFT}--\eqref{Pi3_CFT}
satisfy the commuting diagram property
\eqref{CD} 
in a distribution sense,
with periodic distributions as domain spaces:
$$
V^0 = V^3 = \cD'_{\rm per}
\qquad \text{ and } \qquad
V^1 = V^2 = (\cD'_{\rm per})^3.
$$
\end{lemma}

\bproof
We may consider that $\gamma_\bk = 1$, as the general case follows easily.
The $L^2$ projections over truncated Fourier spaces classically extend
to periodic distributions \cite{Gasquet.Witomski.1999}, writing e.g.
$$
\cF_\bk\big((\partial_\alpha)^{a} \delta_{\px_p}\big)
= \frac{(-1)^a}{L^3} \int_{[0,L]^3} \delta_{\px_p} (\partial_\alpha)^{a} \overline{\Lambda_\bk(\bx)} \dd \bx
= \Big(\frac 1L\Big)^3 \Big(\frac{2\ii \pi k_\alpha}{L}\Big)^a \ee^{-\frac{2\ii \pi \bk \cdot \px_p}{L}}
$$
for $\alpha \in \range{1}{3}$ and $a\in \NN$. The commuting diagram property is then derived
from the symmetry of the discrete de Rham sequence. For instance,
using that $\Div V^1 = \Div V^2 \subset V^3 = V^0$ allows to write for all $\bF \in V^1$
$$ 
\int (\nabla \Pi^0 G) \cdot \bF
  =
  - \int (\Pi^0 G) \Div \bF
  =
  - \int  G \Div \bF
  =
  \int (\nabla G) \cdot \bF
  =
  \int  (\Pi^1 \nabla G) \cdot \bF
$$ 
for any periodic distribution $G \in \cD'_{\rm per}$. Since $\nabla V^0 \subset V^1$ this shows that $\nabla \Pi^0 G = \Pi^1 \nabla G$.
The other relations $\curl \Pi^1 \bG = \Pi^2 \curl \bG$ and $\Div \Pi^2 \bG = \Pi^3 \Div \bG$ are proven
in the same way.
\eproof

\subsection{Commuting projections based on pseudo-differential DFT} 
\label{sec:psd}

In our framework, spectral PIC methods are obtained with
commuting projection operators that involve discrete Fourier transforms on a finite grid.
In \cite{CPKS_variational_2020} we have described a set of projections which
rely on geometric degrees of freedom, namely nodal interpolations for $V^0$ and
edge, face, and volume ``histopolations'' for $V^1$, $V^2$, and $V^3$, respectively.
As a result the current deposition involves face integrals which need to be integrated over the particle
trajectories, which somehow deviates from standard PIC methods where common deposition procedures
are based on point evaluations of the particle shapes.
For this reason we consider here an alternate construction based on a new sequence
of projection operators, obtained by combining
DFT coefficients on a grid with $M \ge 2K+1$ nodes as in Section~\ref{sec:f-pic} with 
standard derivatives and anti-derivatives in Fourier variables. 
As we will see, these new projections will lead to deposition methods that only involve
pointwise evaluations of the particle shapes.
Following the terminology of \cite{kraus2016gempic}, we will refer to the resulting methods
as {\em Fourier-Geometric Electromagnetic Particle-in-Cell} (Fourier-GEMPIC) methods.
Beginning with the space $V^3$, we let 
\begin{equation} \label{Pi3}
\Pi^3 G = \sum_{\bk \in \range{-K}{K}^3} (\Pi^3 G)_{\bk} \Lambda_\bk
\end{equation}
be defined on $\cC^0_{\rm per}$ the space of continuous, $L$-periodic functions, by its coefficients
\begin{equation} \label{cDFT-1}
(\Pi^3 G)_{\bk} := \gamma_\bk \tilde \cF_{M, \bk}(G) :=
\gamma_\bk \tilde \cF_{M, k_1} \otimes \tilde \cF_{M, k_2} \otimes \tilde \cF_{M, k_3} (G).
\end{equation}
Here, the values $\gamma_\bk = \overline{\gamma_{-\bk}} \in \CC$ are again Hermitian filters,
and the univariate operators $\tilde \cF_{M, k_\alpha}$ are defined as
\begin{equation} \label{cDFT-2}
  \tilde \cF_{M, k_\alpha}(G) :=
  \begin{cases}
    \frac{1}{L} \int_0^L G(x_\alpha) \dd x_\alpha
    \quad &\text{if } k_\alpha = 0
  \\
    \cF_{M, k_\alpha}(G) = \frac{1}{M} \sum_{m_\alpha = 1}^M G(m_\alpha h) \ee^{-\frac{2\ii\pi k_\alpha m_\alpha}{M}}
    \quad &\text{if } k_\alpha \neq 0.
  \end{cases}
\end{equation}
We note that for $\gamma_{\bs 0} = 1$ the operator $\Pi^3$ is a {\em conservative}
discrete Fourier transform, 
indeed
$$ 
 \int_{[0,L]^3} \Pi^3 G(\bx) \dd \bx
 = L^3 (\Pi^3 G)_{\bs 0}
 = L^3 \tilde \cF_{M, {\bs 0}}(G) = \int_{[0,L]^3} G(\bx) \dd \bx.
$$ 
For the vector-valued $V^2$ we then set 
$$
\Pi^2 \bG =  \sum_{\alpha = 1}^3 \sum_{\bk \in \range{-K}{K}^3}(\Pi^2_\alpha G_\alpha)_{\bk} \Lambda_\bk \uvec_\alpha
\quad \text{ with } \quad
(\Pi^2_\alpha G_\alpha)_{\bk}
:=
\gamma_\bk(\hat D_{\bk, \alpha})^{-1} \tilde \cF_{M, \bk}(\tilde D_{\bk, \alpha} G_\alpha),
$$
with pseudo-differential operators defined as
\begin{equation} \label{hDtD}
  \hat D_{\bk, \alpha} c_\bk :=
  \begin{cases}
    c_\bk  &\text{if } k_\alpha = 0
    \\
    \frac{2\ii\pi k_\alpha}{L} c_\bk \quad &\text{if } k_\alpha \neq 0
  \end{cases}
  \qquad \text{ and } \qquad
  \tilde D_{\bk, \alpha} G :=
  \begin{cases}
    G &\text{if } k_\alpha = 0
    \\
    \partial_\alpha G \quad &\text{if } k_\alpha \neq 0
  \end{cases}
\end{equation}
for any $c_\bk \in \CC$ and any function $\bG$ in the anisotropic regularity space
$\cC^1_{\rm per, 1} \times \cC^1_{\rm per, 2} \times \cC^1_{\rm per, 3}$, where we have denoted
\begin{equation} \label{C1a}
  \cC^1_{\rm per, \alpha} := \{ G \in \cC^0_{\rm per} : \partial_\alpha G \in \cC^0_{\rm per}\}.
\end{equation}
Similarly for the vector-valued space $V^1$ we define
(using a circular convention $\alpha \equiv \alpha + 3$ for the dimension indices)
$$
\left\{\begin{aligned}
&\Pi^1 \bG =  \sum_{\alpha = 1}^3 \sum_{\bk \in \range{-K}{K}^3}(\Pi^1_\alpha G_\alpha)_{\bk} \Lambda_\bk \uvec_\alpha
\\
&\text{with } \quad(\Pi^1_\alpha G_\alpha)_{\bk} :=
\gamma_\bk (\hat D_{\bk, \alpha+1}\hat D_{\bk, \alpha+2})^{-1}\tilde \cF_{M, \bk}(\tilde D_{\bk, \alpha+1} \tilde D_{\bk, \alpha+2} G_\alpha)
\end{aligned}\right.
$$
for any function $\bG$ in the anisotropic regularity space
$\cC^1_{\rm per, 2,3} \times \cC^1_{\rm per, 1,3} \times \cC^1_{\rm per, 1,2}$, where
\begin{equation} \label{C1ab}
  \cC^1_{\rm per, \alpha,\beta} := \{ G \in \cC^0_{\rm per} : \partial_\alpha \partial_\beta G \in \cC^0_{\rm per}\}.
\end{equation}
Finally for the scalar-valued space $V^0$ we define
$$
\left\{\begin{aligned}
&\Pi^0 G = \sum_{\bk \in \range{-K}{K}^3} (\Pi^0 G)_{\bk} \Lambda_\bk
\\
&\text{with } \quad (\Pi^0 G)_{\bk} :=
\gamma_\bk (\hat D_{\bk, 1}\hat D_{\bk, 2}\hat D_{\bk, 3})^{-1}\tilde \cF_{M, \bk}(\tilde D_{\bk,1} \tilde D_{\bk,2} \tilde D_{\bk,3} G),
\end{aligned}\right.
$$
for all $G$ in the anisotropic regularity space
\begin{equation} \label{C1abc}
  \cC^1_{\rm per, 1,2,3} := \{ G \in \cC^0_{\rm per} : \partial_1 \partial_2 \partial_3 G \in \cC^0_{\rm per}\}.
\end{equation}

\begin{lemma}
The above pseudo-differential operators $\Pi^\ell$ satisfy the commuting diagram property
\eqref{CD} 
with domains defined as
$$
V^0 = \cC^1_{\rm per, 1,2,3},
\quad
V^1 = \cC^1_{\rm per, 2,3} \times \cC^1_{\rm per, 1,3} \times \cC^1_{\rm per, 1,2},
\quad
V^2 = \cC^1_{\rm per, 1} \times \cC^1_{\rm per, 2} \times \cC^1_{\rm per, 3},
\quad 
V^3 = \cC^0_{\rm per}.
$$
Moreover if $\gamma_\bk = 1$, they are projection operators on their respective Fourier spaces.
\end{lemma}
\bproof
The projection property can be checked by direct computation.
To verify the commuting diagram property we consider again the case $\gamma_\bk = 1$,
as the general case follows easily. We begin with the last relation and observe that
$$
\tilde \cF_{M,\bk}(\partial_\alpha G_\alpha) = \tfrac{2\ii\pi k_\alpha}{L} (\Pi^2\bG)_{\bk,\alpha}
$$
holds for every dimension $\alpha \in \range{1}{3}$ and mode $\bk \in \ZZ^3$:
this follows from the definition of $\Pi^2$ if $k_\alpha \neq 0$, and from that of the conservative Fourier transform
$\tilde \cF_{M,\bk}$ if $k_\alpha = 0$.
We then have, for an arbitrary $\bk$,
$$
(\Pi^3 \Div \bG)_\bk = \sum_\alpha (\Pi^3( \partial_\alpha G_\alpha))_\bk = \sum_\alpha \tilde \cF_{M, \bk}(\partial_\alpha G_\alpha)
  = \sum_\alpha \tfrac{2\ii\pi k_\alpha}{L} (\Pi^2\bG)_{\bk,\alpha}
  = (\Div \Pi^2 \bG)_\bk
$$
which shows that $\Pi^3 \Div = \Div \Pi^2$ holds on $V^2$. 
Similarly, for all $\bk$ and $\alpha$, we have
$$
\begin{aligned}
  (\Pi^2 \curl \bG)_{\bk,\alpha}
    &=
    \big(\Pi^2_\alpha (\partial_{\alpha+1}G_{\alpha+2} - \partial_{\alpha+2}G_{\alpha+1})\big)_{\bk}
  \\
    &=
    (\hat D_{\bk, \alpha})^{-1}\tilde \cF_{M, \bk}\big(\tilde D_{\bk, \alpha} (\partial_{\alpha+1}G_{\alpha+2} - \partial_{\alpha+2}G_{\alpha+1})\big)
  \\
    &= \tfrac{2\ii\pi k_{\alpha+1}}{L} (\Pi^1 \bG)_{\bk,\alpha+2}
          -   \tfrac{2\ii\pi k_{\alpha+2}}{L} (\Pi^1 \bG)_{\bk,\alpha+1}
  = (\curl \Pi^1 \bG)_{\bk,\alpha}
  \end{aligned}
$$
so that $\Pi^2 \curl = \curl \Pi^1$ holds on $V^1$. 
Finally we compute, again for all $\bk, \alpha$,
$$
\begin{aligned}
(\Pi^1 \nabla G)_{\bk,\alpha}
  &=
  (\Pi^1_\alpha \partial_\alpha G)_{\bk}
  = (\hat D_{\bk, \alpha+1} \hat D_{\bk, \alpha+2})^{-1}\tilde \cF_{M, \bk}(\tilde D_{\bk, \alpha+1}\tilde D_{\bk, \alpha+2}\partial_{\alpha}G)
  \\
  &= \tfrac{2\ii\pi k_{\alpha}}{L} (\Pi^0 G)_{\bk}
= (\nabla \Pi^0 G)_{\bk,\alpha}
\end{aligned}
$$
which shows that $\Pi^1 \nabla = \nabla \Pi^0$ holds on $V^0$ 
and ends the proof.
\eproof


\subsection{GEMPIF and Fourier-GEMPIC methods}
\label{sec:f-gempic}

Now that we have defined two sets of projection operators $\Pi^\ell$ with commuting diagram properties,
we may specify the spectral variational scheme~\eqref{gfp-solve}--\eqref{gfp-push}.
As announced above, the Geometric Particle-in-Fourier ({GEMPIF}) method corresponds
to the case where the operators $\Pi^\ell$ are defined as the $L^2$ projections of Section~\ref{sec:L2}.
Then any periodic distribution $S \in \cD'_{\rm per}$
is admissible and the coupling terms 
take the form
\begin{equation} \label{Jk_gempif}
    \bJ^S_\bk 
    = \gamma_\bk \sum_{p = 1 \cdots N} q_p \pv_p \cF_\bk(S_{\px_p})
    = \gamma_\bk\cF_\bk(S) \sum_{p = 1 \cdots N} q_p \pv_p \ee^{-\frac{2\ii\pi \bk\cdot\px_p}{L}}
\end{equation}
and
\begin{equation} \label{EBS_gempif}
  \left\{
  \begin{aligned}
    &E^S_{\alpha} (\px_p) 
    = L^3 \sum_{\bk \in \range{-K}{K}^3} \overline{\gamma_\bk} E_{\alpha,\bk} \overline{\cF_\bk(S_{\px_p})}
    = L^3 \sum_{\bk \in \range{-K}{K}^3} \overline{\gamma_\bk \cF_\bk(S)} E_{\alpha,\bk} \ee^{\frac{2\ii\pi \bk\cdot\px_p}{L}}
  \\
    &B^S_{\alpha} (\px_p) 
    = L^3 \sum_{\bk \in \range{-K}{K}^3} \overline{\gamma_\bk} B_{\alpha,\bk} \overline{\cF_\bk(S_{\px_p})}
    = L^3 \sum_{\bk \in \range{-K}{K}^3} \overline{\gamma_\bk \cF_\bk(S)} B_{\alpha,\bk} \ee^{\frac{2\ii\pi \bk\cdot\px_p}{L}}.
  \end{aligned}
  \right.
\end{equation}
Up to the arbitrary filter coefficients $\gamma_\bk$, this corresponds to the gridless coupling
of Section~\ref{sec:pif}.

%
With the pseudo-differential operators defined in Section~\ref{sec:psd}, the coupling terms
involve modified Fourier coefficients such as
$(\Pi^2_\alpha S_{\px_p})_\bk = (\hat D_{\bk, \alpha})^{-1} \tilde \cF_{M, \bk}(\tilde D_{\bk, \alpha} S_{\px_p})$,
see \eqref{hDtD}, and similar coefficients for $\Pi^1_\alpha S_{\px_p}$, involving derivatives along
dimensions $\alpha+1$ and $\alpha-1$.
The resulting coupling terms read then
\begin{equation} \label{Jk_fgempic}
    J^S_{\alpha,\bk}
    = \gamma_\bk \Big(\frac{2\ii\pi k_\alpha}{L}\Big)^{-1}
      \Big(\frac 1 M \Big)^3 \sum_{p, \bsm} q_p v_{p,\alpha} \partial_\alpha S(\bsm h -\px_p) \ee^{-\frac{2\ii\pi \bk\cdot \bsm}{M}}
\end{equation}
and
\begin{equation} \label{EBS_fgempic}
  \left\{
  \begin{aligned}
    &E^S_{\alpha} (\px_p) 
    = \overline{\gamma_\bk} h^3 \sum_{\bk, \bsm} 
      E_{\alpha,\bk} \Big(\frac{-2\ii\pi k_\alpha}{L}\Big)^{-1} \partial_\alpha S(\bsm h -\px_p) \ee^{\frac{2\ii\pi \bk\cdot \bsm}{M}}
  \\
    &B^S_{\alpha} (\px_p) 
    = \overline{\gamma_\bk} h^3 \sum_{\bk, \bsm} 
      B_{\alpha,\bk} \Big(\frac{-2\ii\pi k_{\alpha+1}}{L}\Big)^{-1} \Big(\frac{-2\ii\pi k_{\alpha-1}}{L}\Big)^{-1}
      \partial_{\alpha+1} \partial_{\alpha-1} S(\bsm h -\px_p) \ee^{\frac{2\ii\pi \bk\cdot \bsm}{M}}
  \end{aligned}
  \right.
\end{equation}
with the modifications specified in \eqref{cDFT-1}--\eqref{hDtD} when $k_\alpha = 0$ or $k_{\alpha\pm 1} = 0$.
We observe that in order to be admissible, shape functions need to be in the mixed $C^1$ space \eqref{C1abc}.
For tensor-product B-splines this corresponds to using at least quadratic splines in each dimension.

Although these terms involve a DFT grid, they differ from
the standard spectral PIC coupling terms recalled in Section~\ref{sec:f-pic}, which are not associated with a
variational principle.
For the purpose of comparison we rewrite the latter in the case of a general $\gamma$-filtering,
\begin{equation} \label{Jk_pic_bis}
    J^S_{\alpha,\bk} = \gamma_\bk \Big(\frac 1 M \Big)^3 \sum_{p, \bsm} q_p v_{p,\alpha} S(\bsm h -\px_p) \ee^{-\frac{2\ii\pi \bk\cdot \bsm}{M}}
\end{equation}
and
\begin{equation} \label{EBS_pic_bis}
  \left\{
  \begin{aligned}
    &E^S_{\alpha} (\px_p)
    = \overline{\gamma_\bk} h^3 \sum_{\bk, \bsm} E_{\alpha,\bk} S(\bsm h -\px_p) \ee^\frac{2\ii \pi \bk \cdot \bsm}{M}
  \\
    &B^S_{\alpha} (\px_p)
    = \overline{\gamma_\bk} h^3 \sum_{\bk, \bsm} B_{\alpha,\bk} S(\bsm h -\px_p) \ee^\frac{2\ii \pi \bk \cdot \bsm}{M}.
  \end{aligned}
  \right.
\end{equation}
Below we will see how these differences translate in a fully discrete setting.

\section{Fully discrete schemes based on Hamiltonian time splitting}
\label{sec:fulldis}

In this section we specify some fully discrete schemes for the
variational semi-discrete systems described above. 
As these schemes are derived from a Hamiltonian splitting procedure,
it will be convenient to first rewrite the general
equations~\eqref{gfp-solve}--\eqref{gfp-push} in a matrix form.

\subsection{Matrix formulation of the semi-discrete Hamiltonian system}

Similarly as in \cite{CPKS_variational_2020, kraus2016gempic},
we gather the degrees of freedom of the discrete solution into multi-index arrays.
Particle unknowns will be written as
$$
\arrX = (X_{p,\alpha})_{(p, \alpha) \in  \range{1}{N} \times \range{1}{3}},
\quad
\arrV = (V_{p,\alpha})_{(p, \alpha) \in \range{1}{N} \times \range{1}{3}}
\quad \in ~~ \RR^{3N},
$$
(with an implicit time-dependance) and electro-magnetic field coefficients will be denoted by
$$
\arrE = (E_{\alpha, \bk, \sigma})_{(\alpha, \bk, \sigma) \in \range{1}{3} \times \range{-K}{K}^3 \times \range{0}{1}},
\quad
\arrB = (B_{\alpha, \bk, \sigma})_{(\alpha, \bk, \sigma) \in \range{1}{3} \times \range{-K}{K}^3 \times \range{0}{1}}
\quad \in ~~ \RR^{6(2K+1)^3}
$$
where each component-wise Fourier coefficient is decomposed into its real and imaginary parts,
$$
G_{\alpha, \bk} = G_{\alpha, \bk, 0} + \ii G_{\alpha, \bk, 1} \qquad \text{ with }~~  G = E \text{ or } B.
$$
For notational reasons it is convenient to see these arrays as column vectors (with an arbitrary ordering of the multi-indices),
and to gather them into a global array of time-dependent unknowns,
\begin{equation} \label{u}
  \arrU = \begin{pmatrix}
  \arrX
  \\
  \arrV
  \\
  \arrE
  \\
  \arrB
  \end{pmatrix}.
\end{equation}
We may then rewrite the equations of the geometric
Fourier-particle method \eqref{gfp-solve}--\eqref{gfp-push}
in terms of these coefficients.
Using the fact that the vector-valued operators $\Pi^\ell$
are defined component-wise, in the sense that
their component along any dimension $\alpha \in \range{1}{3}$ reads
\begin{equation} \label{Pi_a}
(\Pi^\ell(\bG))_\alpha = \Pi^\ell_\alpha(G_\alpha)
\end{equation}
for some scalar operators $\Pi^\ell_\alpha$,
we may rewrite the particle trajectories as
\begin{equation} \label{gfp-push-k}
\left\{
\begin{aligned}
  \tfrac{\dd}{\dd t}  X_{p,\alpha} &= V_{p,\alpha}
  \\
  \tfrac{\dd}{\dd t}  V_{p,\alpha} &=
  L^3 \frac{q_p}{m_p} \sum_{\bk}
  \Big( E_{\alpha, \bk} \overline{(\Pi^2_\alpha S_{\px_p})_\bk}
  \pm V_{p,\alpha\pm 1} B_{\alpha\mp 1, \bk} \overline{(\Pi^1_{\alpha \mp 1} S_{\px_p})_\bk}
  \Big)
  \\
  & =
  L^3 \frac{q_p}{m_p} \sum_{\bk, \sigma}
  \Big( E_{\alpha, \bk, \sigma} (\Pi^2_\alpha S_{\px_p})_{\bk,\sigma}
  \pm V_{p,\alpha\pm 1} B_{\alpha\mp 1, \bk, \sigma} (\Pi^1_{\alpha \mp 1} S_{\px_p})_{\bk,\sigma}
  \Big)
\end{aligned}
\right.
\end{equation}
for all $p \in \range{1}{N}$ and $\alpha \in \range{1}{3}$.
Here we have used the fact that the right-hand side is real,
and we recall the circular convention ($\alpha \equiv \alpha + 3$) for the dimension indices.
In terms of the above arrays, this gives
$$
\left\{
\begin{aligned}
  \tfrac{\dd}{\dd t}  \arrX &= \arrV
  \\
  \tfrac{\dd}{\dd t}  \arrV &= \matW_{\frac qm} \big(\matS_2(\arrX) \matM^2 \arrE + \matR(\arrX,\arrB) \arrV \big)
\end{aligned}
\right.
$$
where
$\matW_{\frac qm} = \diag(\frac{q_p}{m_p} : p \in \range{1}{N})$
is the diagonal weighting matrix carrying the particles charge to mass ratios,
$\matS^2(\arrX)$ is the $\Pi^2$ Fourier-particle coupling matrix,
$$
\matS^2(\arrX)_{(p,\beta),(\alpha,\bk,\sigma)} = \delta_{\alpha,\beta}(\Pi^2_\alpha S_{\px_p})_{\bk,\sigma},
$$
$\matM^\ell = L^3 \II_{N_\ell}$ is the (diagonal) finite element ``mass'' matrix
for the Fourier basis in $V^\ell_K$,
and $\matR(\arrX,\arrB)$ is the skew-symmetric matrix corresponding to magnetic rotation,
\begin{equation} \label{matR}
  \begin{aligned}
  \matR(\arrX, \arrB)_{(p,\alpha), (p',\beta)}
    &= \delta_{p,p'}
    \sum_{\bk,\sigma} L^3 \big( \delta_{\alpha+1,\beta} B_{\alpha - 1, \bk, \sigma} (\Pi^1_{\alpha-1} S_{\px_p})_{\bk, \sigma}
    - \delta_{\alpha-1,\beta} B_{\alpha + 1, \bk,\sigma} (\Pi^1_{\alpha+1} S_{\px_p})_{\bk, \sigma} \big)
    \\
    &= \delta_{p,p'}
    \sum_{\nu = \pm 1} \delta_{\alpha+\nu,\beta}  \int \nu B_{\alpha - \nu}(\bx) (\Pi^1_{\alpha-\nu} S_{\px_p})(\bx) \dd \bx.
  \end{aligned}
\end{equation}
As for the Maxwell solver, we first rewrite \eqref{gfp-solve} in terms of (complex) Fourier coefficients, namely
\begin{equation} \label{gfp-solve-k}
  \left\{
  \begin{aligned}
    \tfrac{\dd}{\dd t} E_{\alpha,\bk} &= \sum_{\nu = \pm 1} \nu \tfrac{2\ii \pi }{L} k_{\alpha + \nu} B_{\alpha - \nu,\bk}
      - \sum_{p} q_p v_{p,\alpha} \Pi^2_\alpha(S_{\px_p})_\bk
    \\
    \tfrac{\dd}{\dd t} B_{\alpha,\bk} &= - \sum_{\nu = \pm 1} \nu\tfrac{2\ii \pi }{L} k_{\alpha + \nu}E_{\alpha-\nu,\bk}
  \end{aligned}
  \right.
\end{equation}
for $\bk \in \range{-K}{K}^3$ and $\alpha \in \range{1}{3}$. In terms of the above arrays, this rewrites as
\begin{equation} \label{gfp-solve-mat}
  \left\{
  \begin{aligned}
    \tfrac{\dd}{\dd t} \arrE &= \matC \arrB - \matS_2(\arrX)^T \matW_{q} \arrV
    \\
    \tfrac{\dd}{\dd t} \arrB &= - \matC \arrE
  \end{aligned}
  \right.
\end{equation}
with $\matC$ the symmetric matrix of the curl operator, which writes here
$$
\matC_{(\alpha,\bk,\sigma),(\beta,\bell,\tau)} =
\delta_{\bk, \bell} (1-\delta_{\sigma, \tau}) (-1)^{\sigma+1} \frac{2\pi}{L}
  \big(\delta_{\alpha-1,\beta} k_{\alpha+1} - \delta_{\alpha+1,\beta} k_{\alpha-1} \big).
$$
Writing next the discrete Hamiltonian~\eqref{H} as a function of the array variable \eqref{u},
\begin{equation} \label{Hu}
  \cH(\arrU) =
  \tfrac 12 \arrV^\top \matW_m \arrV + \tfrac 12 \arrE^\top \matM^2 \arrE + \tfrac 12 \arrB^\top \matM^1 \arrB
\end{equation}
we find for the corresponding derivatives
$$
\nabla_{\arrU} \cH(\arrU) = \begin{pmatrix}
  \nabla_{\arrX} \cH
  \\
  \nabla_{\arrV} \cH
  \\
  \nabla_{\arrE} \cH
  \\
  \nabla_{\arrB} \cH
\end{pmatrix}(\arrU)
= \begin{pmatrix}
  \arr{0}
  \\
  \matW_m \arrV
  \\
  \matM^2 \arrE
  \\
  \matM^1 \arrB
  \end{pmatrix}
$$
and this allows us to rewrite the abstract spectral particle method \eqref{gfp-solve}--\eqref{gfp-push}
in the form of a non-canonical Hamiltonian system
\begin{equation} \label{ham}
\tfrac{\dd}{\dd t} \arrU = \matJ(\arrU) \nabla_{\arrU} \cH(\arrU)
\end{equation}
with
\begin{equation} \label{JJ}
  \matJ(\arrU)
  =
  \matJ(\arrX,\arrB)
  = \begin{pmatrix}
    0 & \matW_{\frac 1m} & 0 & 0
    \\
    -\matW_{\frac 1m}
      & \matW_{\frac qm} \matR(\arrX,\arrB) \matW_{\frac 1m}
      & \matW_{\frac qm} \matS_2(\arrX) & 0
    \\
    0 & -\matS_2(\arrX)^T \matW_{\frac qm} & 0 & \matC \mat(M^1)^{-1}
    \\
    0 & 0 & -\mat(M^1)^{-1} \matC & 0
  \end{pmatrix}.
\end{equation}
In \cite{CPKS_variational_2020} we have shown that $\matJ$ is a Poisson matrix
in the sense of \cite[Def.~VII.2.4]{HairerLubichWanner:2006},
i.e., it is skew-symmetric and it satisfies the matrix Jacobi identity.
In particular, System \eqref{ham} may be rewritten in the usual form
$$
\tfrac{\dd}{\dd t} \arrU = \{\arrU,\cH\}
$$
with a discrete Poisson bracket given by
$  \{ \cF, \cG \} = (\nabla_\arrU \cF)^T \matJ \nabla_{\arrU} \cG.$

\subsection{Hamiltonian splitting time discretization}
\label{sec:split}

Following \cite{HairerLubichWanner:2006,Crouseilles:2015}, we now apply a splitting procedure
to the semi-discrete Hamiltonian system~\eqref{ham}. This will provide us
with a series of Hamiltonian structure-preserving schemes of various orders in time,
for the variational Fourier-particle equations \eqref{gfp-solve}--\eqref{gfp-push}.


Similarly as in \cite{He:2015, kraus2016gempic}, we split the kinetic part along the three different dimensions,
but keep together the electric and magnetic parts as we are solving the Maxwell equations in Fourier spaces.
This leads us to the following Hamiltonian splitting,
\begin{equation} \label{Hsplit}
  \cH =  \sum_{\alpha=1}^3 \cH_{v_\alpha} + \cH_{EB}
\end{equation}
with
$$
\cH_{v_\alpha}(\arrU) := \sum_{p=1}^N \frac {m_p}{2}  \abs{V_{p,\alpha}}^2
\qquad \text{ and } \qquad
\cH_{EB}(\arrU) := \frac 12 \int_\Omega \Big(\abs{\bE_K(\bx)}^2 + \abs{\bB_K(\bx)}^2\Big) \dd \bx
$$
where we remind that $\arrU$ carries the time dependent coefficients of the full solution, see \eqref{u}.
This splitting has two key properties. It leads to split steps that can all be solved exactly,
and it preserve the fact that $\matJ(\arrX,\arrB)$ is a Poisson matrix, see Theorem~\ref{th:consplit}.
As a consequence, we know that any combination of the split steps will provide a Hamiltonian
time scheme which preserves the Casimir invariants and the total energy up to some constant
time discretization error, see e.g. \cite{HairerLubichWanner:2006}.
Specifically, we decompose System~\eqref{ham} into the subsystems,
\begin{equation} \label{ham_kin}
\tfrac{\dd}{\dd t} \arrU(t) = \matJ(\arrU) \nabla_{\arrU} \cH_{v_\alpha}(\arrU)
\qquad \text{ for } \quad \alpha \in \range{1}{3}
\end{equation}
and
\begin{equation} \label{ham_em}
\tfrac{\dd}{\dd t} \arrU(t) = \matJ(\arrU) \nabla_{\arrU} \cH_{EB}(\arrU).
\end{equation}
Denoting by
$\vp_{\tau, v_\alpha}$ and $\vp_{\tau,EB}$ the corresponding solution flow maps,
we can use standard composition methods to obtain time integrators of various orders,
as described e.g. in \cite{kraus2016gempic}: either a first order Lie-Trotter scheme
$$
\Phi_{\Dt,L} = \vp_{\Dt,EB} \circ \vp_{\Dt,v_3} \circ \vp_{\Dt,v_2} \circ \vp_{\Dt,v_1}
$$
or a second-order Strang scheme
\begin{equation} \label{S2}
\Phi_{\Dt,S2} = \vp_{\Dt/2,L} \circ \vp_{\Dt/2,L}^*
\end{equation}
where $\vp_{\tau,L}^* := \vp_{-\tau,L}^{-1}$ denotes the adjoint flow. Since each
split flow is the exact solution of an autonomous system they are all self-adjoint, \textit{i.e.} symmetric, which yields
$$ 
\Phi_{\Dt,S2} = \vp_{\Dt/2,EB} \circ \vp_{\Dt/2,v_3} \circ \vp_{\Dt/2,v_2} \circ \vp_{\Dt,v_1}
\circ \vp_{\Dt/2,v_2} \circ \vp_{\Dt/2,v_3}\circ \vp_{\Dt/2,EB}.
$$ 
Similarly we can use a fourth-order Suzuki-Yoshida scheme
\begin{equation} \label{S4}
  \vp_{\Dt,S4} = \vp_{\gamma_1 \Dt, S2} \circ \vp_{\gamma_2 \Dt, S2} \circ \vp_{\gamma_1 \Dt, S2}
\end{equation}
with $\gamma_1 = 1/(2-2^{1/3})$ and $\gamma_1 = -2^{1/3}/(2-2^{1/3})$,
or higher-order composition methods, see e.g. \cite{McLachlanQuispel:2002,HairerLubichWanner:2006}.

In the sections below we specify the resulting equations for each split step,
and we provide an explicit solution for the GEMPIF and Fourier-GEMPIC methods
described in Section~\ref{sec:f-gempic}.

\subsection{Discrete Hamiltonian subsystems}

Before giving the solutions we detail the subsystems \eqref{ham_kin} and \eqref{ham_em}.

\paragraph{Kinetic $\cH_{v_\alpha}$ subsystems.}
For each dimension $\alpha \in \range{1}{3}$, System~\eqref{ham_kin} reads
$$
\left\{
\begin{aligned}
  &\tfrac{\dd }{\dd t} \arrX = \arrV^{[\alpha]}
  \\
  &\tfrac{\dd }{\dd t} \arrV = \matW_{\frac qm} \matR(\arrX,\arrB) \arrV^{[\alpha]}
  \\
  &\tfrac{\dd }{\dd t} \arrE = -\matS_2(\arrX)^T  \matW_{q} \arrV^{[\alpha]}
  \\
  &\tfrac{\dd }{\dd t} \arrB = 0
\end{aligned}
\right.
\qquad \text{i.e., } \qquad
\left\{
\begin{aligned}
  &\tfrac{\dd }{\dd t} \px_p = \pv^{[\alpha]}_p
  \\
  &\tfrac{\dd }{\dd t} \pv_p = \frac{q_p}{m_p}\pv^{[\alpha]}_p \times \bB^S(\px_p)
  \\
  &\tfrac{\dd }{\dd t} \bE_K = - \Pi^2(\bJ^{S,[\alpha]}_N)
  \\
  &\tfrac{\dd }{\dd t} \bB_K = 0
\end{aligned}
\right.
$$
(for all $p$), where we have denoted $\pv^{[\alpha]}_p = \uvec_\alpha V_{p,\alpha}$ and similarly
for $\arrV^{[\alpha]}$, $\bJ^{S,[\alpha]}_N$.
Expressed in scalar coefficients and using the definition \eqref{gfp-push} of $\bB^S$, this gives
(again with the circular convention $\alpha \equiv \alpha + 3$ on dimension indices)
\begin{equation} \label{Hkin-f}
\left\{
\begin{aligned}
  &\tfrac{\dd }{\dd t} X_{p,\alpha} = V_{p,\alpha}
  \\
  &\tfrac{\dd }{\dd t} X_{p,\alpha + \nu} = 0
  \\
  &\tfrac{\dd }{\dd t} V_{p,\alpha} = 0
  \\
  &\tfrac{\dd }{\dd t} V_{p,\alpha + \nu} = -\nu \frac{q_p}{m_p} L^3
    \sum_{\bk} V_{p,\alpha} B_{\alpha - \nu,\bk}\overline{(\Pi^1_{\alpha -\nu} S_{\px_p})_{\bk}}
  \\
  &\tfrac{\dd }{\dd t} E_{\alpha,\bk} = - \sum_p q_p V_{p,\alpha} (\Pi^2_\alpha S_{\px_p})_{\bk}
  \\
  &\tfrac{\dd }{\dd t} E_{\alpha + \nu,\bk} = 0
  \\
  &\tfrac{\dd }{\dd t} \bB_\bk = 0
\end{aligned}
\right.
\end{equation}
for all $p \in \range{1}{N}$, $\bk \in \range{-K}{K}^3$ and $\nu = \pm 1$.

\paragraph{Electromagnetic $\cH_{EB}$ subsystems.}
For the electromagnetic part, System~\eqref{ham_em} reads
$$
\left\{
\begin{aligned}
  &\tfrac{\dd }{\dd t} \arrX = 0
  \\
  &\tfrac{\dd }{\dd t} \arrV = \matW_{\frac qm} \matS_2(\arrX) \matM^2 \arrE
  \\
  &\tfrac{\dd }{\dd t} \arrE = \matC \arrB
  \\
  &\tfrac{\dd }{\dd t} \arrB = -\matC \arrE
\end{aligned}
\right.
\qquad \text{i.e., } \qquad
\left\{
\begin{aligned}
  &\tfrac{\dd }{\dd t} \px_p = 0
  \\
  &\tfrac{\dd }{\dd t} \pv_p = \frac{q_p}{m_p} \bE^S(\px_p)
  \\
  &\tfrac{\dd }{\dd t} \bE_K = \curl \bB_K
  \\
  &\tfrac{\dd }{\dd t} \bB_K = - \curl \bE_K
\end{aligned}
\right.
$$
for all $p \in \range{1}{N}$.
Using scalar coefficients and the definition \eqref{gfp-push} of $\bE^S$, this reads
\begin{equation} \label{Hem-f}
\left\{
\begin{aligned}
  &\tfrac{\dd }{\dd t} \px_p = 0
  \\
  &\tfrac{\dd }{\dd t} V_{p, \alpha} = \frac{q_p}{m_p} L^3 \sum_\bk E_{\alpha,\bk} \overline{(\Pi^2_\alpha S_{\px_p})_{\bk}}
  \\
  &\tfrac{\dd }{\dd t} \bE_\bk = \tfrac{2\ii\pi\bk}{L} \times \bB_\bk
  \\
  &\tfrac{\dd }{\dd t} \bB_\bk = - \tfrac{2\ii\pi\bk}{L} \times \bE_\bk
  \end{aligned}
  \right.
\end{equation}
for all $\alpha \in \range{1}{3}$, $p \in \range{1}{N}$ and $\bk \in \range{-K}{K}^3$.
This splitting enjoys the following properties.
\begin{theorem}
  \label{th:consplit}
  All the split steps above preserve the discrete Gauss laws \eqref{gfp-Gauss}, namely
  \begin{equation} \label{var-Gauss}
    \Div \bE_K = \Pi^3 \rho_N = \sum_{p = 1}^N q_p \Pi^3 S_{\px_p}
      \qquad \text{ and } \qquad
      \Div \bB_K = 0
  \end{equation}
  and the fact that $\matJ(\arrX,\arrB)$ is a Poisson bracket.
  In particular, any combination of the individual flows $\vp_{\tau,v_\alpha}$ and $\vp_{\tau, EB}$
  preserves the discrete Hamiltonian structure.
  If in addition the operators $\Pi^\ell$ satisfy \eqref{allPi}, then the discrete momentum \eqref{P}
  is also preserved exactly.
\end{theorem}

\bproof
By applying the computations from the proof of Theorem~\ref{th:cons}
to any of the split steps, one verifies that the discrete Gauss laws are preserved,
as well as the discrete momentum in the case where \eqref{allPi} holds.
In particular the Gauss law $\Div \bB_K = 0$ is preserved,
so that Theorem~1 from \cite{CPKS_variational_2020} applies and this shows that
$\matJ$ defines a discrete Poisson bracket at each split step.
The standard theory of Hamiltonian splitting schemes then applies, see \cite{HairerLubichWanner:2006}.
\eproof

\subsection{Explicit steps for the GEMPIF method}
\label{sec:gempif-steps}

In the GEMPIF method the operators $\Pi^\ell$ are defined as $L^2$ projections with general filter coefficients $\gamma_\bk$,
see Section~\ref{sec:L2}. In particular we have
\begin{equation} \label{PiSL2}
(\Pi^1_{\alpha} S_{\px_p})_{\bk} = (\Pi^2_\alpha S_{\px_p})_{\bk}
= \Big(\frac 1L \Big)^3 \gamma_\bk \int_\Omega S(\bx-\px_p) \ee^{-\frac{2\ii\pi \bk\cdot\bx}{L}} \dd \bx
= \gamma_\bk\cF_\bk(S) \ee^{-\frac{2\ii\pi \bk\cdot\px_p}{L}}
\end{equation}
for any direction $\alpha \in \range{1}{3}$. This allows us to give explicit solutions for each split subsystem.

\paragraph{Directional kinetic steps.}

\begin{lemma}
In the GEMPIF method,
the exact solution
$\vp_{\tau,v_\alpha}: \arrU^0 \to \arrU(\tau)$
of the kinetic split step \eqref{Hkin-f} in a direction
$\alpha \in \range{1}{3}$ is given by the explicit expressions 
$$
\vp_{\tau,v_\alpha}: \qquad
\left\{
\begin{aligned}
  X_{p,\alpha}(\tau) &= X_{p,\alpha}^0 + \tau V_{p,\alpha}
  \\ \noalign{\smallskip}
  X_{p,\alpha + \nu}(\tau) &= X_{p,\alpha + \nu}^0
  \\ \noalign{\smallskip}
  V_{p,\alpha}(\tau) &= V_{p,\alpha}^0
  \\
  V_{p,\alpha + \nu}(\tau) &= V_{p,\alpha + \nu}^0
    - \nu \tau \frac{q_p}{m_p} L^3 \sum_{\bk \in \range{-K}{K}^3} B_{\alpha - \nu,\bk}
      \overline{\Big(\gamma_\bk\cF_\bk(S) \hat V_{p, \alpha, \bk}(\tau)\Big)}
  \\
  E_{\alpha,\bk}(\tau) &= E_{\alpha,\bk}^0 - \tau \gamma_\bk\cF_\bk(S) \sum_{p = 1 \cdots N} q_p \hat V_{p, \alpha, \bk}(\tau)
  \\
  E_{\alpha + \nu,\bk}(\tau) &= E_{\alpha + \nu,\bk}^0
  \\ \noalign{\smallskip}
  \bB_K(\tau) &= \bB_K^0.
\end{aligned}
\right.
$$
for all $p \in \range{1}{N}$, $\bk \in \range{-K}{K}^3$ and $\nu = \pm 1$,
with
\begin{equation} \label{hv}
\hat V_{p, \alpha, \bk}(\tau)
:= \begin{cases}
  V_{p,\alpha} \ee^{- \frac{2\ii\pi \bk\cdot\px_p^0}{L}}
  &\text{ if } k_\alpha = 0
  \\
  \Big(-\frac{2\ii\pi k_\alpha \tau}{L}\Big)^{-1} \Big(\ee^{- \frac{2\ii\pi k_\alpha \tau V_{p,\alpha}}{L}}-1\Big) \ee^{- \frac{2\ii\pi \bk\cdot\px_p^0}{L}}
  &\text{ if } k_\alpha \neq 0.
\end{cases}
\end{equation}
\end{lemma}

\bproof
In \eqref{Hkin-f}, the velocity equations read
$$
\frac{\dd }{\dd t} V_{p,\alpha + \nu}
  = -\nu \frac{q_p}{m_p} L^3 \sum_{\bk}
  B_{\alpha - \nu,\bk}
  \overline{\Big(\gamma_\bk\cF_\bk(S) V_{p,\alpha} \ee^{-\frac{2\ii\pi \bk\cdot\px_p}{L}}\Big)}
$$
and for the electric field we have
$$
\frac{\dd }{\dd t} E_{\bk,\alpha}
  = - \gamma_\bk\cF_\bk(S) \sum_p q_p V_{p,\alpha} \ee^{-\frac{2\ii\pi \bk\cdot\px_p}{L}}.
$$
Since the only time-varying term in the right hand sides is
$\frac{\dd}{\dd t}\px_p(t) = \pv_p^{[\alpha]} = V_{p,\alpha}\uvec_\alpha $, we integrate
$$
\int_0^\tau V_{p,\alpha} \ee^{- \frac{2\ii\pi \bk\cdot\px_p(t)}{L}} \dd t
= \tau \hat V_{p, \alpha, \bk}(\tau)
$$
with the expression in \eqref{hv}, which proves the lemma.
\eproof

\paragraph{Electromagnetic step.}

In \eqref{Hem-f} the source-free Maxwell equations have an explicit solution (see e.g. \cite{Vay.Haber.Godfrey.2013.jcp}),
which allows to solve also for the particles.
The resulting flow takes the form
\begin{equation} \label{Hem_sol}
  \vp_{\tau,EB}: \qquad
  \left\{
  \begin{aligned}
    \px_p(\tau) &= \px_p^0
    \\
    V_{p,\alpha}(\tau) &= V_{p,\alpha}^0 + \frac{q_p}{m_p} L^3
        \sum_{\bk \in \range{-K}{K}^3} \Big( \int_0^\tau E_{\alpha,\bk}(t) \dd t \Big) \overline{(\Pi^2_\alpha S_{\px_p^0})_{\bk}}
    \\
    \bE_\bk(\tau) &= \bE_\bk^0 + \hat \bk \times \Big( (1 - c(\tau)) (\hat \bk \times \bE^0_\bk) +  \ii s(\tau) \bB^0_\bk \Big)
    \\
    \bB_\bk(\tau) &= \bB_\bk^0 + \hat \bk \times \Big( (1 - c(\tau)) (\hat \bk \times \bB^0_\bk) -  \ii s(\tau) \bE^0_\bk \Big)
  \end{aligned}
 \right.
\end{equation}
for all $p \in \range{1}{N}$, $\alpha \in \range{1}{3}$ and $\bk \in \range{-K}{K}^3$.
Here the projected shaped particle is given by \eqref{PiSL2},
and the integrated electric field reads
\begin{equation} \label{Hem_sol_2}
\int_0^\tau \bE_{\bk}(t) \dd t
= \tau \bE^0_\bk + \hat \bk \times \Big( \big(\tau - \tfrac{L}{2\pi \abs{\bk}} s(\tau)\big) (\hat \bk \times \bE^0_\bk)
              +  \ii \tfrac{L}{2\pi \abs{\bk}} (1-c(\tau)) \bB^0_\bk \Big)
\end{equation}
where we have set
\begin{equation} \label{Hem_sol_3}
  c(\tau) = \cos\big(\tfrac{2\pi \abs{\bk}\tau}{L}\big),
  \qquad
  s(\tau) = \sin\big(\tfrac{2\pi \abs{\bk}\tau}{L}\big)
  \qquad
  \text{with} \quad
  \abs{\bk} = (\bk\cdot\bk)^{\frac 12},
\end{equation}
and $\hat \bk := \bk /\abs{\bk}$ if $\bk \neq 0$, otherwise $\hat \bk := 0$.

\subsection{Explicit steps for the Fourier-GEMPIC method}
\label{sec:f-gempic-steps}

We now consider the Fourier-GEMPIC method defined by the
pseudo-differential DFT operators from Section~\ref{sec:psd}.

\paragraph{Directional kinetic steps.}

We have the following result.

\begin{lemma} \label{lem:sol_kin_dft}
For the operators $\Pi^\ell$ defined as in Section~\ref{sec:psd},
the exact solution $\vp_{\tau,v_\alpha}: \arrU^0 \to \arrU(\tau)$
to the kinetic split step \eqref{Hkin-f} in a direction
$\alpha \in \range{1}{3}$ is given by the explicit expressions
\begin{equation} \label{f-gempic_kin}
  \vp_{\tau,v_\alpha}: \qquad
  \left\{
  \begin{aligned}
  X_{p,\alpha}(\tau) &= X_{p,\alpha}^0 + \tau V_{p,\alpha}
  \\ \noalign{\smallskip}
  X_{p,\alpha + \nu}(\tau) &= X_{p,\alpha + \nu}^0
  \\ \noalign{\smallskip}
  V_{p,\alpha}(\tau) &= V_{p,\alpha}^0
  \\
  V_{p,\alpha + \nu}(\tau) &= V_{p,\alpha + \nu}^0
  - \nu \tau \frac{q_p}{m_p} L^3 \sum_{\bk \in \range{-K}{K}^3} B_{\alpha -\nu,\bk} \, \overline{\hat V^{2,\nu}_{p, \alpha, \bk}(\tau)}
  \\
  E_{\alpha,\bk}(\tau) &= E_{\alpha,\bk}^0  - \tau \sum_{p = 1\cdots N} q_p \hat V^{3}_{p, \alpha, \bk}(\tau)
  \\
  E_{\alpha + \nu,\bk}(\tau) &= E_{\alpha + \nu,\bk}^0
  \\ \noalign{\smallskip}
  \bB_\bk(\tau) &= \bB_\bk^0
  \end{aligned}
  \right.
\end{equation}
for all $p \in \range{1}{N}$, $\bk \in \range{-K}{K}^3$ and $\nu = \pm 1$, with
\begin{equation} \label{f-gempic_hv2}
  \hat V_{p, \alpha, \bk}^{2,\nu}(\tau)
    :=
  \begin{cases}
    V_{p,\alpha}( \Pi^2_{\alpha + \nu} S_{\px_p^0} )_{\bk} 
    &\text{ if } k_\alpha = 0
    \\ \noalign{\medskip}
    - \Big(\tfrac{2\ii\pi k_\alpha}{L}\Big)^{-1} \frac 1 \tau \big[( \Pi^2_{\alpha + \nu} S_{\px_p} )_{\bk}\big]^\tau_0
    &\text{ if } k_\alpha \neq 0
  \end{cases}
\end{equation}
and
\begin{equation} \label{f-gempic_hv3}
  \hat V_{p, \alpha, \bk}^{3}(\tau)
    :=
  \begin{cases}
    V_{p,\alpha}( \Pi^3 S_{\px_p^0} )_{\bk} 
    &\text{ if } k_\alpha = 0
    \\ \noalign{\medskip}
    - \Big(\tfrac{2\ii\pi k_\alpha}{L}\Big)^{-1} \frac 1 \tau\big[( \Pi^3 S_{\px_p} )_{\bk}\big]^\tau_0
    &\text{ if } k_\alpha \neq 0.
  \end{cases}
\end{equation}
\end{lemma}

\bproof
We remind the velocity equation from \eqref{Hkin-f},
$$
\frac{\dd }{\dd t} V_{p,\alpha + \nu}
  = -\nu  \frac{q_p}{m_p} L^3 \sum_{\bk}
  B_{\alpha -\nu, \bk}
V_{p,\alpha}
\overline{(\Pi^1_{\alpha - \nu} S_{\px_p})_{\bk}}
$$
and the electric one,
$$
\frac{\dd }{\dd t} E_{\alpha,\bk}
  = - \sum_p q_p V_{p,\alpha}
  (\Pi^2_\alpha S_{\px_p})_{\bk}.
$$
Here by definition of the pseudo-differential projection operators, we have
\begin{equation} \label{Pi1aS-psd}
(\Pi^1_{\alpha -\nu } S_{\px_p})_{\bk}
= \gamma_\bk \hat D_{\bk,\alpha}^{-1} \hat D_{\bk,\alpha + \nu}^{-1} \tilde \cF_{M,\bk} (\tilde D_{\bk,\alpha}\tilde D_{\bk,\alpha + \nu} S_{\px_p})
\end{equation}
and
\begin{equation} \label{Pi2aS-psd}
(\Pi^2_\alpha S_{\px_p})_{\bk} = \gamma_\bk  \hat D_{\bk,\alpha}^{-1} \tilde \cF_{M,\bk} (\tilde D_{\bk,\alpha} S_{\px_p}).
\end{equation}
We begin with the term involving $\Pi^2$ as the computations are simpler. We first observe that
if $k_\alpha = 0$, then
$$
V_{p,\alpha} (\Pi^2_{\alpha} S_{\px_p})_{\bk}
  =  \gamma_\bk V_{p,\alpha} \tilde \cF_{ M,\bk }( S_{\px_p})
  =  V_{p,\alpha} (\Pi^3 S_{\px_p})_{\bk}
$$
and this quantity is constant, since the conservative DFT coefficient is an integral along $\uvec_\alpha$.
Next if $k_\alpha \neq 0$, we compute
$$
\begin{aligned}
  V_{p,\alpha} (\Pi^2_{\alpha} S_{\px_p})_{\bk}
    &= \gamma_\bk \Big(\tfrac{2\ii\pi k_\alpha}{L}\Big)^{-1} V_{p,\alpha} \tilde \cF_{M,\bk}(\partial_\alpha S_{\px_p})
    \\
    & = \gamma_\bk \Big(\tfrac{2\ii\pi k_\alpha}{L}\Big)^{-1} \frac {V_{p,\alpha}}{M^3}
              \sum_{\bsm} \partial_\alpha S(\bsm h - \px_p(t)) \ee^{-\frac{2\ii\pi \bk \cdot \bsm}{M}}
    \\
    & = -\gamma_\bk \Big(\tfrac{2\ii\pi k_\alpha}{L}\Big)^{-1} \frac {1}{M^3} \sum_{\bsm}
      \frac{\dd}{\dd t} \Big(S(\bsm h - \px_p(t))\Big) \ee^{-\frac{2\ii\pi \bk \cdot \bsm}{M}}
    \\
    & = - \Big(\tfrac{2\ii\pi k_\alpha}{L}\Big)^{-1} \frac{\dd}{\dd t} ( \Pi^3 S_{\px_p} )_{\bk}.
\end{aligned}
$$
It follows that we can integrate exactly
$
\int_0^\tau V_{p,\alpha} (\Pi^2_{\alpha} S_{\px_p})_{\bk} \dd t
=
\tau \hat V^{3}_{p, \alpha, \bk}(\tau)
$
with the expression from \eqref{f-gempic_hv3}, which provides the update of $E_{\alpha,\bk}$.
Turning to the terms involving $\Pi^1$ we observe that if $k_\alpha = 0$, then
$$
V_{p,\alpha} (\Pi^1_{\alpha - \nu} S_{\px_p})_{\bk}
  =  \gamma_\bk V_{p,\alpha} \hat D_{\bk,\alpha+\nu}^{-1} \tilde \cF_{ M,\bk }( \tilde D_{\bk,\alpha+\nu} S_{\px_p})
  =  V_{p,\alpha} (\Pi^2_{\alpha + \nu} S_{\px_p})_{\bk}
$$
and again this quantity is constant, for the same reason as above.
Now if $k_\alpha \neq 0$, then using $\frac{\dd}{\dd t} \px_p(t) = \pv_p^{[\alpha]} = V_{p,\alpha}\uvec_\alpha $ we compute
$$
\begin{aligned}
  V_{p,\alpha} (\Pi^1_{\alpha-\nu} S_{\px_p})_{\bk}
    &= \gamma_\bk \Big(\tfrac{2\ii\pi k_\alpha}{L}\Big)^{-1} V_{p,\alpha} \hat D_{\bk,\alpha + \nu}^{-1} \tilde \cF_{M,\bk}(\partial_\alpha \tilde D_{\bk,\alpha + \nu} S_{\px_p})
    \\
    & = \gamma_\bk \Big(\tfrac{2\ii\pi k_\alpha}{L}\Big)^{-1} \hat D_{\bk,\alpha +\nu}^{-1} \frac {V_{p,\alpha}}{M^3}
              \sum_{\bsm} \partial_\alpha \tilde D_{\bk,\alpha + \nu} S(\bsm h - \px_p(t)) \ee^{-\frac{2\ii\pi \bk \cdot \bsm}{M}}
    \\
    & = -\gamma_\bk\Big(\tfrac{2\ii\pi k_\alpha}{L}\Big)^{-1} \hat D_{\bk,\alpha+\nu}^{-1} \frac {1}{M^3} \sum_{\bsm}
      \frac{\dd}{\dd t} \Big(\tilde D_{\bk,\alpha+\nu} S(\bsm h - \px_p(t))\Big) \ee^{-\frac{2\ii\pi \bk \cdot \bsm}{M}}
    \\
    & = - \Big(\tfrac{2\ii\pi k_\alpha}{L}\Big)^{-1} \frac{\dd}{\dd t} ( \Pi^2_{\alpha+\nu} S_{\px_p} )_{\bk}.
\end{aligned}
$$
It follows that
$
\int_0^\tau V_{p,\alpha} (\Pi^1_{\alpha-\nu} S_{\px_p})_{\bk} \dd t
=
\tau \hat V^{2,\nu}_{p, \alpha, \bk}(\tau),
$
now with the expression from \eqref{f-gempic_hv2}. This gives the update for $V_{p,\alpha + \nu}$.
\eproof

\paragraph{Electromagnetic step.}
The exact solution flow $\vp_{\tau,EB}$ is again given by \eqref{Hem_sol}--\eqref{Hem_sol_3}, up to
replacing the $\Pi^2$ projected term in \eqref{Hem_sol} by its specific expression, see \eqref{Pi2aS-psd}.

\subsection{A Gauss and momentum preserving Fourier-PIC scheme}
\label{sec:f-pic-steps}

By applying the same splitting method as above to the momentum and Gauss preserving scheme presented in
Section~\ref{sec:mom-gauss-variant}, we obtain a standard Fourier-PIC coupling as described in Section~\ref{sec:f-pic},
as follows. For the kinetic subsystem \eqref{Hkin-f} along $\alpha$, the modification from \eqref{EBS-k} to \eqref{EBS_mom-k} applies
to $\alpha \leftarrow \alpha + \nu$ and amounts to replacing the operator $\Pi^1_{\alpha - \nu}$ by $\Pi^2_\alpha$.
For the electromagnetic subsystem \eqref{Hem-f} it just amounts to replacing $\Pi^2_\alpha$ by $\Pi^3$.
The modified equations read then
$$
\left\{
\begin{aligned}
  &\tfrac{\dd }{\dd t} X_{p,\alpha} = V_{p,\alpha}
  \\
  &\tfrac{\dd }{\dd t} X_{p,\alpha + \nu} = 0
  \\
  &\tfrac{\dd }{\dd t} V_{p,\alpha} = 0
  \\
  &\tfrac{\dd }{\dd t} V_{p,\alpha + \nu} = -\nu  \frac{q_p}{m_p} L^3 \sum_{\bk} V_{p,\alpha}
      B_{\alpha - \nu,\bk} \overline{(\Pi^2_{\alpha} S_{\px_p})_{\bk}}
  \\
  &\tfrac{\dd }{\dd t} E_{\alpha,\bk} = - \sum_p q_p V_{p,\alpha} (\Pi^2_\alpha S_{\px_p})_{\bk}
  \\
  &\tfrac{\dd }{\dd t} E_{\alpha + \nu,\bk} = 0
  \\
  &\tfrac{\dd }{\dd t} \bB_\bk = 0
\end{aligned}
\right.
\quad \text{ and } \quad
\left\{
\begin{aligned}
  &\tfrac{\dd }{\dd t} \px_p = 0
  \\
  &\tfrac{\dd }{\dd t} \pv_p = \frac{q_p}{m_p} L^3 \sum_\bk \bE_{\bk} \overline{(\Pi^3 S_{\px_p})_{\bk}}
  \\
  &\tfrac{\dd }{\dd t} \bE_\bk = \tfrac{2\ii\pi\bk}{L} \times \bB_\bk
  \\
  &\tfrac{\dd }{\dd t} \bB_\bk = - \tfrac{2\ii\pi\bk}{L} \times \bE_\bk,
  \end{aligned}
  \right.
$$
and explicit solutions are obtained by computing as in the proof of Lemma~\ref{lem:sol_kin_dft}.
For the kinetic subsystem along $\alpha$, the flow $\vp_{\tau,v_\alpha}$ is given by
$$
\vp_{\tau,v_\alpha}: \qquad
  \left\{
  \begin{aligned}
  X_{p,\alpha}(\tau) &= X_{p,\alpha}^0 + \tau V_{p,\alpha}
  \\ \noalign{\smallskip}
  X_{p,\alpha + \nu }(\tau) &= X_{p,\alpha + \nu }^0
  \\ \noalign{\smallskip}
  V_{p,\alpha}(\tau) &= V_{p,\alpha}^0
  \\
  V_{p,\alpha + \nu}(\tau) &= V_{p,\alpha + \nu}^0
    -\nu \tau \frac{q_p}{m_p} L^3 \sum_{\bk \in \range{-K}{K}^3} B_{\alpha -\nu, \bk} \overline{\hat V^{3}_{p, \alpha, \bk}(\tau)}
  \\
  E_{\alpha,\bk}(\tau) &= E_{\alpha,\bk}^0 - \tau \sum_p q_p \hat V^{3}_{p, \alpha, \bk}(\tau)
  \\
  E_{\alpha + \nu,\bk}(\tau) &= E_{\alpha + \nu,\bk}^0
  \\ \noalign{\smallskip}
  \bB_K(\tau) &= \bB_K^0
  \end{aligned}
  \right.
$$
where we remind that
$
\hat V_{p, \alpha, \bk}^{3}(\tau)
$ is defined in \eqref{f-gempic_hv3}.
For the electromagnetic step the flow
$\vp_{\tau,EB}$ corresponds to \eqref{Hem_sol}--\eqref{Hem_sol_3} with a modified velocity equation, namely
\begin{equation} \label{Hem_xv_sol-1-P3}
V_{p,\alpha}(\tau) = V_{p,\alpha}^0 + \frac{q_p}{m_p} L^3
    \sum_{\bk \in \range{-K}{K}^3} \Big( \int_0^\tau E_{\alpha,\bk}(t) \dd t \Big) \overline{(\Pi^3 S_{\px_p^0})_{\bk}}.
\end{equation}
In particular, we see that the fully discrete steps all involve the operator $\Pi^3$,
see \eqref{cDFT-1}, which corresponds to the standard DFT coupling described in
Section~\ref{sec:f-pic}, up to the use of exact integrals \eqref{cDFT-2} along zero-modes.
This scheme does not have a Hamiltonian structure, but it satisfies some
important conservation properties.

\begin{lemma} \label{spic:cons}
  The Fourier-PIC scheme described above preserves the discrete Gauss laws
  \eqref{gfp-Gauss} and the discrete momentum \eqref{P}.
\end{lemma}
\bproof
The proof is a matter of elementary computations, similar to the ones involved in the proofs of Theorem~\ref{th:cons}.
One key observation is that the above modifications in the velocity equations have no effect
on the Gauss laws being preserved, and they precisely lead to an exact momentum preservation.
\eproof

\subsection{Summary of the proposed methods}

In the above sections we have derived geometric Fourier-particle methods
following the Discrete Action Principle formalized in \cite{CPKS_variational_2020},
where the coupling between the fields and the particles
is represented by abstract projection operators on the truncated Fourier spaces
that satisfy a commuting diagram property.
This results in semi-discrete schemes that preserve the Gauss laws and have a discrete Hamiltonian structure
relying on a non-canonical Poisson bracket.

When the coupling operators are defined as $L^2$ projections, the coupling is gridless
and essentially relies on continuous Fourier coefficients. The resulting method
preserves the total momentum in addition to the charge and energy,
and corresponds to the Particle-In-Fourier (PIF) approach.
Here it is called {GEMPIF} to emphasize its geometric nature.

When the coupling operators are defined as pseudo-differential DFT projections,
the coupling involves a grid and essentially relies on discrete Fourier coefficients.
The resulting method can be seen as a variant of standard spectral PIC methods and is called
{Fourier-GEMPIC}.

Fully discrete schemes of various orders in time have then been constructed by
a Hamiltonian splitting procedure. These schemes are Poisson maps, in particular they
preserve the discrete Poisson bracket and a modified energy
that approximates the exact one to the time order of the splitting.

Finally we have observed that {Fourier-PIC} schemes with standard coupling terms
can be obtained as a variant of the above Fourier-GEMPIC method.
These schemes are not Hamiltonian, but they preserve exactly the Gauss laws and the total momentum.

Since our approach handles general shape functions $S$ and arbitrary filter coefficients $\gamma_\bk$,
we may use high order shapes with anti-aliasing properties in the Fourier-GEMPIC methods,
and the ad-hoc {\em back-filtering} procedure \eqref{Jk_pic_bf}--\eqref{EB_bf} to avoid the
damping of relevant modes in the computational range.

\section{Numerical illustration in 1D2V and 1D1V}
\label{sec:num}

In this section we present some numerical results obtained with several of the methods described above.
Specifically, we will compare
\begin{itemize}
  \item the GEMPIF scheme described in Section~\ref{sec:gempif-steps}
  with point shape functions ($S=\delta$),
  \item the Fourier-GEMPIC scheme described in Section~\ref{sec:f-gempic-steps},
  involving a DFT grid with $M$ points and B-spline shapes of degree $\kappa$.
  To assess the benefits of back-filtering, we will use two versions of this scheme:
  a plain {\em smoothed Fourier} SF-GEMPIC method corresponding to $\gamma_\bk = 1$,
  and a {\em back-filtered Fourier} BFF-GEMPIC method that involves filter coefficients
  \begin{equation} \label{bf_gk}
    \gamma_\bk := \frac{1}{\sigma_\bk} = \prod_{\alpha = 1}^3 \left( \sinc\Big(\frac{\pi k_\alpha}{\mu(2K+1)}\Big)\right)^{-(\kappa+1)}
      \qquad \text{with} \qquad \mu := \frac{M}{2K+1} \ge 1
  \end{equation}
  where we remind that $\mu$ is the oversampling parameter, see \eqref{sigmak}.

  \item the Fourier-PIC scheme described in Section~\ref{sec:f-pic-steps},
  which also uses a DFT grid with $M$ points and B-spline shapes of degree $\kappa$.
  Similarly as above, this scheme will be used in two versions, namely
  a basic {\em smoothed Fourier} SF-PIC method corresponding to $\gamma_\bk = 1$,
  and a {\em back-filtered Fourier} BFF-PIC method
  involving the coefficients \eqref{bf_gk}.
\end{itemize}

\subsection{Periodic plasma test-cases}
\label{sec:cases}

For our numerical experiments we consider a periodic one-species model in 1D2V and 1D1V
similarly as in \cite{CPKS_variational_2020}, with zero-mean current
to preserve the total momentum, i.e.
$$
\dt E_1 = -J_1 + \frac{1}{L_1} \int_0^{L_1} J_1, 
\qquad
\dt E_2 + \partial_{x_1} B_3 = -J_2 + \frac{1}{L_1} \int_0^{L_1} J_1, 
\qquad
\dt B_3 + \partial_{x_1} E_3 = 0,
$$
and standard plasma test-cases:
classical {\bf Landau damping} test-cases corresponding to
$$
f^0(x,v_1) = (1+ \epsilon \cos(\sfk x)) \frac {1}{\sqrt{2\pi}} \exp\Big(-\frac{v_1^2}{2}\Big)
$$
with $\sfk = 0.5$ and $\epsilon = 0.5$ or $0.01$ for the strong and weak damping,
a standard {\bf Weibel instability} \cite{Weibel:1959, kraus2016gempic} where
$$
f^0(x,v_1,v_2) = \frac {1}{2\pi\sigma_1\sigma_2} \exp\Big(-\frac 12 \Big(\frac{v_1^2}{\sigma_1^2}+\frac{v_2^2}{\sigma_2^2}\Big)\Big)
$$
and where the initial magnetic field is $B_3(t=0,x_1) = \beta \cos(\sfk x_1)$,
with $\sigma_1 = 0.02/\sqrt{2}$ and $\sigma_2=\sqrt{12}\sigma_1$, $\beta = 10^{-4}$ and $\sfk=1.25$,
a {\bf two-stream instability}
$$
f^0(x,v_1) = (1+ \epsilon \cos(\sfk x)) \frac {1}{2\sqrt{2\pi}} \Big(\exp\big(-\frac{(v_1+2.4)^2}{2}\big)-\exp\big(-\frac{(v_1-2.4)^2}{2}\big)\Big)
$$
with $\epsilon = 5\cdot 10^{-4}$ and $\sfk = 0.2$,
and a {\bf bump-on-tail instability} corresponding to
$$
f^0(x,v_1) = (1+ \epsilon \cos(\sfk x)) \frac {1}{\sqrt{2\pi}}
\Big(\delta \exp\big(-\frac{v_1^2}{2} \big)
  + 2(1-\delta) \exp\big(-2(v_1-3.5)^2 \big)
\Big)
$$
with $\delta = 0.9$, $\epsilon = 5\cdot10^{-3}$ and $\sfk = 0.3$.
For these test-cases the domain size is $L_1 = \frac{2\pi}{\sfk}$, except for the last one
where we take $L_1 = \frac{4\pi}{\sfk}$. The initial $B_3$ field is zero unless stated otherwise,
and the initial $\bE$ field is computed from the particles by solving the periodic Poisson equation.
In practice we run all these cases using a 1D2V implementation of the above methods.
When the initial distribution is only 1D1V we sample the particles following a Maxwellian
in $v_2$ with thermal velocity $\sigma_2 = 1$.

\subsection{Qualitative energy behaviors}
\label{sec:qualitative}

In order to compare the qualitative response of the different schemes,
we first plot the relevant energy curves for the above test-cases using
rather coarse numerical parameters such as $K$, 
$M$ 
and $\kappa$, but a number of particles $N$ high enough to match the reference damping
or growth rates.

\begin{figure}[tp]
  \begin{subfigure}{.48\textwidth}
    \centering
    \includegraphics[width=.9\linewidth]{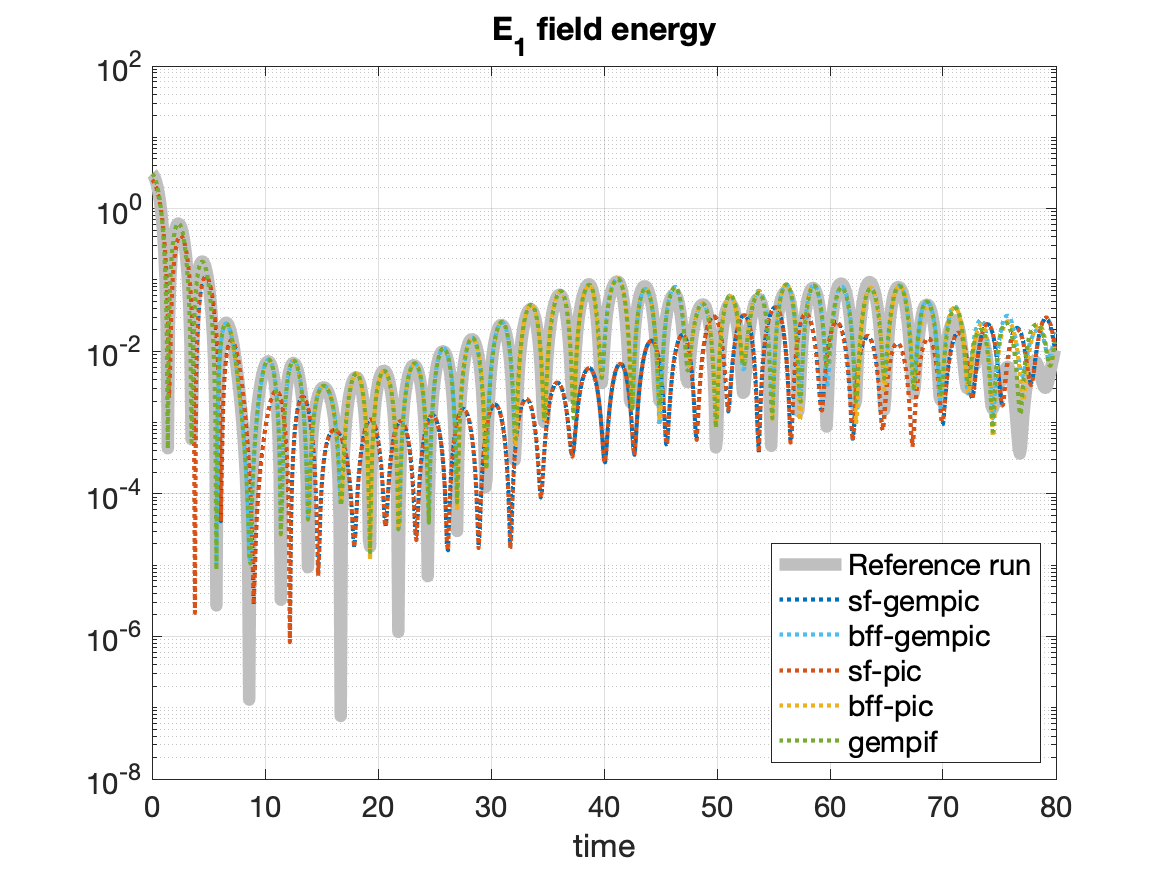}
    \subcaption{strong damping with an $M=8$ grid}
  \end{subfigure}%
  \hspace{10pt}
  \begin{subfigure}{.48\textwidth}
    \centering
    \includegraphics[width=.9\linewidth]{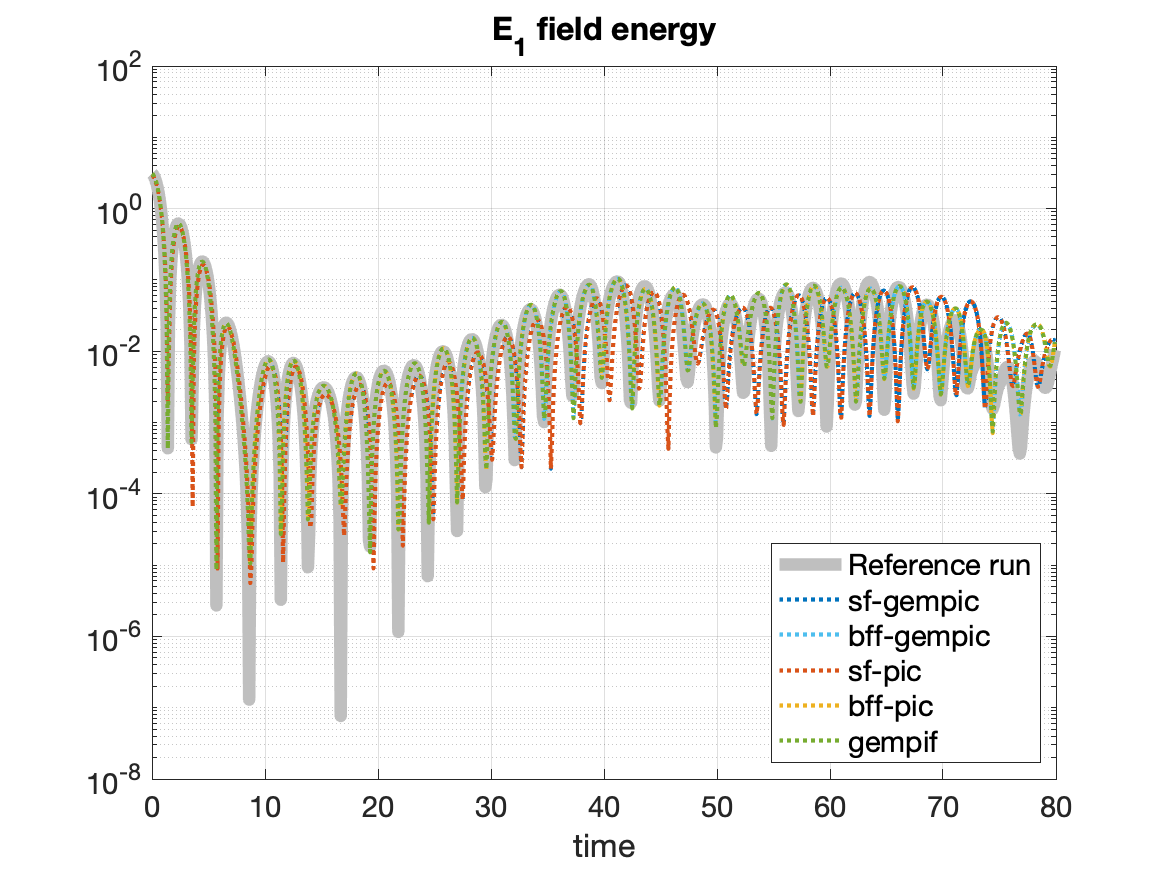}
    \subcaption{strong damping with an $M=16$ grid}
  \end{subfigure}
  \vspace{10pt}
  \\
  \begin{subfigure}{.48\textwidth}
    \centering
    \includegraphics[width=.9\linewidth]{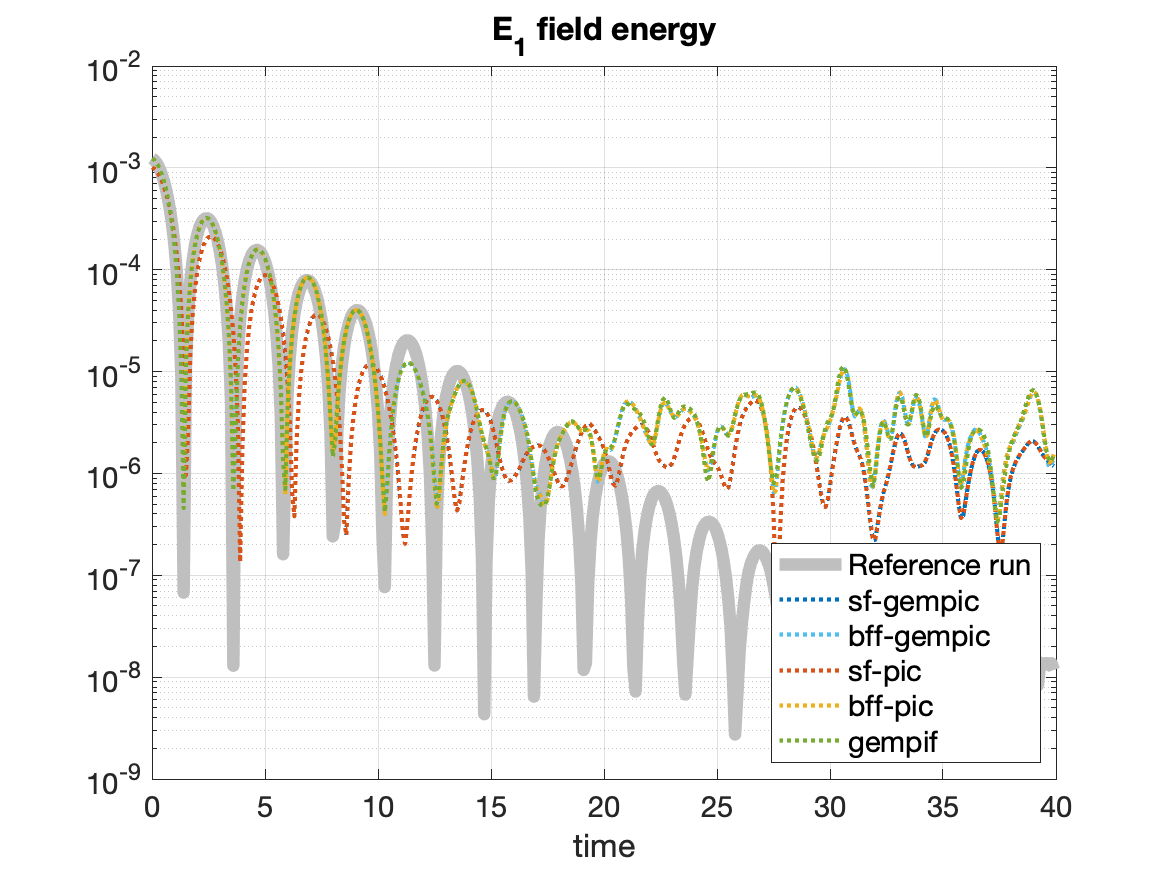}
    \subcaption{weak damping with an $M=8$ grid}
  \end{subfigure}
  \hspace{10pt}
  \begin{subfigure}{.48\textwidth}
    \centering
    \includegraphics[width=.9\linewidth]{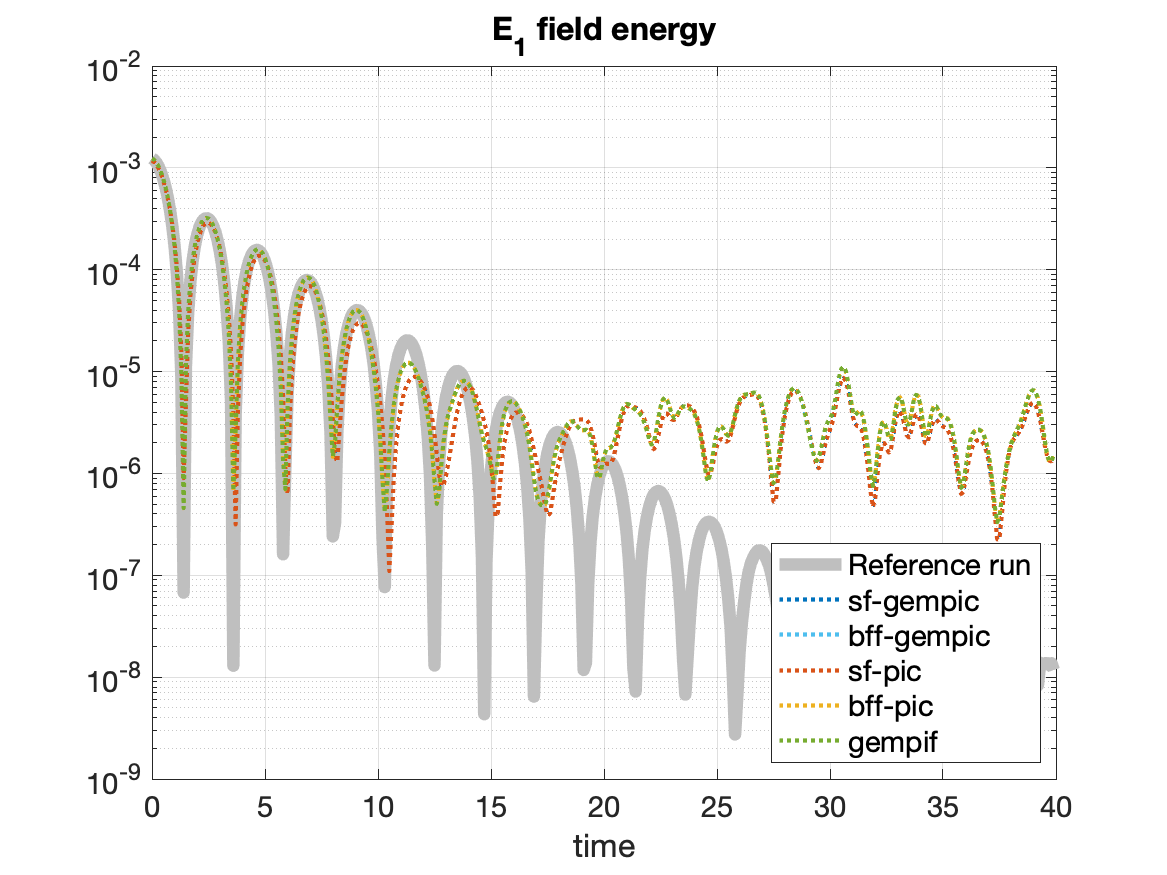}
    \subcaption{weak damping with an $M=16$ grid}
  \end{subfigure}
  \caption{
  Landau damping test-cases: $E_1$ field energy curves, using $K = 3$ Fourier modes,
  $N = 5\cdot 10^4$ particles for the strong damping (top) and $N = 10^5$ for the weak damping (bottom).
  Gridded (GEMPIC and PIC) methods use B-spline shapes of degree $\kappa=3$ and DFT grids with $M = 8$ (left) or $M=16$ points (right).
  }
  \label{fig:landau_diags_deg3}
\end{figure}

\begin{figure}[tp]
  \begin{subfigure}{.48\textwidth}
    \centering
    \includegraphics[width=.9\linewidth]{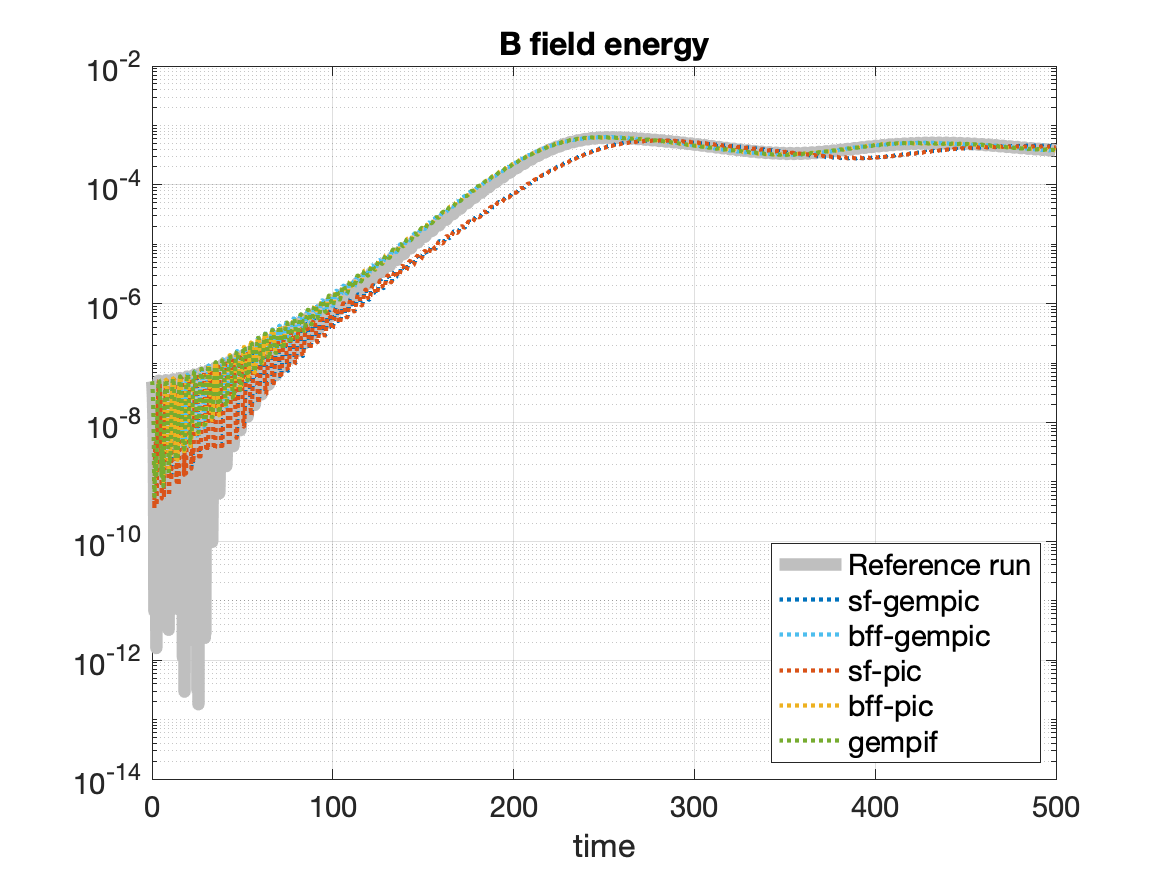}
    \subcaption{Weibel instability with an $M=8$ grid}
  \end{subfigure}
  \hspace{10pt}
  \begin{subfigure}{.48\textwidth}
    \centering
    \includegraphics[width=.9\linewidth]{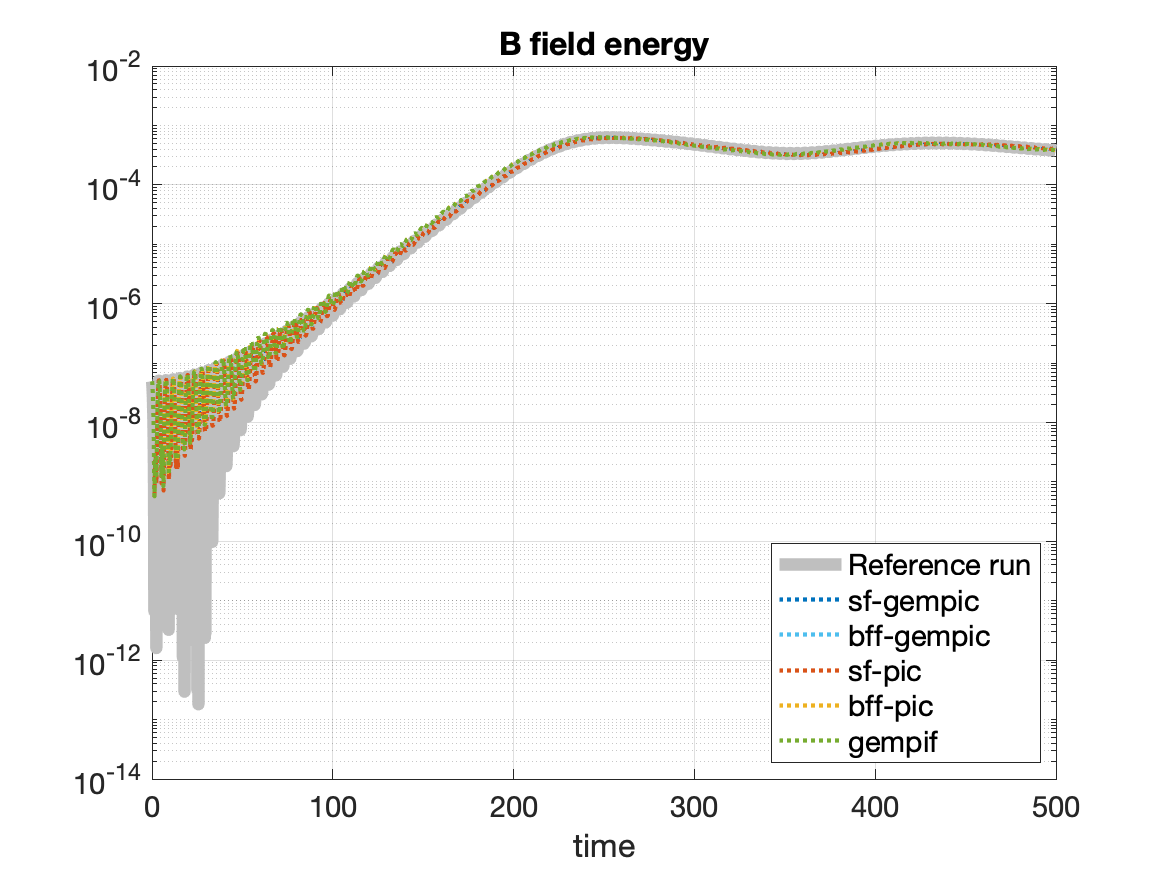}
    \subcaption{Weibel instability with an $M=16$ grid}
  \end{subfigure}
  \vspace{10pt}
  \\
  \begin{subfigure}{.48\textwidth}
    \centering
    \includegraphics[width=.9\linewidth]{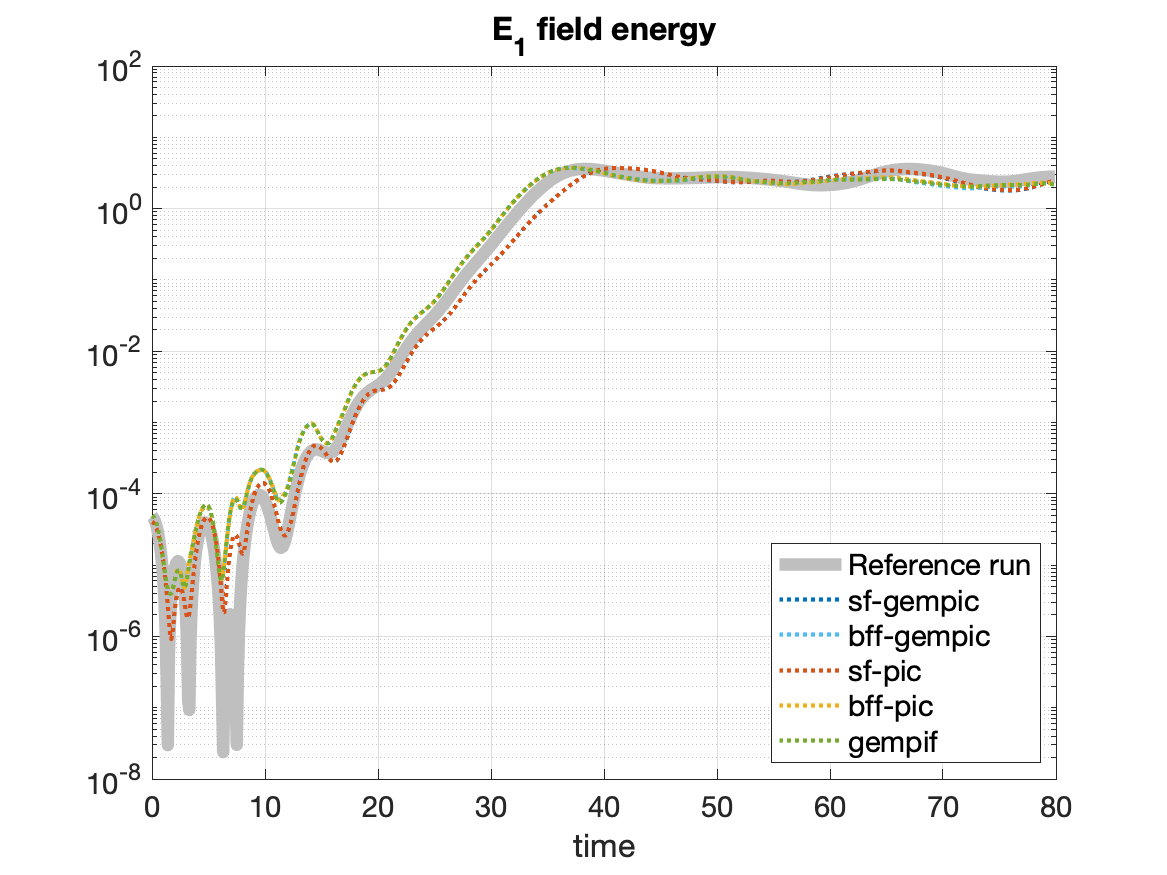}
    \subcaption{two stream instability with an $M=8$ grid}
  \end{subfigure}%
  \hspace{10pt}
  \begin{subfigure}{.48\textwidth}
    \centering
    \includegraphics[width=.9\linewidth]{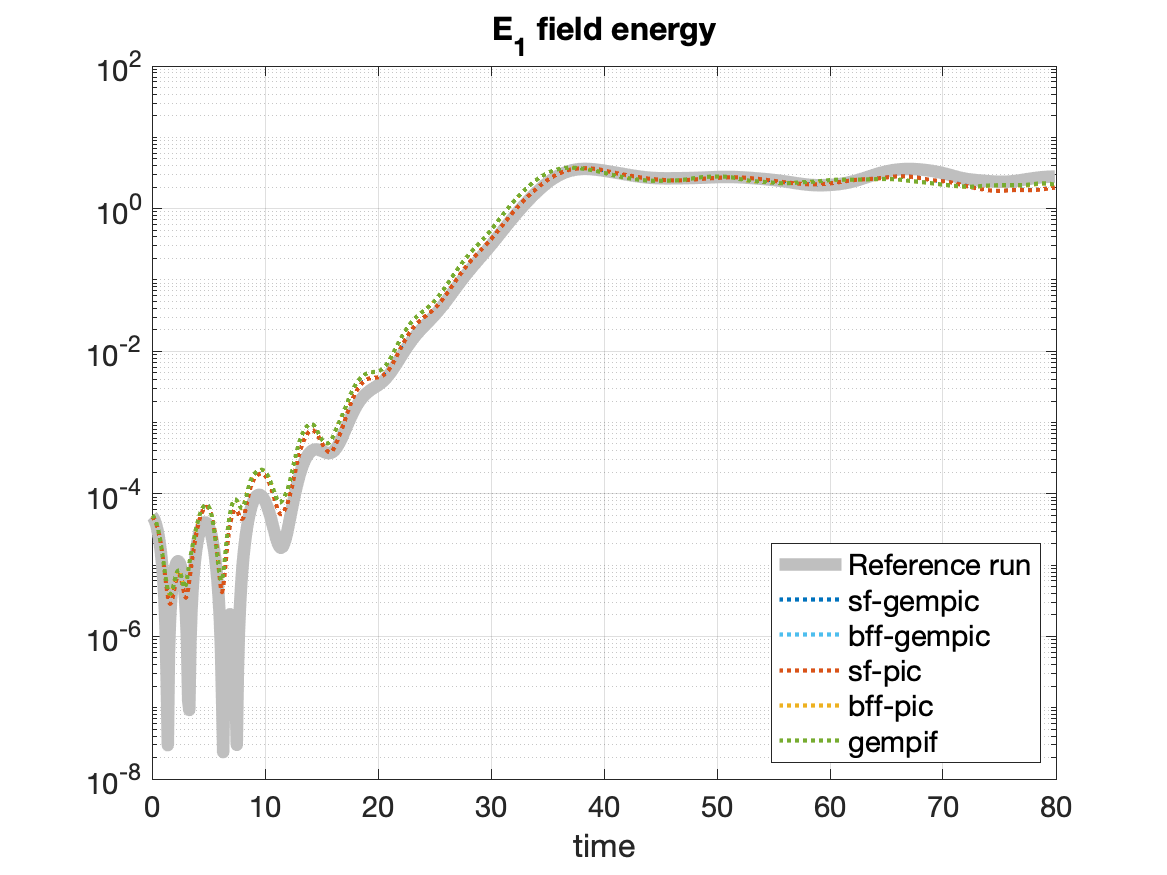}
    \subcaption{two stream instability with an $M=16$ grid}
  \end{subfigure}
  \vspace{10pt}
  \\
  \begin{subfigure}{.48\textwidth}
    \centering
    \includegraphics[width=.9\linewidth]{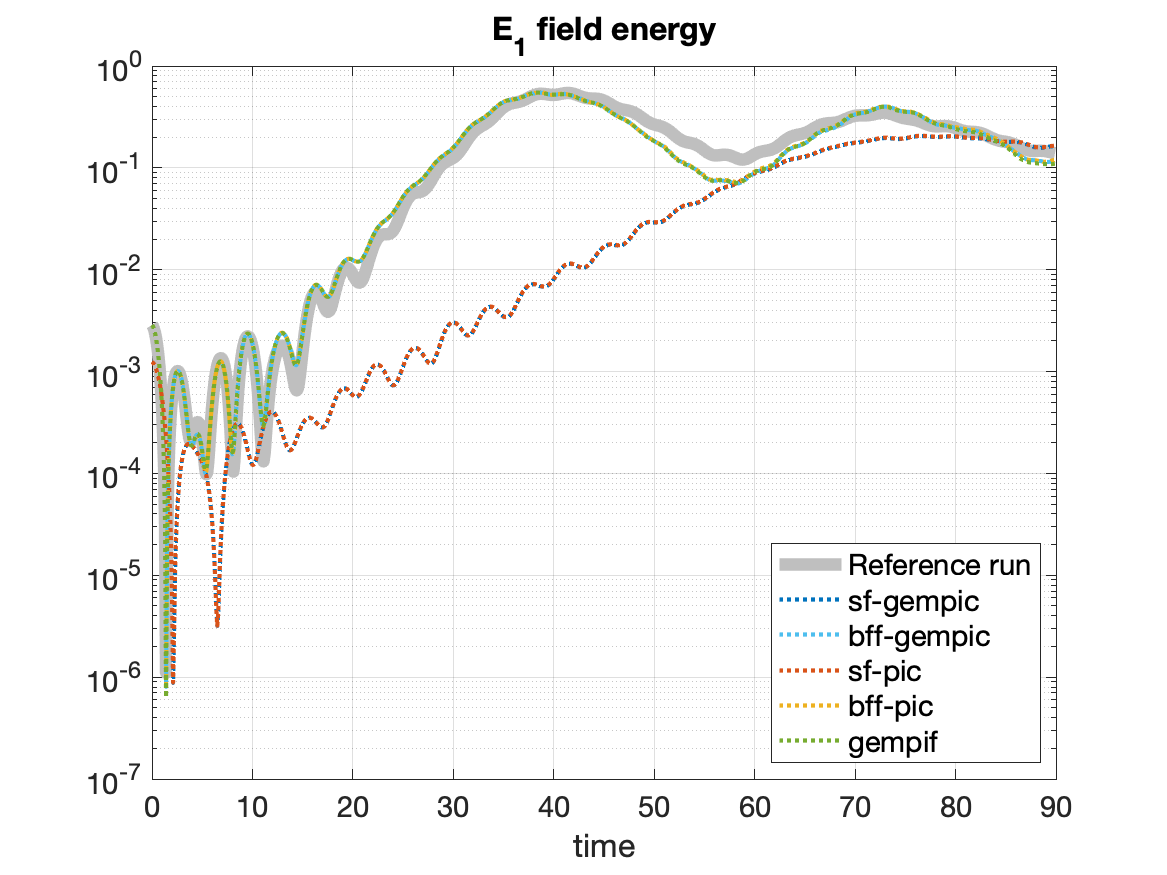}
    \subcaption{bump-on-tail instability with an $M=8$ grid}
  \end{subfigure}
  \hspace{10pt}
  \begin{subfigure}{.48\textwidth}
    \centering
    \includegraphics[width=.9\linewidth]{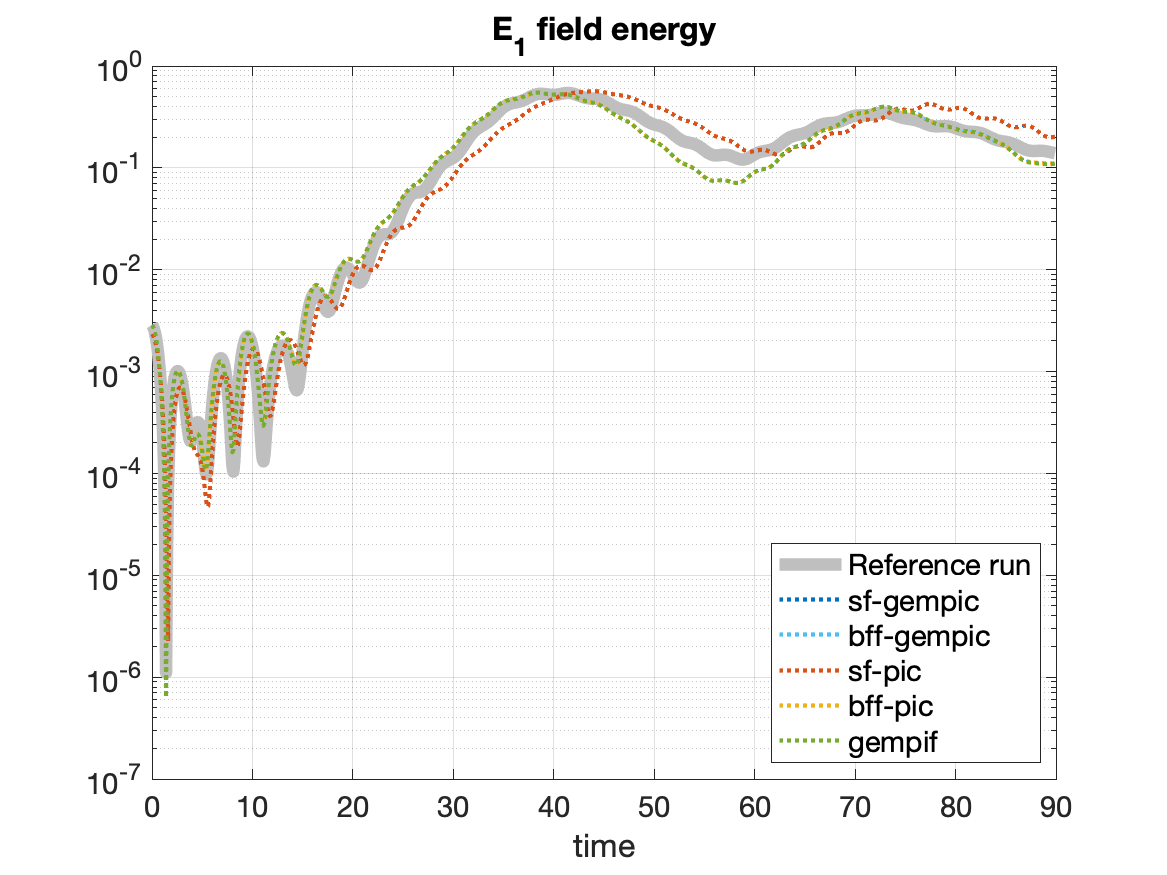}
    \subcaption{bump-on-tail instability with an $M=16$ grid}
  \end{subfigure}
  \caption{
  Energy behavior for instability test-cases: $B_3$ field energy for the Weibel instability (top) using $N = 10^4$ particles,
  and $E_1$ field energy for the two-stream (center) and bump-on-tail instabilities (bottom), using $N=5\cdot 10^4$ and $N=10^5$ particles, respectively.
  All runs use $K = 3$ Fourier modes, and gridded (GEMPIC and PIC) methods use B-spline shapes of degree $\kappa=3$
  and DFT grids with $M = 8$ (left) or $M=16$ points (right).
  }
  \label{fig:instab_diags_deg3}
\end{figure}

In Figure~\ref{fig:landau_diags_deg3} and \ref{fig:instab_diags_deg3} we plot the energy curves for the above test cases,
using $K = 3$ Fourier modes, splines of degree $\kappa=3$
and DFT grids with $M = 8$ (left) or $M=16$ points (right).
For the strong (top) and weak (bottom) Landau damping test-cases in Figure~\ref{fig:landau_diags_deg3}
we use $N = 5\cdot 10^4$ and $N = 10^5$ particles, respectively.
For the Weibel (top), two-stream (center) and bump-on-tail instabilities in Figure~\ref{fig:instab_diags_deg3} we
use $N = 10^4$, $N = 5\cdot 10^4$ and $N = 10^5$ particles, respectively.
The gray curves show reference runs, obtained using the GEMPIF scheme with $5\cdot 10^6$ particles
and $K = 15$ Fourier modes.
In all these runs, the time scheme is given by a fourth-order (Suzuki-Yoshida) composition method \eqref{S4}
with a time step of $\Dt = 0.1$.

Our first observation is that the qualitative energy behaviors of the different simulations essentially depends on whether
the method is back-filtered or not.
Indeed we see that for each test-case the energy curves of the BFF-GEMPIC and BFF-PIC methods are perfectly on top of that
of the gridless GEMPIF method, and they match very well the reference energy curve.
Moreover, their numerical damping and growth rates (not reported here) agree with the reference ones,
either given by dispersion relations from linear theory, or available in reference
results from the literature, as in the strong Landau damping.
(For the bump-on-tail instability the agreement of the numerical growth rates is weaker,
but this test-case is known to be more demanding in terms of particles.)
In contrast, the methods SF-GEMPIC and SF-PIC involving a regular smoothed-Fourier coupling
are also on top of each other, but in many cases they do not match the reference curve.
This is most visible with the coarsest grid or in the bump-on-tail test-case,
where the growth rates are clearly wrong.
With the finer grid (corresponding to an oversampling factor of $\mu = 16/7 \approx 2$),
the results on the right panels show an improvement of the energy accuracy, but not as good as with back-filtering.
Thus, these simulations show a clear benefit of back-filtered Fourier-PIC schemes
in the qualitative energy behavior of low-resolution methods.

\begin{figure}[tp]
  \begin{subfigure}{.48\textwidth}
    \centering
    \includegraphics[width=.9\linewidth]{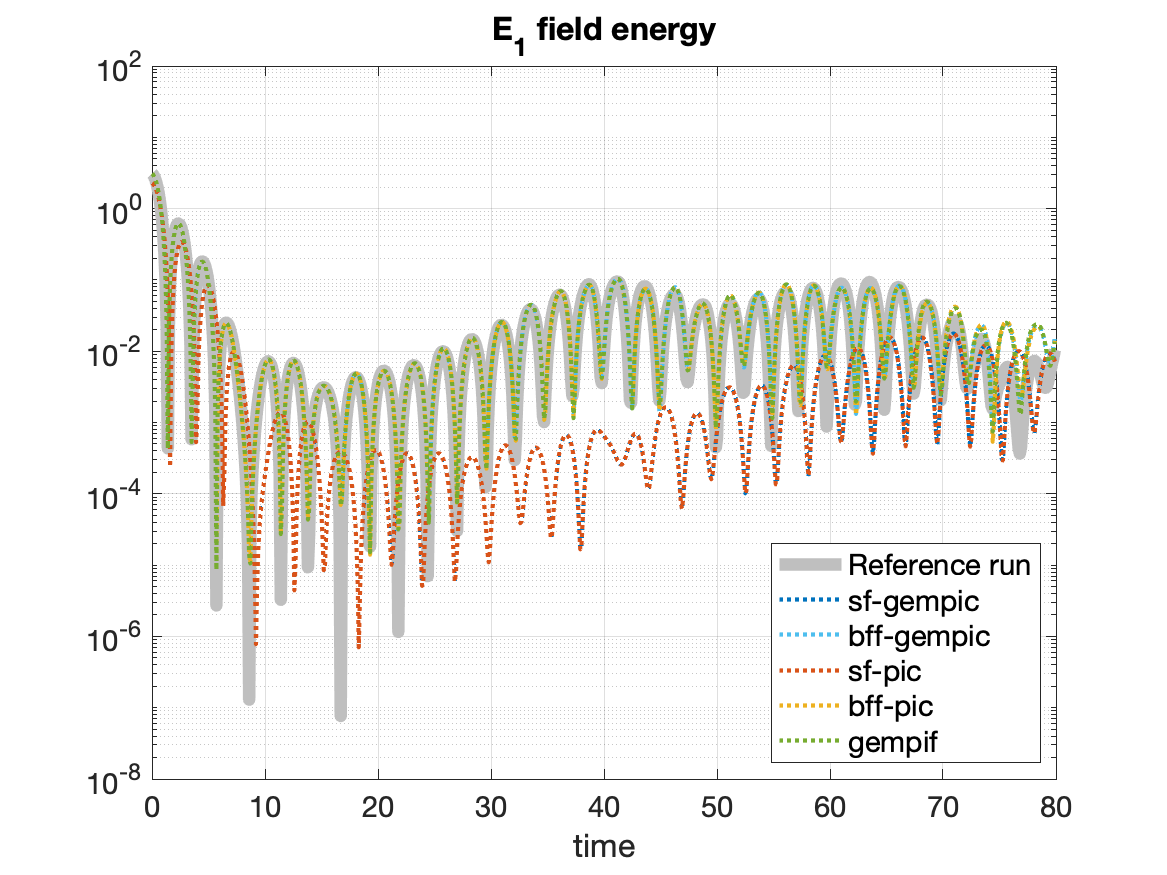}
    \subcaption{strong damping with an $M=8$ grid}
  \end{subfigure}%
  \hspace{10pt}
  \begin{subfigure}{.48\textwidth}
    \centering
    \includegraphics[width=.9\linewidth]{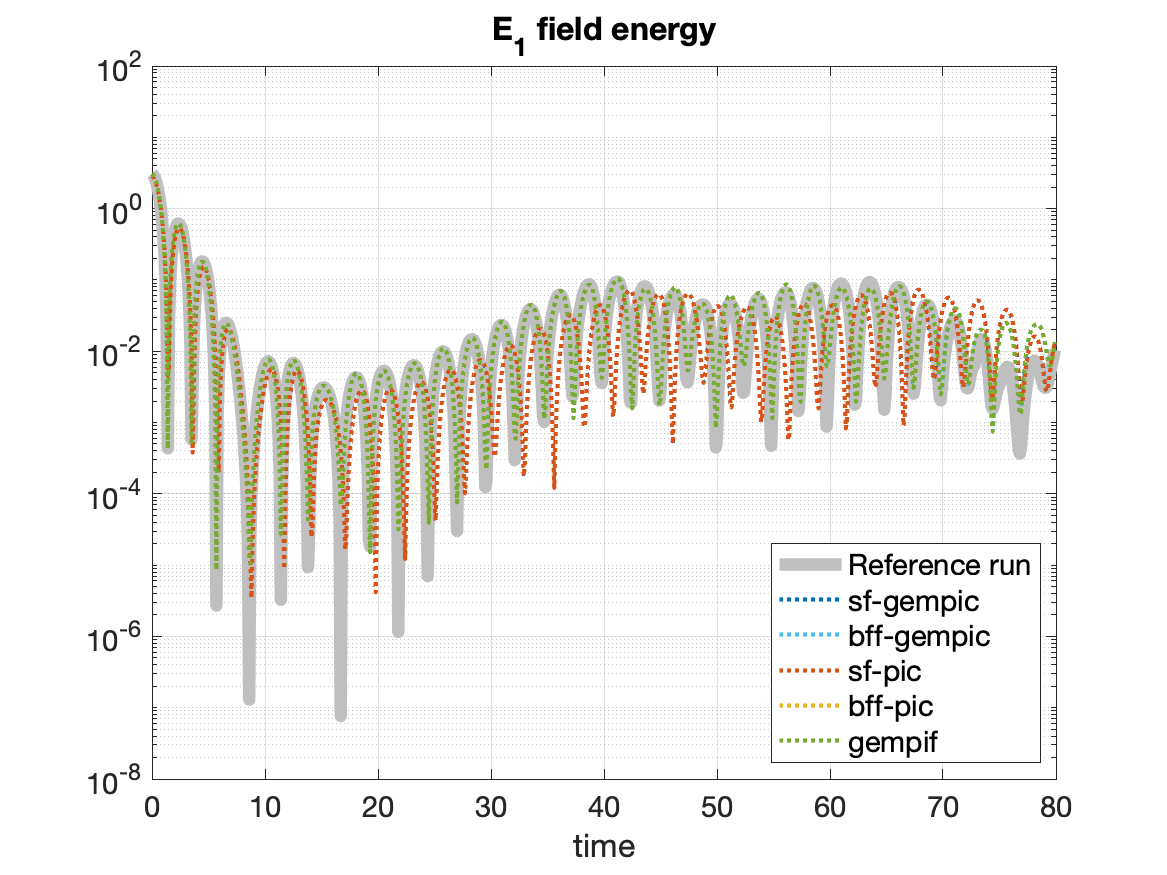}
    \subcaption{strong damping with an $M=16$ grid}
  \end{subfigure}
  \vspace{10pt}
  \\
  \begin{subfigure}{.48\textwidth}
    \centering
    \includegraphics[width=.9\linewidth]{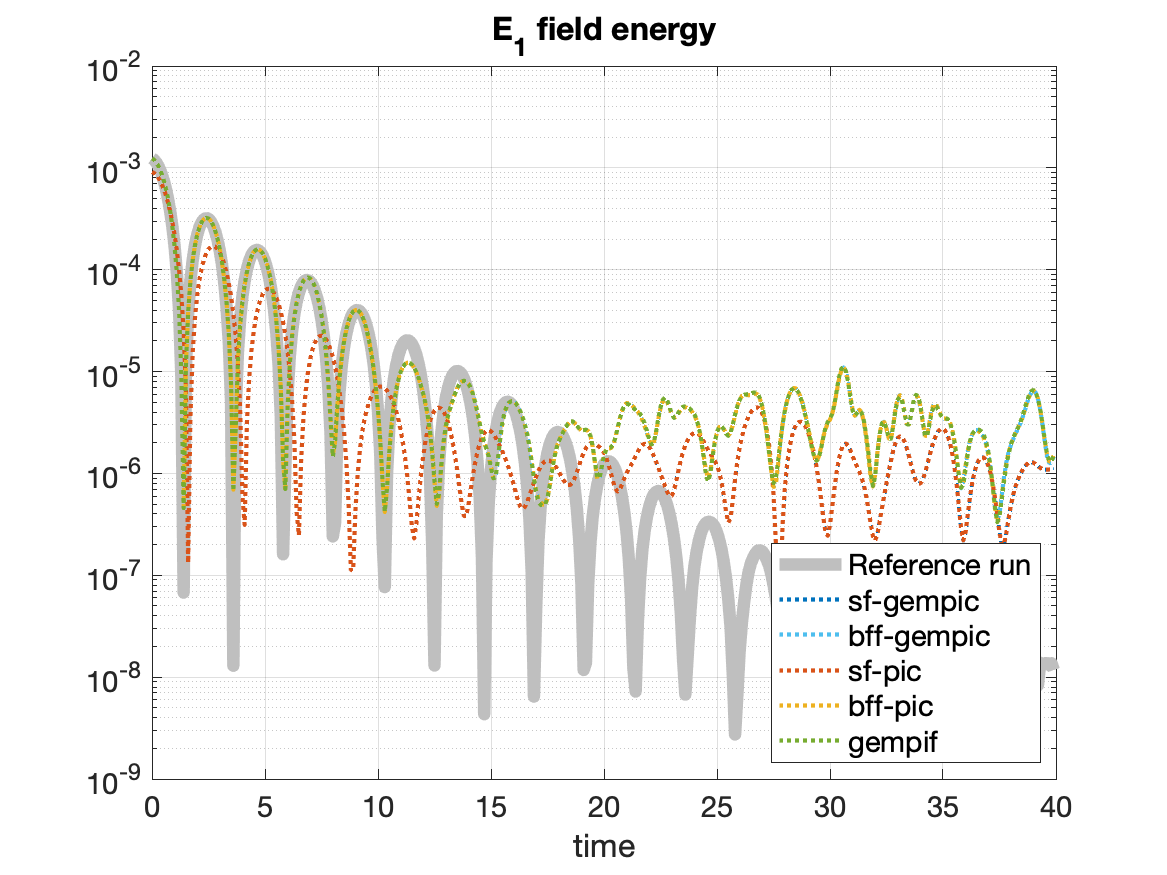}
    \subcaption{weak damping with an $M=8$ grid}
  \end{subfigure}
  \hspace{10pt}
  \begin{subfigure}{.48\textwidth}
    \centering
    \includegraphics[width=.9\linewidth]{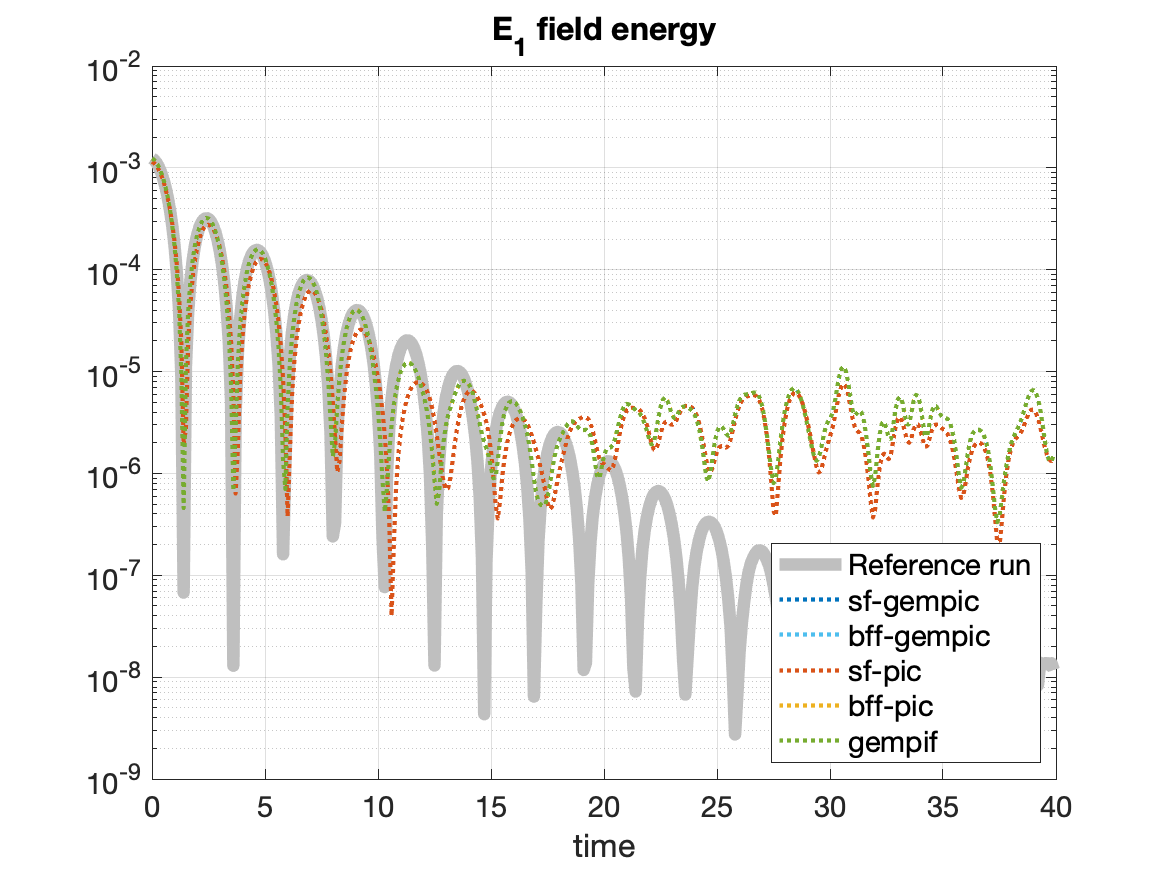}
    \subcaption{weak damping with an $M=16$ grid}
  \end{subfigure}
  \\
  \begin{subfigure}{.48\textwidth}
    \centering
    \includegraphics[width=.9\linewidth]{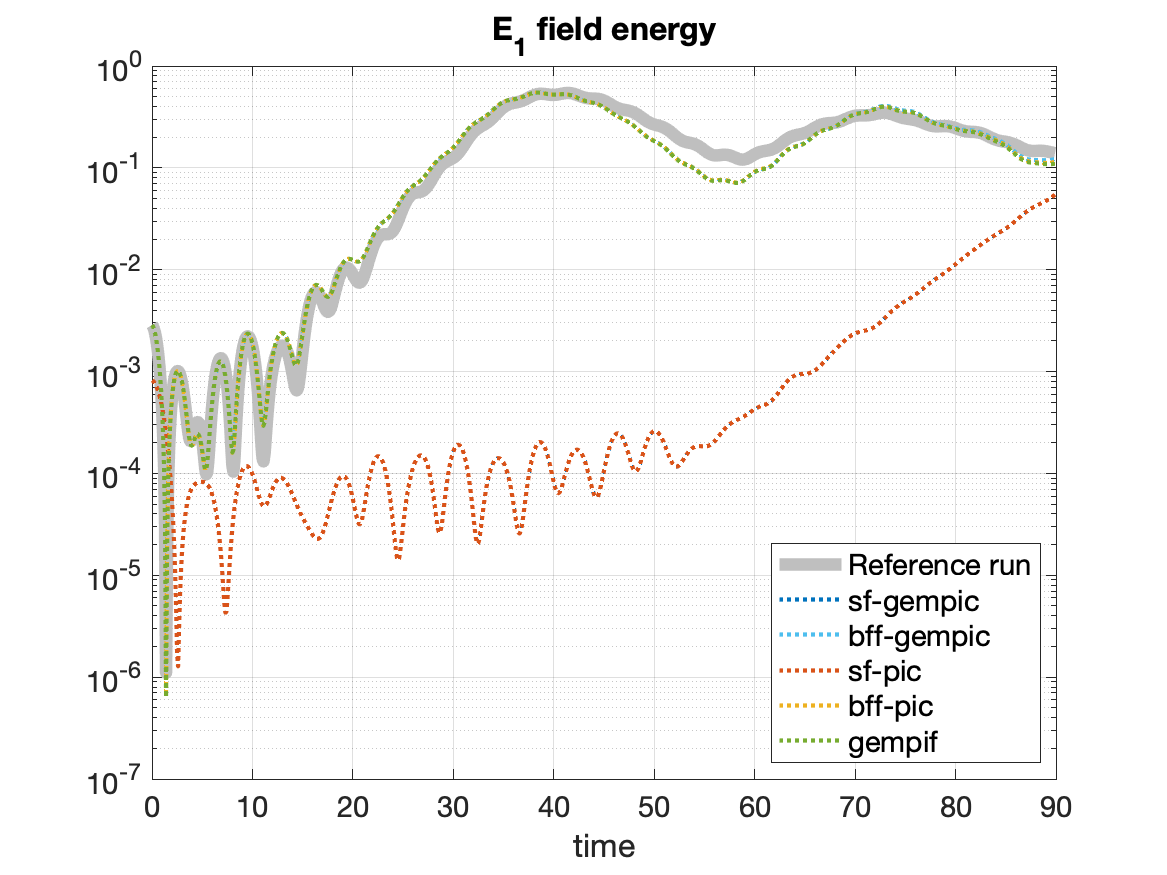}
    \subcaption{bump-on-tail instability with an $M=8$ grid}
  \end{subfigure}
  \hspace{10pt}
  \begin{subfigure}{.48\textwidth}
    \centering
    \includegraphics[width=.9\linewidth]{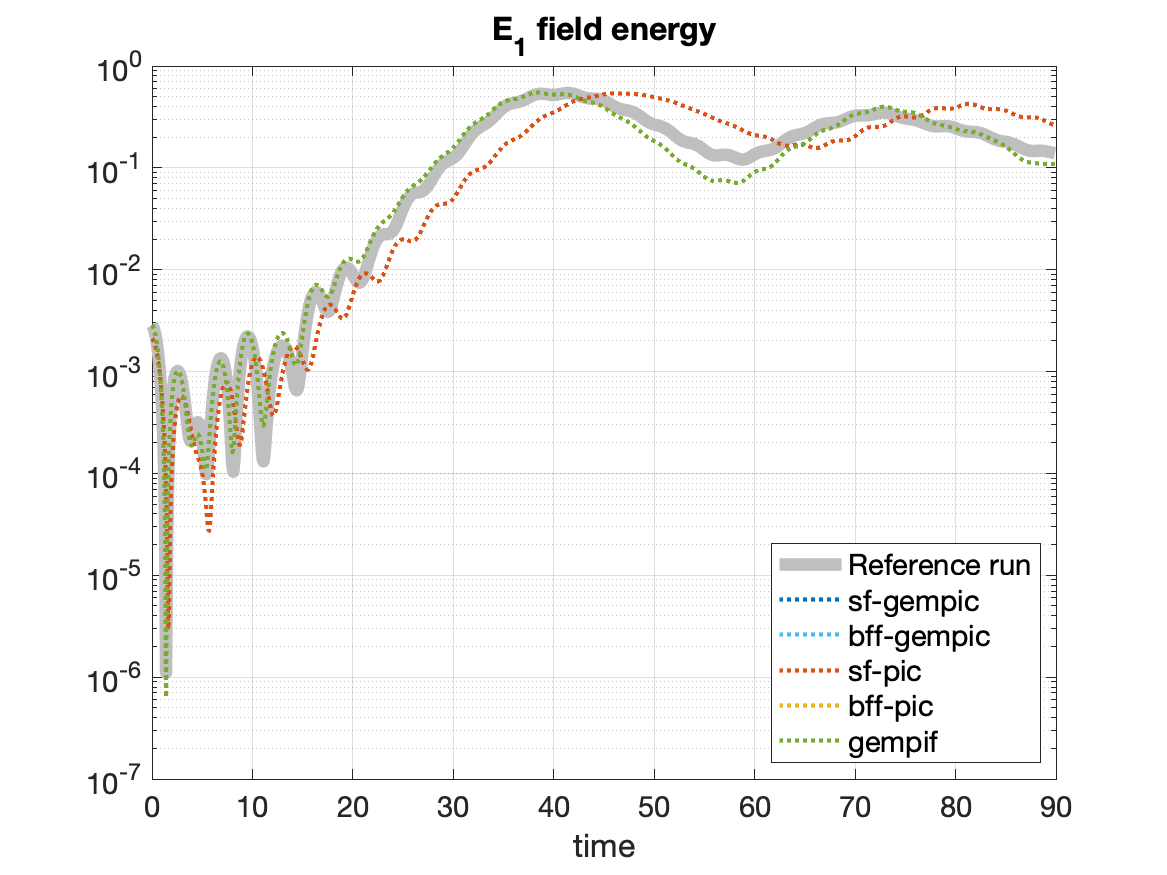}
    \subcaption{bump-on-tail instability with an $M=16$ grid}
  \end{subfigure}
  \caption{
  Landau damping and bump-on-tail instability test-cases, using the same parameters
  as in Figures~\ref{fig:landau_diags_deg3} and \ref{fig:instab_diags_deg3} and
  B-spline shapes of degree $\kappa=5$.
  \\$~$}
  \label{fig:diags_deg5}
\end{figure}

In order to see the effect of using high order splines, we plot in Figure~\ref{fig:diags_deg5} 
the energy curves obtained using splines of degree $\kappa = 5$ for some of the above test cases.
For the two filtered methods, this has no visible effect: they are still on top of the gridless method (which involves no splines).
For the two unfiltered methods this actually degrades the results, as the energy curves are more distant from the reference one than with $\kappa=3$.

\newpage

\subsection{Long-time conservation properties}
\label{sec:long}

\begin{figure}[tp]
  \begin{subfigure}{.48\textwidth}
    \centering
    \includegraphics[width=.9\linewidth]{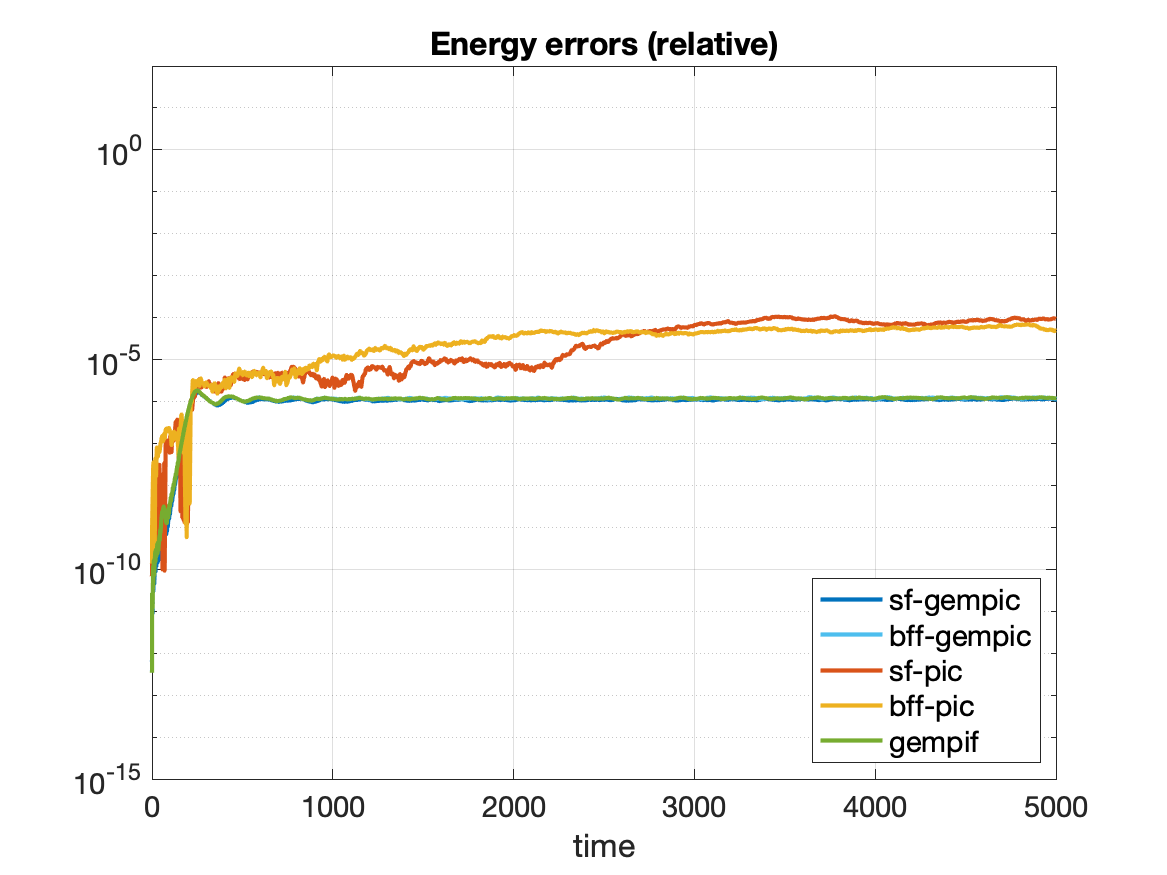}
    \subcaption{Energy errors with shapes of degree $\kappa=3$}
  \end{subfigure}
  \hspace{10pt}
  \begin{subfigure}{.48\textwidth}
    \centering
    \includegraphics[width=.9\linewidth]{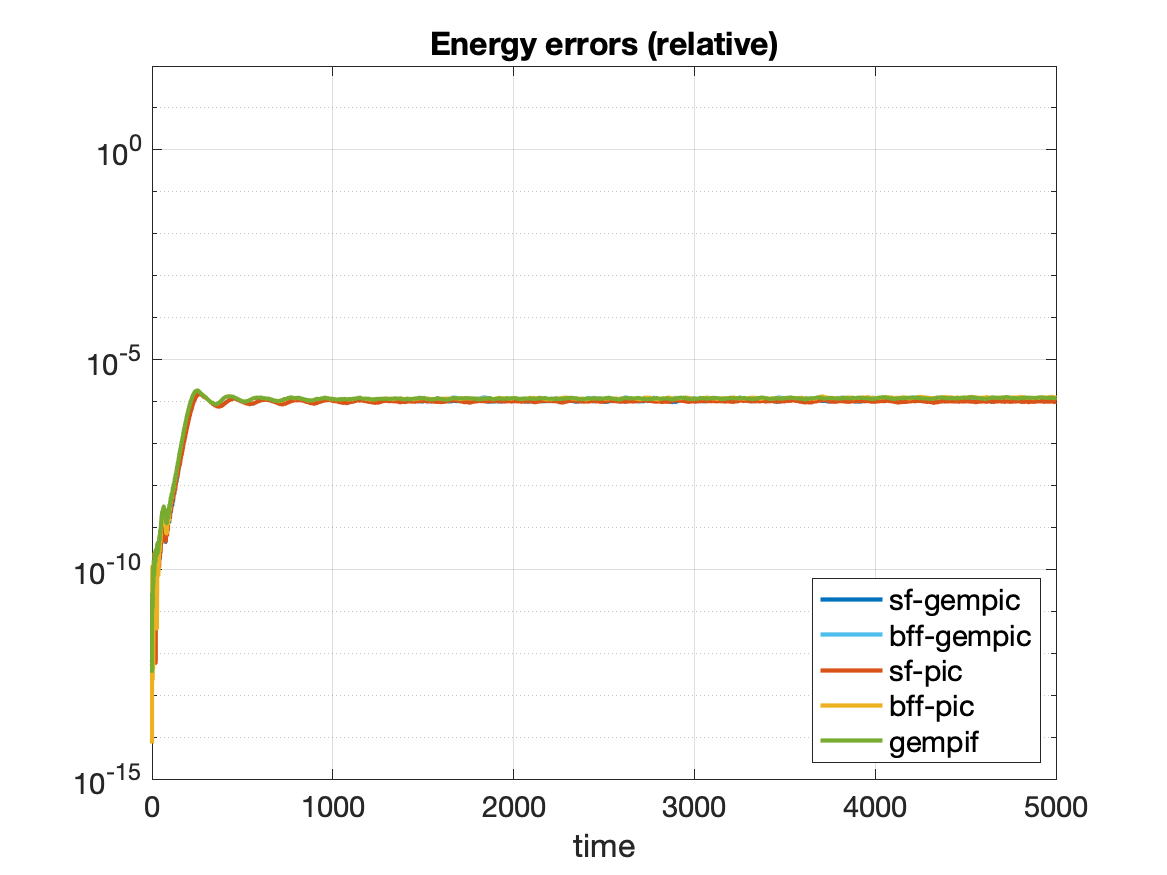}
    \subcaption{Energy errors with shapes of degree $\kappa=7$}
  \end{subfigure}
  \vspace{10pt}
  \\
  \begin{subfigure}{.48\textwidth}
    \centering
    \includegraphics[width=.9\linewidth]{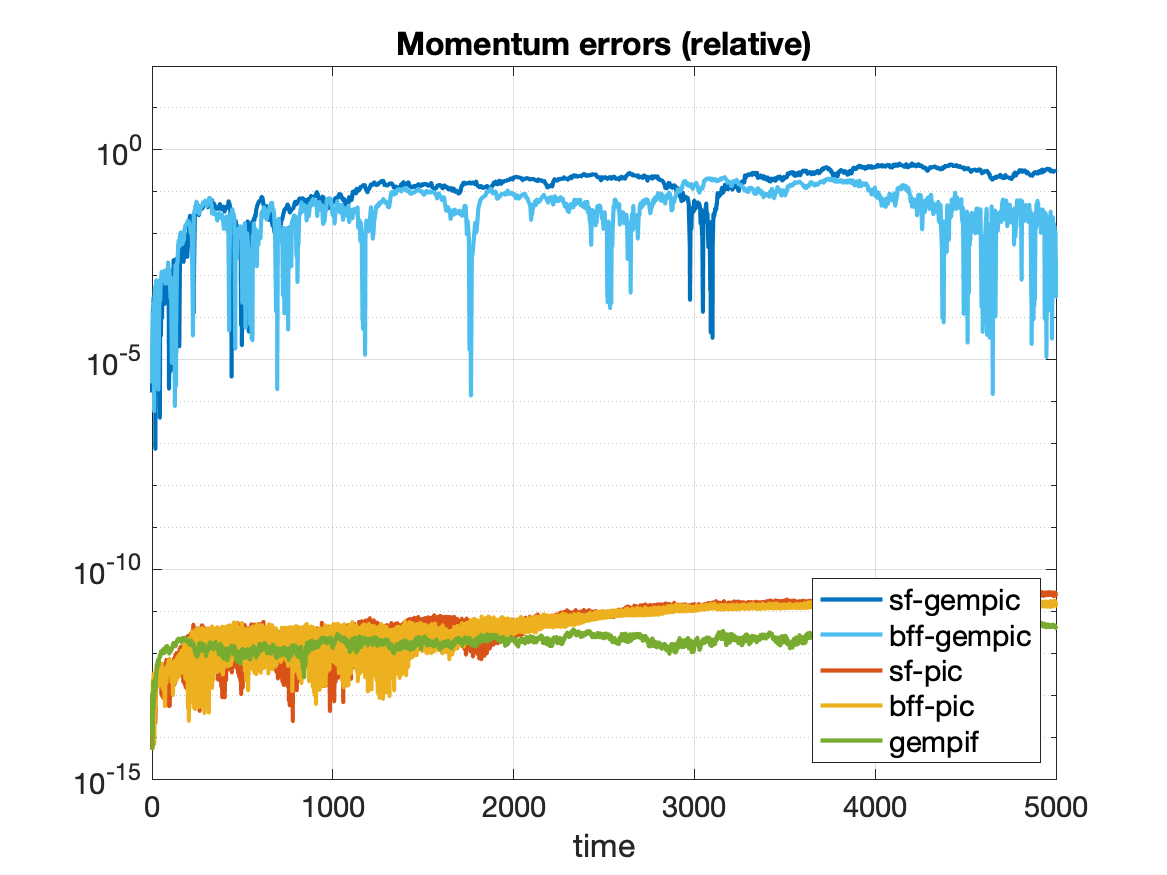}
    \subcaption{Momentum errors with shapes of degree $\kappa=3$}
  \end{subfigure}
  \hspace{10pt}
  \begin{subfigure}{.48\textwidth}
    \centering
    \includegraphics[width=.9\linewidth]{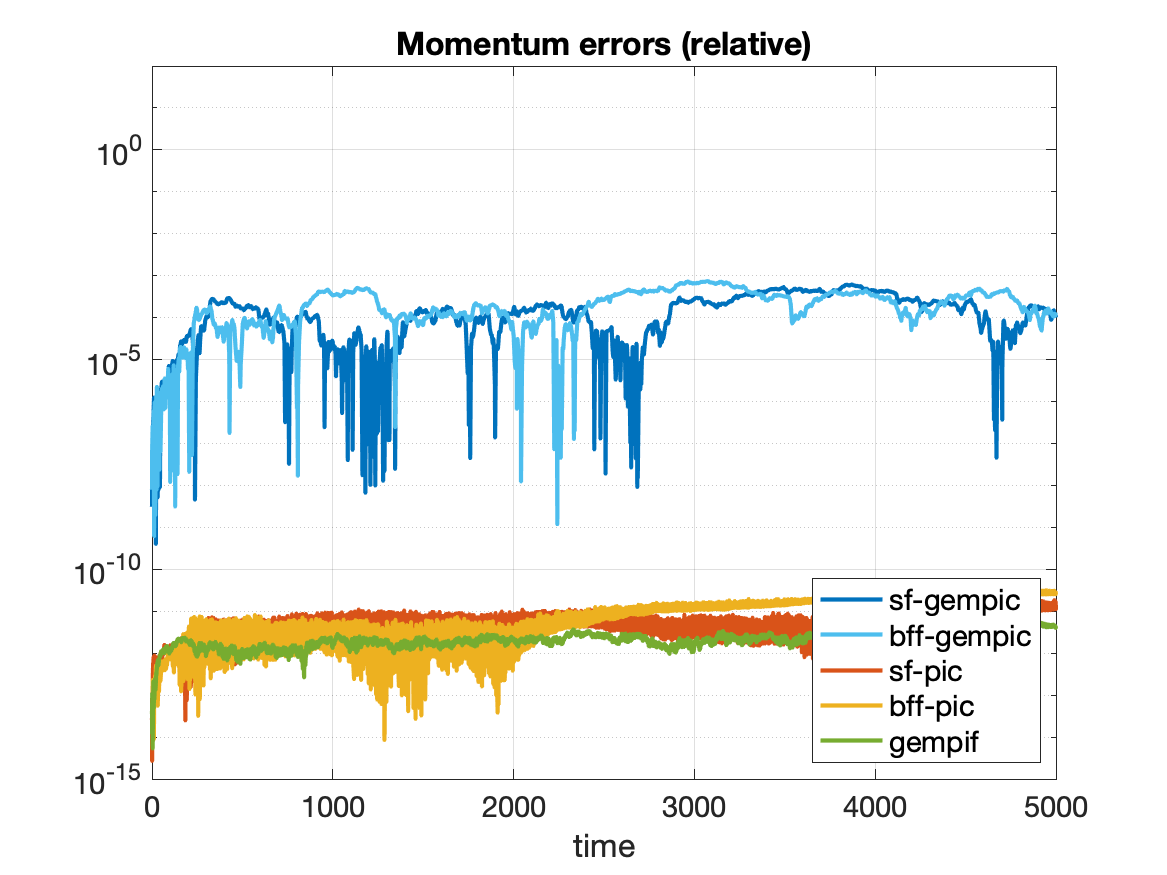}
    \subcaption{Momentum errors with shapes of degree $\kappa=7$}
  \end{subfigure}
  \vspace{10pt}
  \\
  \begin{subfigure}{.48\textwidth}
    \centering
    \includegraphics[width=.9\linewidth]{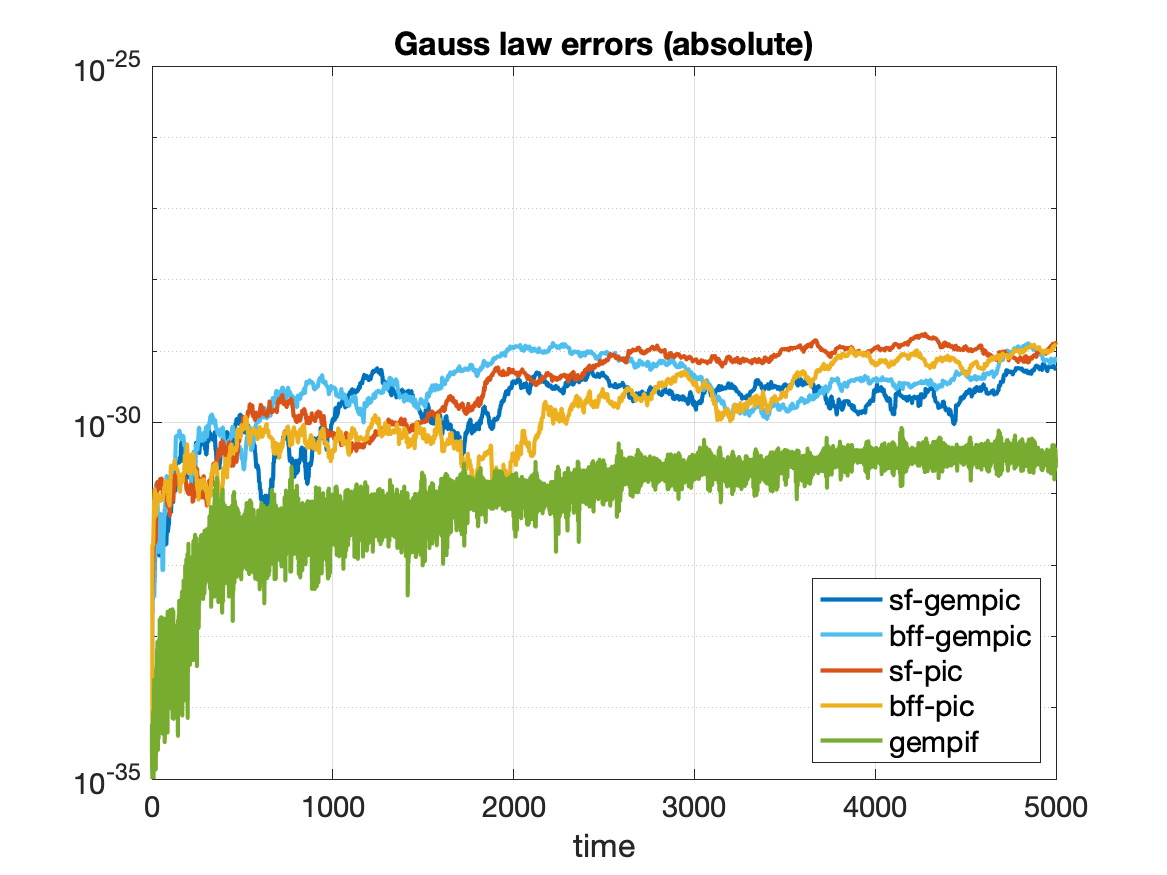}
    \subcaption{Gauss law errors with shapes of degree $\kappa=3$}
  \end{subfigure}
  \hspace{10pt}
  \begin{subfigure}{.48\textwidth}
    \centering
    \includegraphics[width=.9\linewidth]{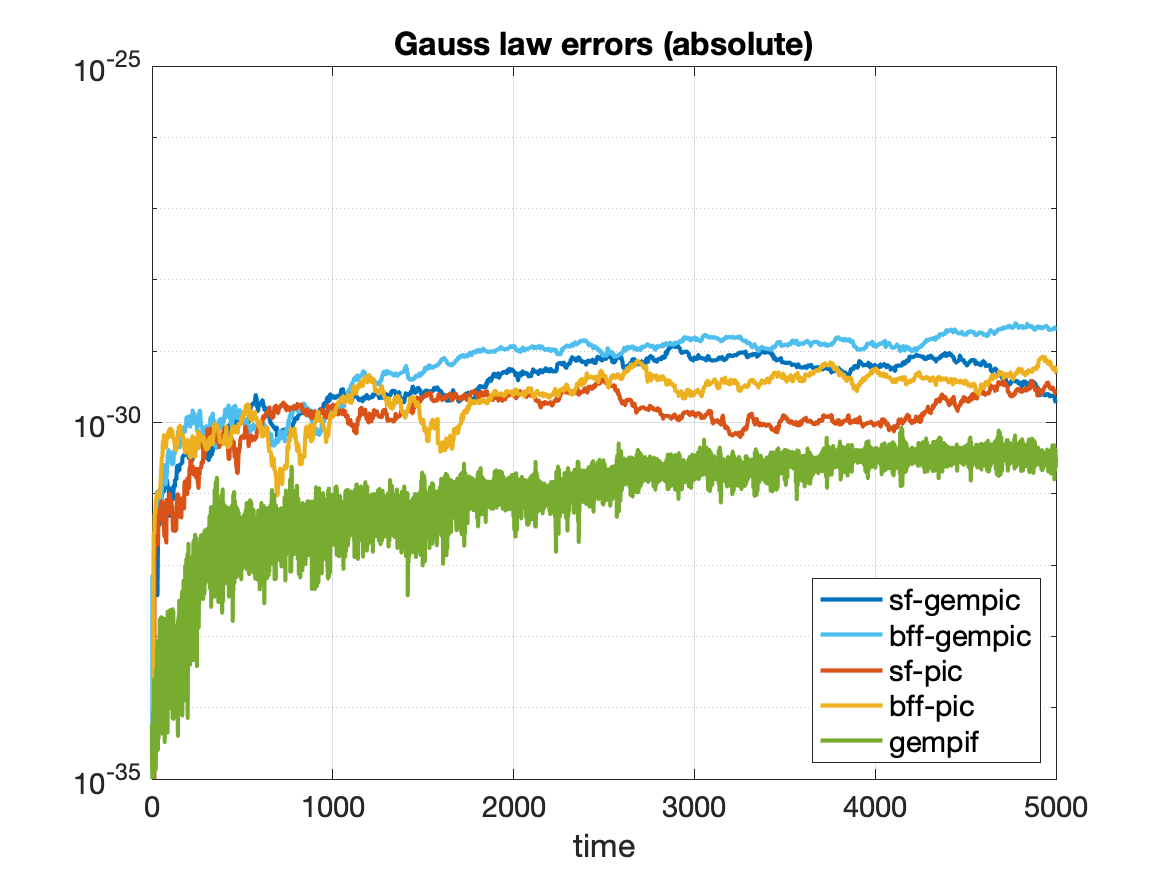}
    \subcaption{Gauss law errors with shapes of degree $\kappa=7$}
  \end{subfigure}
  \caption{
  Long-time conservation properties.
  Energy conservation errors (top), momentum conservation errors (middle) 
  and Gauss law errors (bottom) are shown for Weibel instability runs
  using $N = 10^4$ particles and $K = 3$ Fourier modes. Gridded (Fourier-Gempic and Fourier-PIC) methods use
  a DFT grid with $M = 16$ points corresponding to an oversampling parameter of $\mu = \frac{M}{2K+1} \approx 2.3$
  and B-spline shapes of degree $\kappa=3$ (left) or $\kappa=7$ (right).
  Similar curves have been observed for higher resolution runs with analogous oversampling parameter, such as
  $K = 13$ and $M = 64$.
  Compared to the runs in Fig.~\ref{fig:instab_diags_deg3}, the simulation range is ten times longer.
  }
  \label{fig:long_time_conservation} 
\end{figure}

In order to assess the long-time conservation properties of the different methods,
we show in Figure~\ref{fig:long_time_conservation} 
several errors for the Weibel instability test-case on a time range ten times longer than above.
The plotted errors are in energy conservation,
\begin{equation} \label{Err-H}
  {\rm Err}^n_\cH = \frac{\abs{\cH^n-\cH^0}}{\cH^0}
  \qquad \text{ with } \qquad
  \cH^n = \frac 12 \sum_{p = 1}^N m_p \abs{\pv_p^n}^2 + \frac 12 \int_{[0,L]} \Big(\abs{\bE^n_K(x)}^2 + \abs{\bB^n_K(x)}^2\Big) \dd x
\end{equation}
momentum conservation,
\begin{equation} \label{Err-P}
  {\rm Err}^n_\cP = \frac{\norm{\cP^n - \cP^0}_1}{\norm{\cP^0}_1}
  \qquad \text{ with } \qquad
  \cP^n = \sum_{p = 1}^N m_p \pv_p^n + \int_{[0,L]} \bE^n_K(x) \times \bB^n_K(x) \dd x
\end{equation}
and Gauss law
\begin{equation} \label{Err-G}
  {\rm Err}^n_G = \int_{[0,L]}\Bigabs{\Big(\Div \bE^n_K - \sum_{p=1}^N q_p \Pi^3 S_{\px_p^n}\Big)(x)}^2 \dd x.
\end{equation}
For each method we use the same numerical parameters as in Figure~\ref{fig:instab_diags_deg3},
and low-resolution runs with spline shapes of degree $\kappa = 3$ and $\kappa = 7$.

For the energy conservation we observe overall a good behavior for all the methods. Here the main
observation is that a clear separation is visible between the three geometric methods
and the two non-geometric ones. While the former show a very good stability of the energy conservation over
these long-time ranges, the latter sometimes display sensible growth in the energy error, characteristic of a less stable behavior.
Another interesting difference is that when the DFT grid is refined (right plots), the energy errors made by the non-geometric
methods get reduced, while those of the geometric methods hardly change.
This behavior is consistent with the backward error analysis which states that
Hamiltonian splitting time integrators of order $q$ preserve exactly a modified energy which approximates
the exact one with the same order.

As for the momentum conservation, the first observation is that the curves confirm the exact conservation of the GEMPIF method and the two
non-geometric schemes (SF-PIC and BFF-PIC). For the Fourier-GEMPIC methods the conservation is only approximate, however we observe
a very good stability over these long-time simulations. We also observe that the accuracy to which the momentum is preserved is improved
by refining grid or increasing the spline order.

Finally the Gauss error curves confirm the exact charge conserving property of all methods.

\newpage

\subsection{Convergence studies}

We finally study how the conservation of energy and momentum is improved by refining the numerical parameters.

\begin{figure}[tp]
  \begin{subfigure}{.48\textwidth}
    \centering
    \includegraphics[width=.9\linewidth]{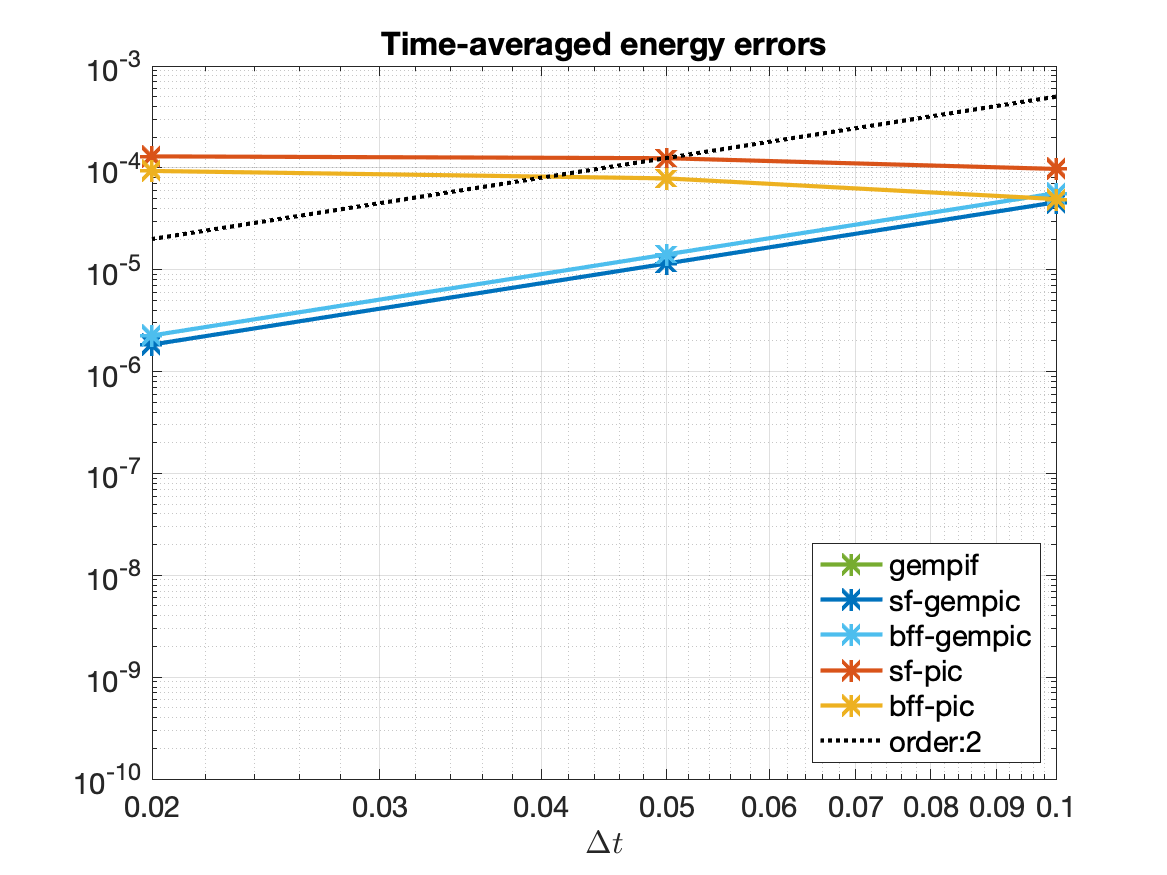}
    \subcaption{$M = 8$ and $\kappa=3$}
  \end{subfigure}%
  \hspace{10pt}
  \begin{subfigure}{.48\textwidth}
    \centering
    \includegraphics[width=.9\linewidth]{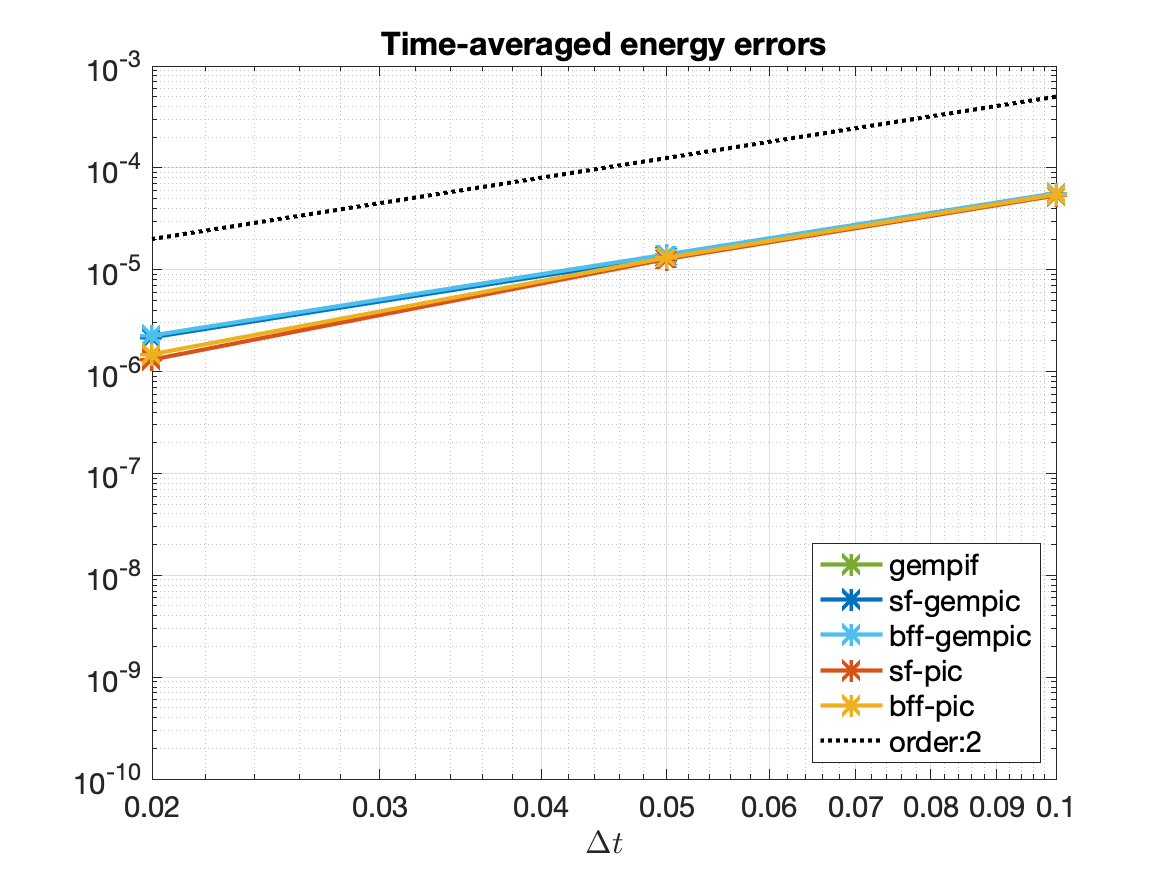}
    \subcaption{$M = 16$ and $\kappa=3$}
  \end{subfigure}
  \vspace{10pt}%
  \\
  \begin{subfigure}{.48\textwidth}
    \centering
    \includegraphics[width=.9\linewidth]{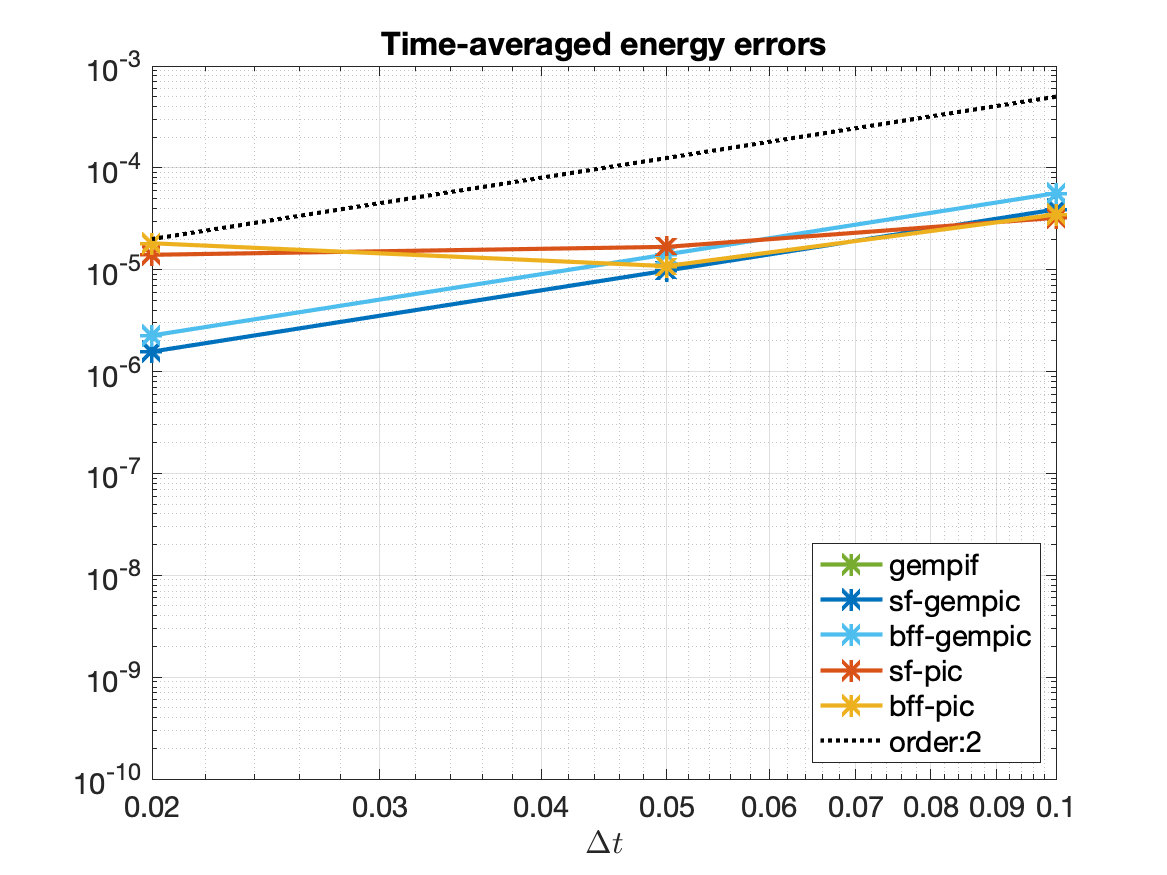}
    \subcaption{$M = 8$ and $\kappa=5$}
  \end{subfigure}%
  \hspace{10pt}
  \begin{subfigure}{.48\textwidth}
    \centering
    \includegraphics[width=.9\linewidth]{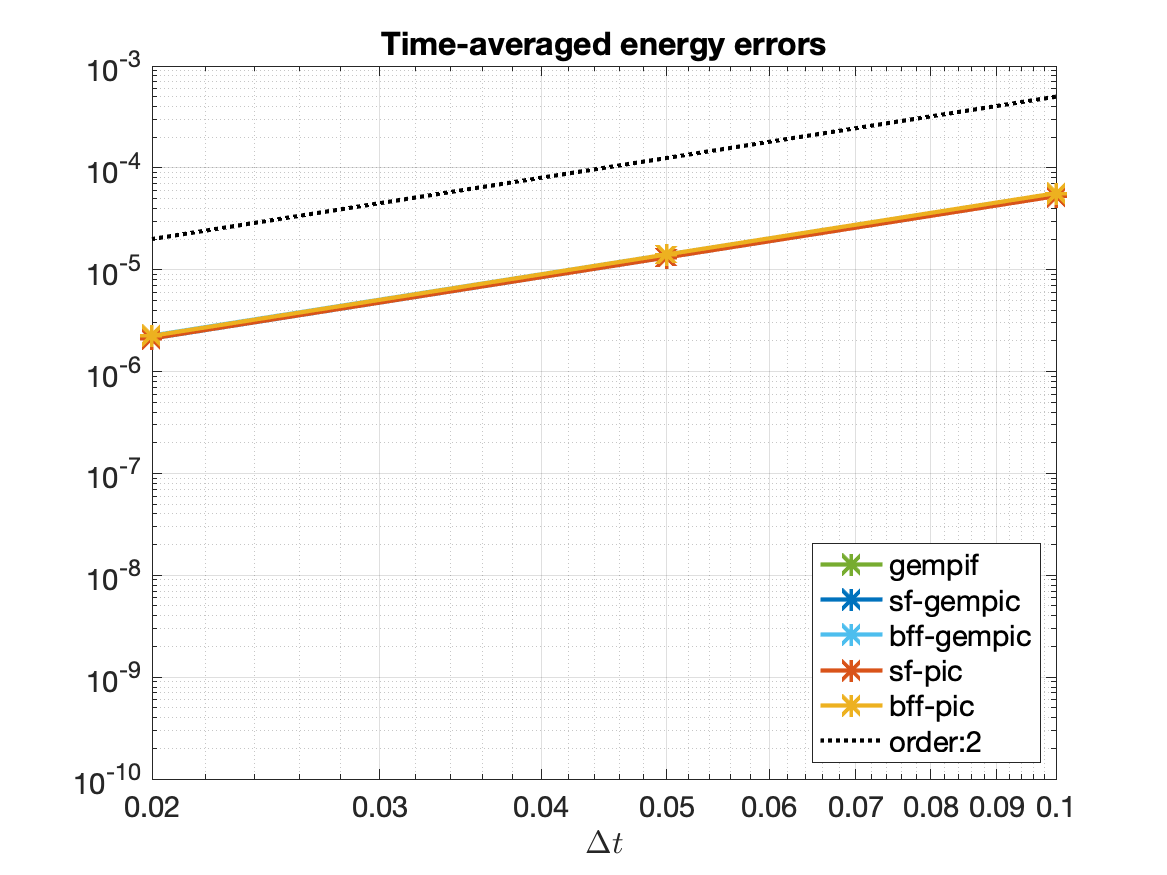}
    \subcaption{$M = 16$ and $\kappa=5$}
  \end{subfigure}%
  \caption{Energy error convergence: time-averaged energy conservation errors as a function of the time step $\Dt$,
  for a Weibel instability test-case with $K = 3$ Fourier modes, $N = 5\cdot 10^4$ particles and various grids and spline degrees, as indicated.
  Results reported here use a second-order (Strang) time scheme.
  }
  \label{fig:weibel_energy_error_order2}
\end{figure}

\begin{figure}[tp]
  \begin{subfigure}{.48\textwidth}
    \centering
    \includegraphics[width=.9\linewidth]{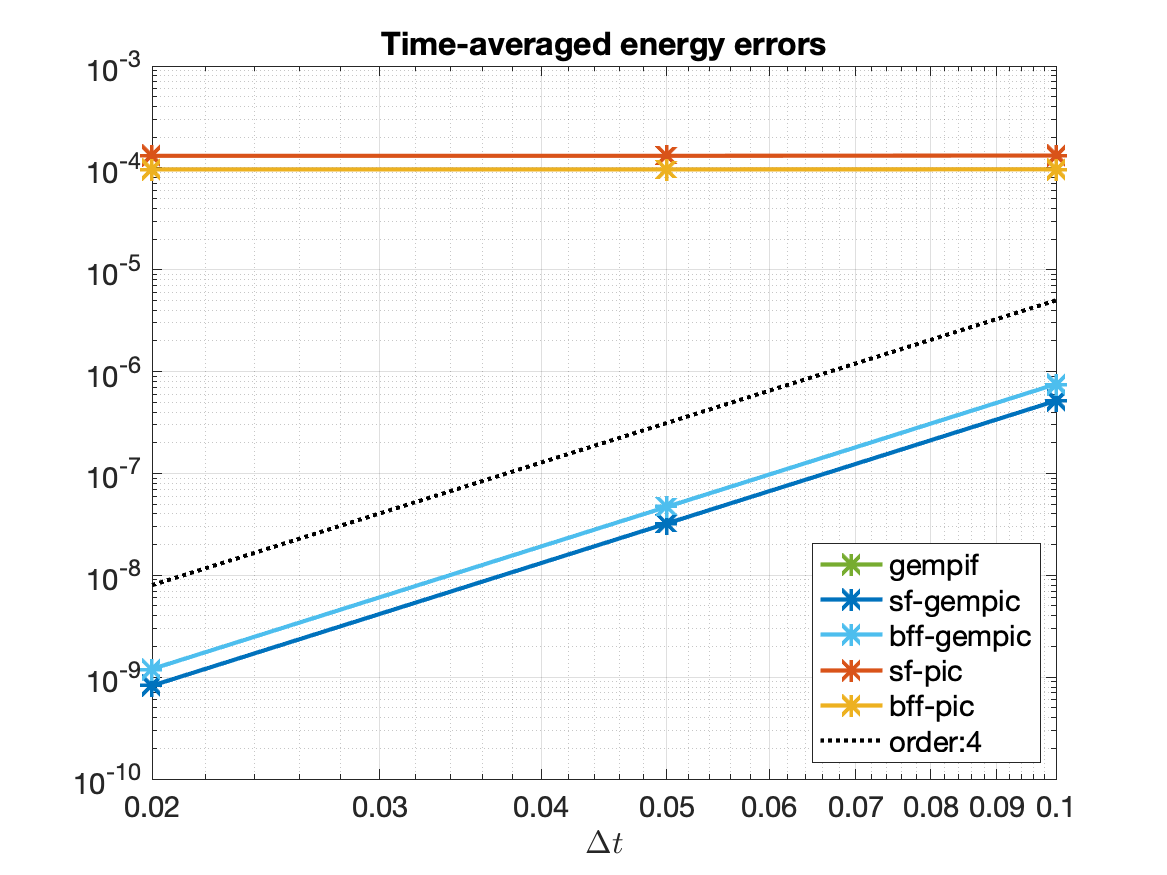}
    \subcaption{$M = 8$ and $\kappa=3$}
  \end{subfigure}%
  \hspace{10pt}
  \begin{subfigure}{.48\textwidth}
    \centering
    \includegraphics[width=.9\linewidth]{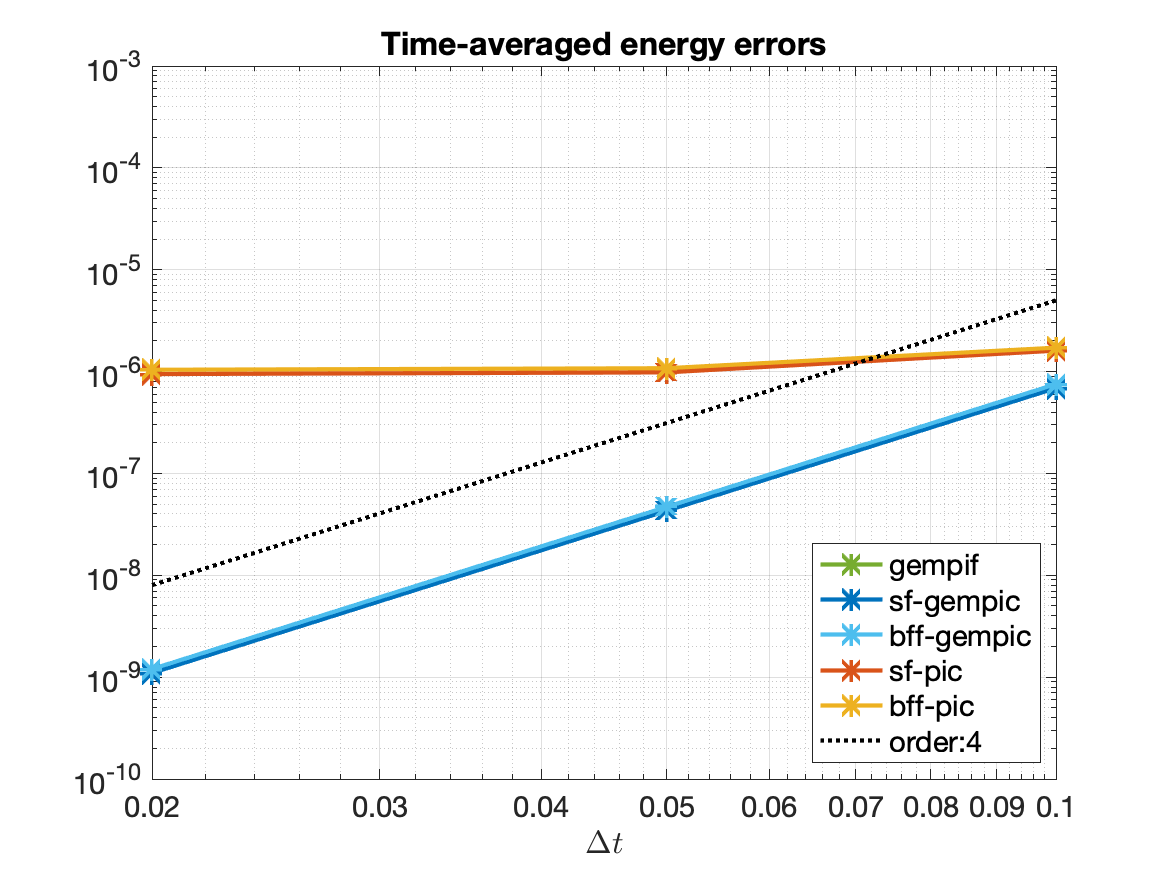}
    \subcaption{$M = 16$ and $\kappa=3$}
  \end{subfigure}
  \vspace{10pt}%
  \\
  \begin{subfigure}{.48\textwidth}
    \centering
    \includegraphics[width=.9\linewidth]{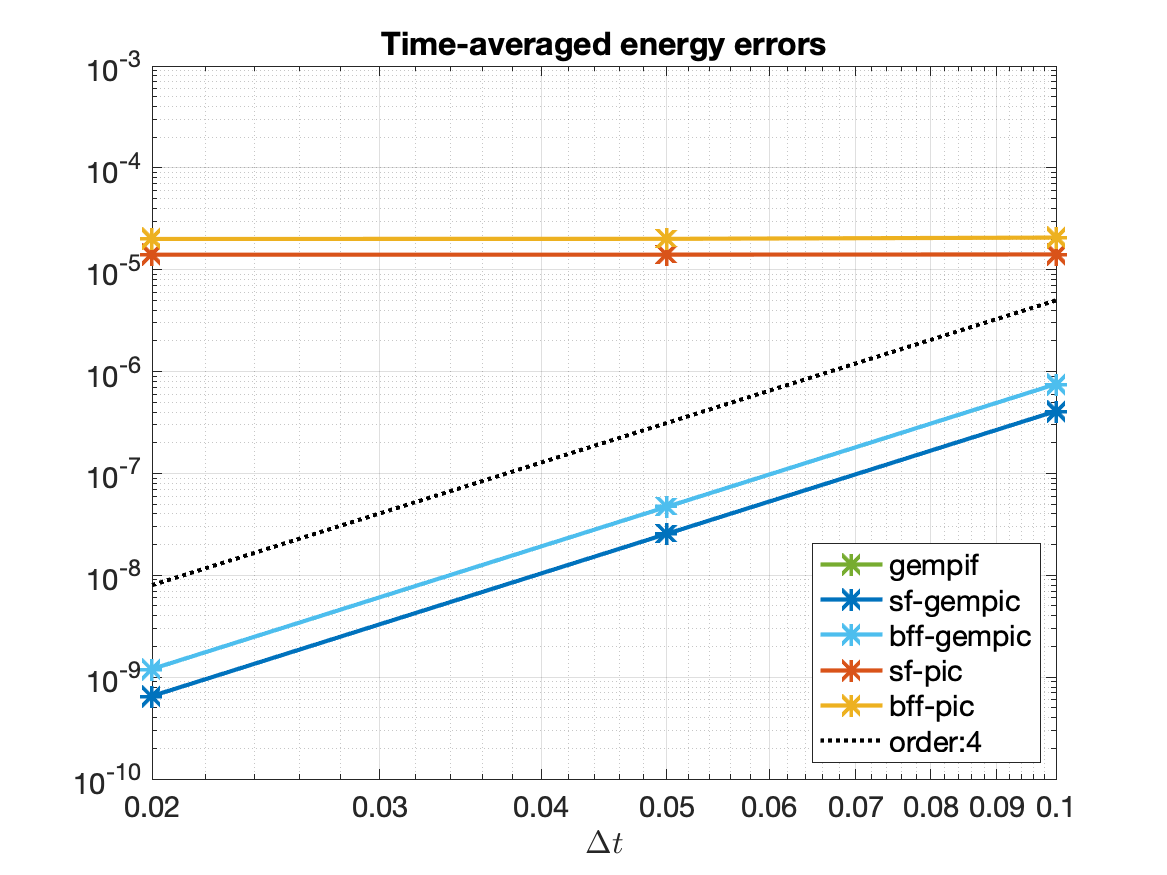}
    \subcaption{$M = 8$ and $\kappa=5$}
  \end{subfigure}%
  \hspace{10pt}
  \begin{subfigure}{.48\textwidth}
    \centering
    \includegraphics[width=.9\linewidth]{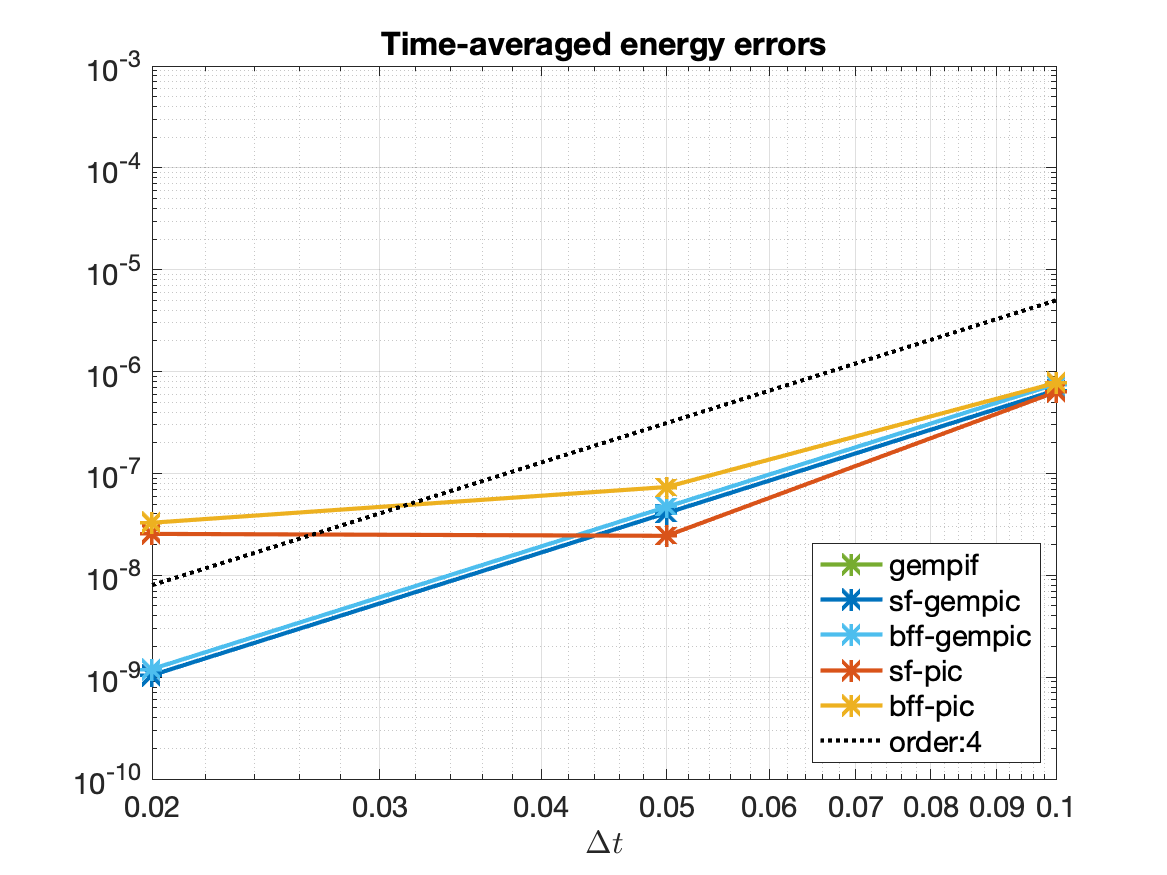}
    \subcaption{$M = 16$ and $\kappa=5$}
  \end{subfigure}%
  \caption{Energy error convergence curves, using the same parameters as in Figure~\ref{fig:weibel_energy_error_order2},
  and a fourth-order (Suzuki-Yoshida) time scheme.
  $$~$$
  }
  \label{fig:weibel_energy_error_order4}
\end{figure}

In Figure~\ref{fig:weibel_energy_error_order2} and \ref{fig:weibel_energy_error_order4}
we plot the time-averaged energy errors
$
\frac {1}{N_t} \sum_{n = 0}^{N_t-1} {\rm Err}^n_\cH,
$
for the five methods (using the same color key as above)
as a function of the time step, for various DFT grids and spline degrees.
Results shown in Figure~\ref{fig:weibel_energy_error_order2} are obtained with
the second order Strang scheme \eqref{S2}, while those in Figure~\ref{fig:weibel_energy_error_order4}
correspond to the fourth-order Suzuki-Yoshida scheme \eqref{S4}. For an easier comparison,
both figures use the same scale.

For the non-geometric methods we observe an improvement in the energy errors when the time step decreases,
but the convergence is limited by the resolution of the grid and the degree of the splines.
This artifact is not present with the geometric methods, where the convergence of the energy errors holds
almost independently of the grid resolution and spline degree.
This confirms the behavior already observed in the long-time runs, see Figure~\ref{fig:long_time_conservation}.
It is also in strong agreement with the backward error analysis,
which predicts a convergence of the energy errors of the same order as the time scheme.
Here the runs correspond to the Weibel instability with a moderate time range ($T=500$),
but results obtained with other test cases have showed similar behavior.

\begin{figure}[tp]
  \begin{subfigure}{.48\textwidth}
    \centering
    \includegraphics[width=.9\linewidth]{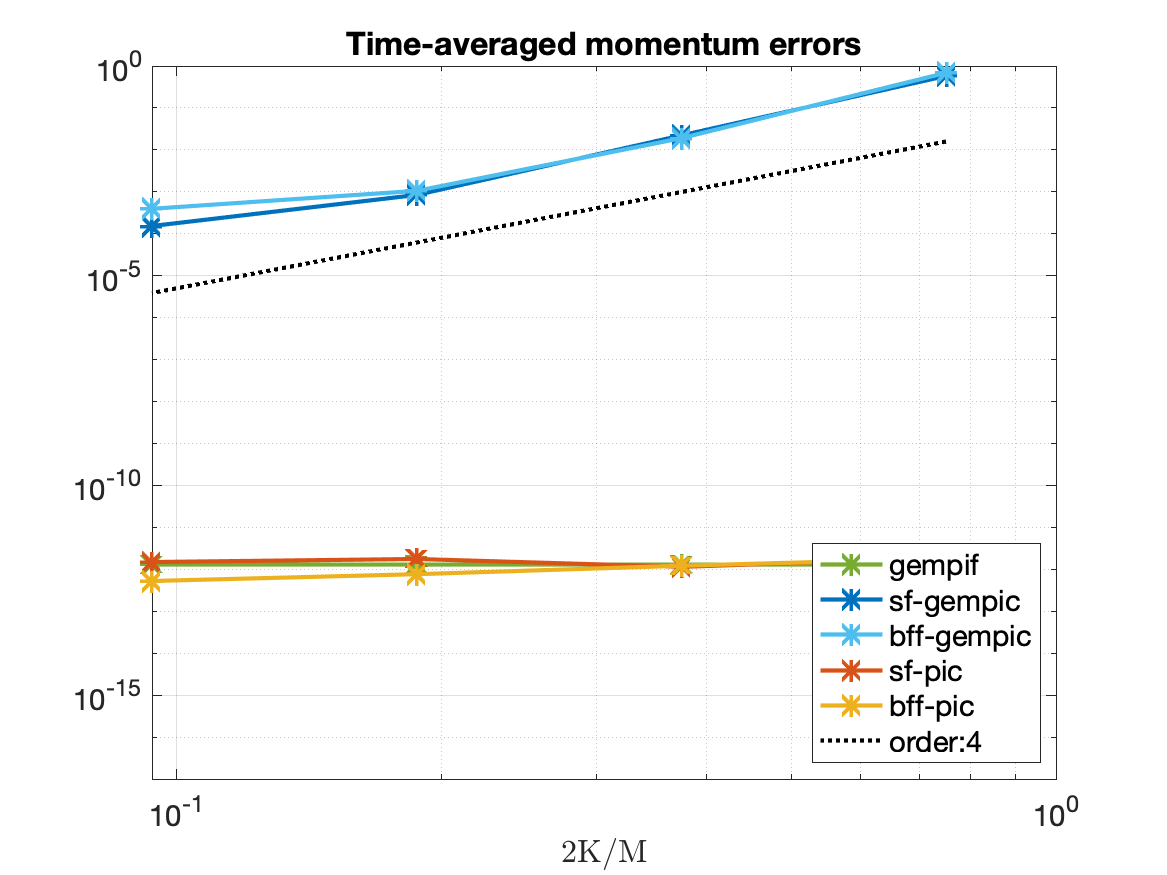}
    \subcaption{Spline degree $\kappa=3$}
  \end{subfigure}%
  \hspace{10pt}
  \begin{subfigure}{.48\textwidth}
    \centering
    \includegraphics[width=.9\linewidth]{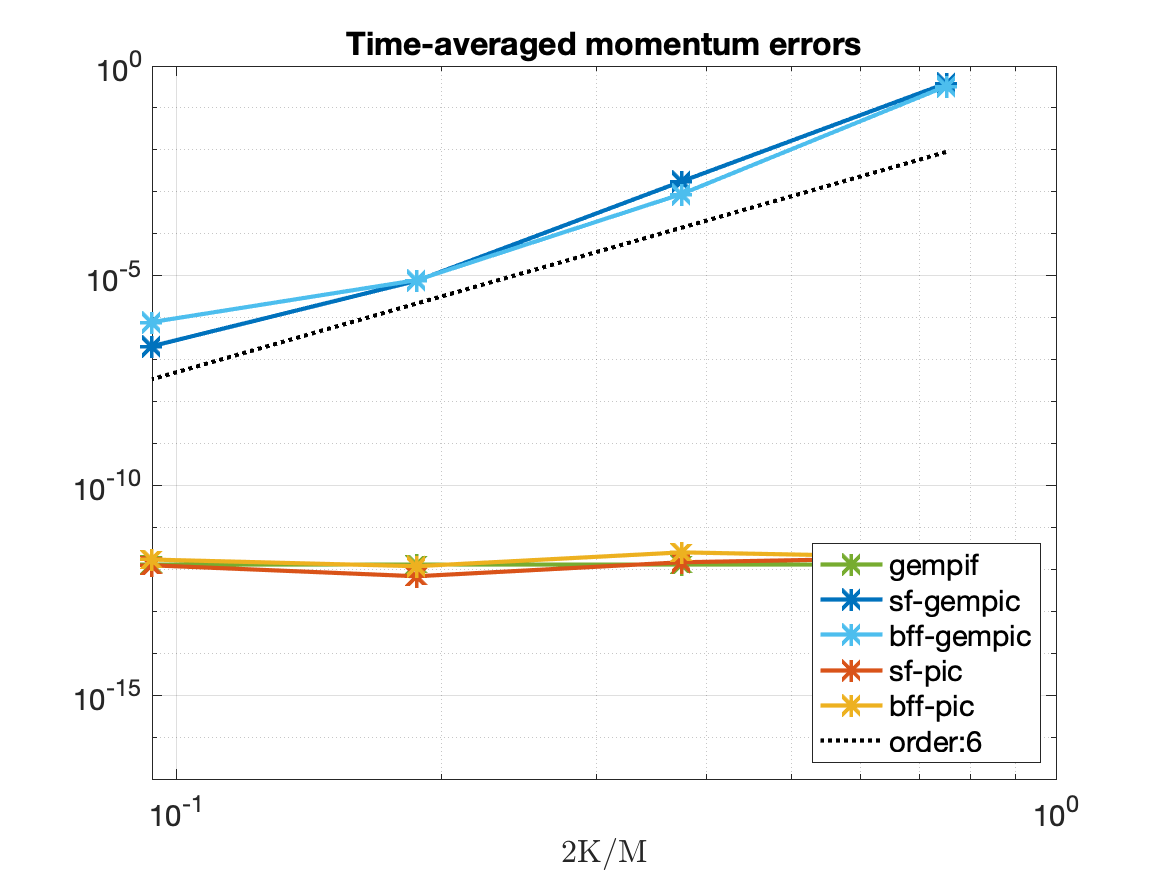}
    \subcaption{Spline degree $\kappa=5$}
  \end{subfigure}
  \vspace{10pt}%
  \\
  \begin{subfigure}{.48\textwidth}
    \centering
    \includegraphics[width=.9\linewidth]{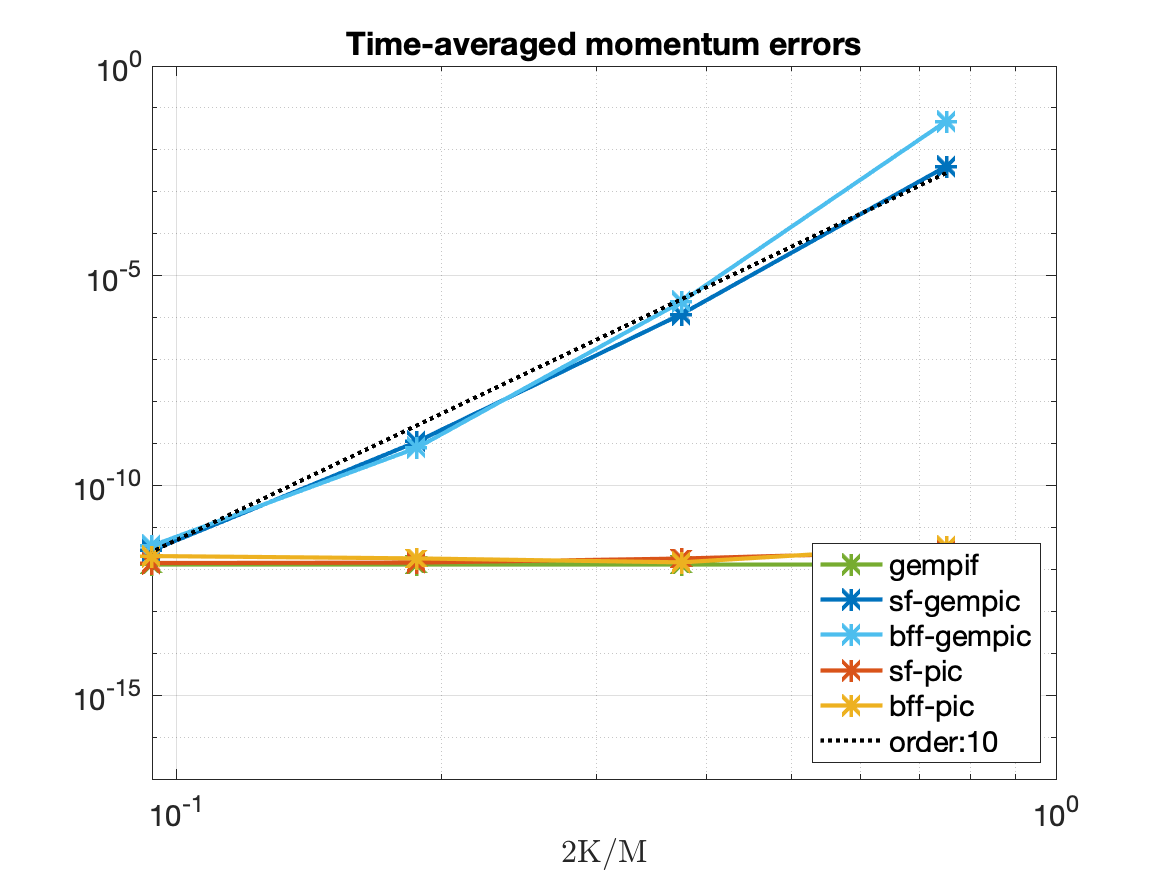}
    \subcaption{Spline degree $\kappa=9$}
  \end{subfigure}%
  \hspace{10pt}
  \begin{subfigure}{.48\textwidth}
    \centering
    \includegraphics[width=.9\linewidth]{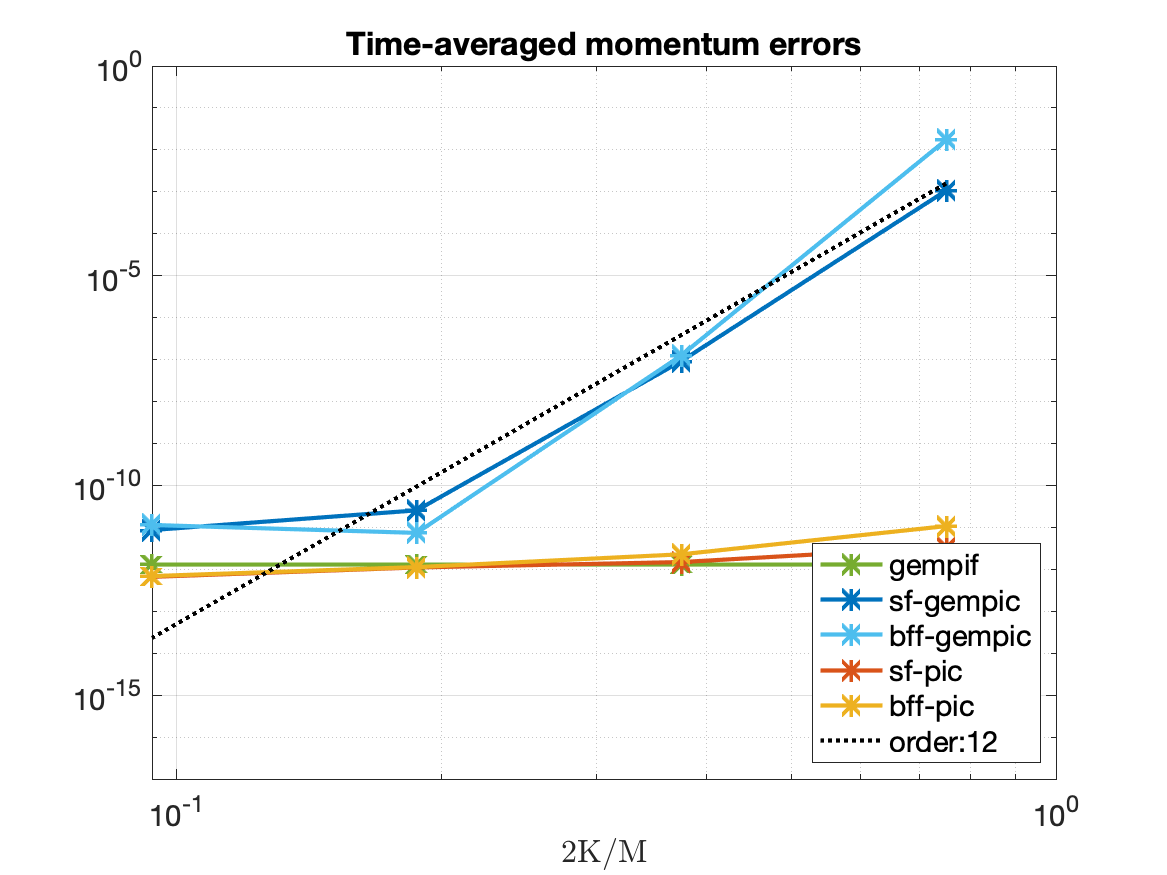}
    \subcaption{Spline degree $\kappa=11$}
  \end{subfigure}%
  \caption{Momentum error convergence: time-averaged momentum conservation errors as a function of the ratio $2K/M$,
  using the Weibel instability test-case with $K = 3$ Fourier modes, $N = 5\cdot 10^4$ particles and spline degrees as indicated.
  Results reported here use a fourth-order (Suzuki-Yoshida) time scheme.}
  \label{fig:weibel_K3_mean_momentum}
\end{figure}

In Figure~\ref{fig:weibel_K3_mean_momentum} we plot the time-averaged
momentum errors
$
\frac {1}{N_t} \sum_{n = 0}^{N_t-1} {\rm Err}^n_\cP,
$
for the five methods (again with the same color key),
as a function of the ratio $2K/M \approx 1/\mu$, for various spline degrees.
For the GEMPIF and Fourier-PIC methods the error is close to machine accuracy, since
the methods are exactly momentum conserving.
For the Fourier-GEMPIC methods we observe that in every case they converge to 0, with a rate close to $\cO(K/M)^{\kappa+1}$.
Here the test-case is again the Weibel instability with $K=3$ modes as in Section~\ref{sec:qualitative},
but the same convergence behavior was observed with other test-cases and a higher number of Fourier modes.

\section{Acknowledgements}
This work has been carried out within the framework of the EUROfusion Consortium and has received funding from the Euratom research and training programme 2014-2018 and 2019-2020 under grant agreement No 633053. The views and opinions expressed herein do not necessarily reflect those of the European Commission.

\newpage

\bibliographystyle{plainnat}
\bibliography{fourier_gempic_arxiv}

\end{document}